\documentclass[twocolumn,twocolappendix]{aastex631}

\received{January 23, 2023}
\revised{March 20, 2023}
\accepted{April 3, 2023}

\submitjournal{ApJ}

\shorttitle{Outflows from Star Forming Dwarf Galaxies}
\shortauthors{Aravindan et al.}

\graphicspath{{./}{figures/}}

\begin{document}

\title{A Comparison of Outflow Properties in AGN Dwarfs vs. Star Forming Dwarfs}

\correspondingauthor{Archana Aravindan}
\email{aarav005@ucr.edu}

\author[0000-0001-7578-2412]{Archana Aravindan}
\affiliation{Department of Physics and Astronomy, University of California, Riverside, 900 University Ave, Riverside CA 92521, USA}

\author[0000-0003-3762-7344]{Weizhe Liu}
\affiliation{Steward Observatory, University of Arizona, 933 N Cherry Ave, Tucson, AZ 85719}

\author[0000-0003-4693-6157]{Gabriela Canalizo}
\affiliation{Department of Physics and Astronomy, University of California, Riverside, 900 University Ave, Riverside CA 92521, USA}

\author[0000-0002-3158-6820]{Sylvain Veilleux}
\affiliation{Department of Astronomy, University of Maryland, College Park, MD 20742, USA}
\affiliation{Joint Space-Science Institute, University of Maryland, College Park, MD 20742, USA}

\author[0000-0002-4375-254X]{Thomas Bohn}
\affiliation{Hiroshima Astrophysical Science Center, Hiroshima University, 1-3-1 Kagamiyama, Higashi-Hiroshima, Hiroshima 739-8526, Japan}

\author[0000-0003-3432-2094]{Remington O. Sexton}
\affiliation{George Mason University, Department of Physics and Astronomy, MS3F3, 4400 University Drive, Fairfax, VA 22030, USA}
\affiliation{U.S. Naval Observatory, 3450 Massachusetts Avenue NW, Washington, DC 20392-5420, USA}

\author[0000-0002-1608-7564]{David S.N. Rupke}
\affiliation{Department of Physics, Rhodes College, Memphis, TN 38112, USA}

\author[0000-0002-1912-0024]{Vivian U}
\affiliation{Department of Physics and Astronomy, 4129 Frederick Reines Hall, University of California, Irvine, CA 92697, USA}

\begin{abstract}
Feedback likely plays a crucial role in resolving discrepancies between observed and theoretical predictions of dwarf galaxy properties. Stellar feedback was once believed to be sufficient to explain these discrepancies, but it has thus far failed to fully reconcile theory and observations. The recent discovery of energetic galaxy-wide outflows in dwarf galaxies hosting Active Galactic Nuclei (AGN) suggests that AGN feedback may have a larger role in the evolution of dwarf galaxies than previously suspected. In order to assess the relative importance of stellar versus AGN feedback in these galaxies, we perform a detailed Keck/KCWI optical integral field spectroscopic study of a sample of low-redshift star-forming (SF) dwarf galaxies that show outflows in ionized gas in their SDSS spectra. We characterize the outflows and compare them to observations of AGN-driven outflows in dwarfs. We find that SF dwarfs have outflow components that have comparable widths (W$_{80}$) to those of outflows in AGN dwarfs, but are much less blue-shifted, indicating that SF dwarfs have significantly slower outflows than their AGN counterparts. The outflows in SF dwarfs are spatially resolved and significantly more extended than those in AGN dwarfs. The mass loss rates, momentum and energy rates of SF-driven outflows are much lower than those of AGN-driven outflows. Our results indicate that AGN feedback in the form of gas outflows may play an important role in dwarf galaxies and should be considered along with SF feedback in models of dwarf galaxy evolution. 
\keywords{Dwarf Galaxies (416), Extragalactic Astronomy (506), Stellar winds (1636), Active galactic nuclei (16), Galaxy evolution (594)}
\end{abstract}

\section{Introduction} \label{sec:intro}
Dwarf galaxies are some of the most abundant objects in the universe with stellar masses $< \sim$ 10$^{10}$ M$_{\odot}$. They form an important part of cosmological simulations used to study the early universe. However, observations of dwarf galaxies have posed several challenges, particularly because some of these observations are at odds with the $\Lambda$CDM models. Examples of these discrepancies include the core-cusp problem \citep{1991ApJ...378..496D, 1996ApJ...462..563N}, the missing satellites problem \citep{1999ApJ...522...82K, 1999ApJ...524L..19M}, the too-big-to-fail problem \citep{2011MNRAS.415L..40B}, among several others (For a comprehensive review on the challenges for cosmological models of dwarf galaxies see \cite{2022NatAs...6..897S}). The most widely accepted solution to all these issues has been to incorporate the effects of feedback in the models.

\begin{figure}[ht!]
\includegraphics[width=0.45\textwidth]{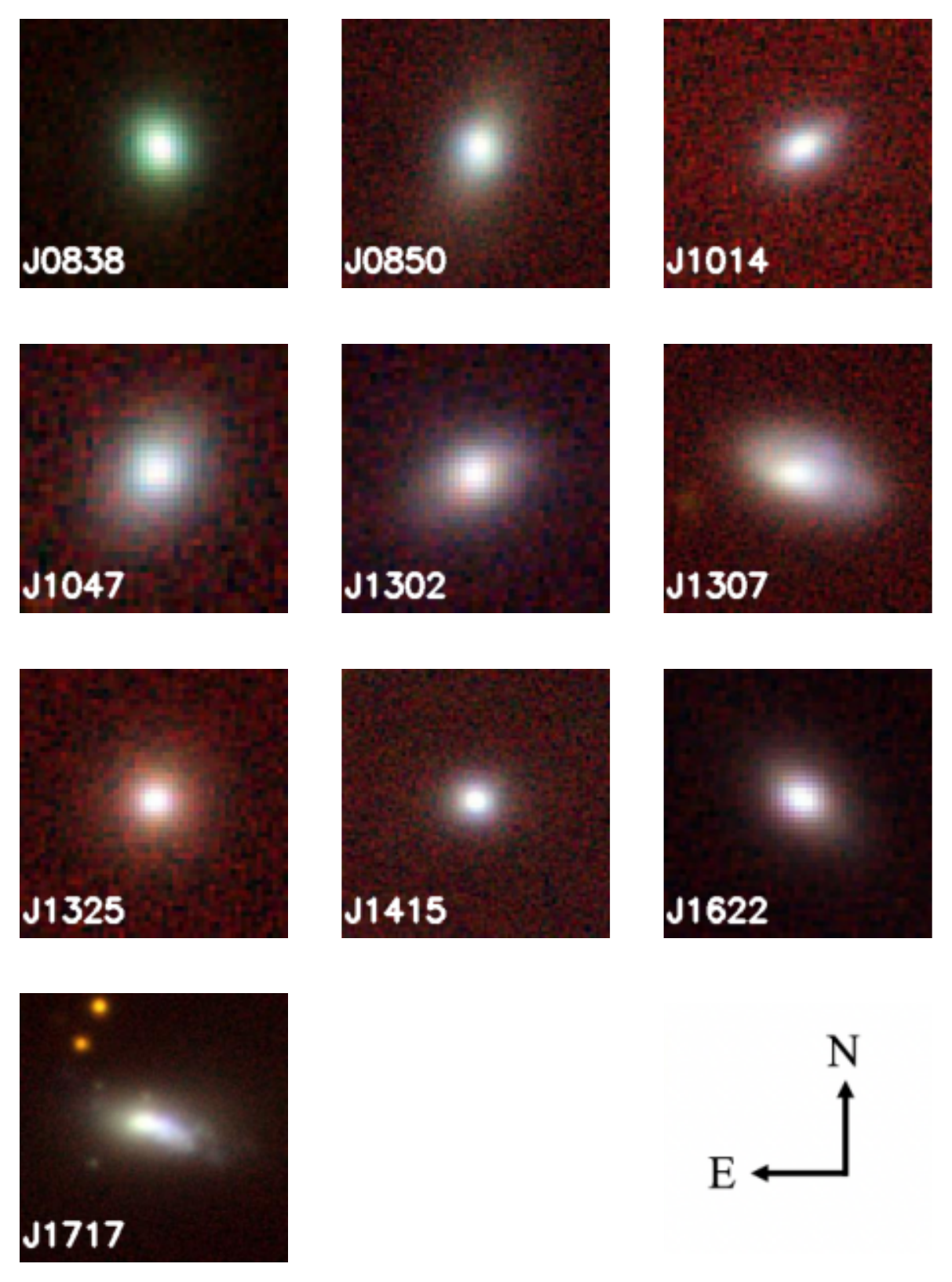}
\caption{ \footnotesize Pan-STARRS \citep{2016arXiv161205560C} color images of the star-forming dwarfs, generated using the PS1 Image Cutout Server by combining images from the y, i, and g filters. Each image is scaled to 10 kpc on a side.\label{fig:sdss}}
\end{figure}

Stellar feedback, believed to be the dominant source of feedback in dwarf galaxies \citep{1974MNRAS.169..229L, 2005ARA&A..43..769V, 2018ApJ...855L..20M} usually originates from outflows and fast winds from star-forming regions and supernovae (SNe) explosions. There have been contrasting results regarding the role of stellar feedback in resolving the discrepancies between predictions with models. For example, \citet{2008Sci...319..174M} indicated that random bulk motions of gas driven by SNe explosions would result in the flattening of the central dark matter cusp while \citet{2011ApJ...736L...2O} argued that the mass loss driven by stellar feedback might not be an effective mechanism to flatten the central cusp. \citet{2016MNRAS.457.1931S} showed that SNe feedback is sufficient to explain both the missing satellites problem and the too-big-to-fail problem, while \citet{2012MNRAS.422.1203B} and \citet{2013MNRAS.433.3539G} modeled the effects of SNe feedback using high-resolution numerical simulations and concluded that SNe feedback alone is unlikely to explain the too-big-to-fail problem in Milky Way subhaloes.

Such contrasting results could possibly indicate that stellar feedback alone may not be sufficient to explain the discrepancies between models and observations. 
The recent increase in the number of active black holes (i.e., Active Galactic Nuclei or AGN) detected in dwarf galaxies \citep[e.g.,][]{2007ApJ...670...92G, 2013ApJ...775..116R,2014AJ....148..136M, 2021ApJ...922..155M,2020ApJ...898L..30M} has raised the possibility that AGN feedback could play a prominent role in fixing the unresolved discrepancies. The effect of AGN-powered feedback is well-known in massive galaxies \citep{ 2013ApJ...775L..15R, 2013ApJ...768...75R, 2014ApJ...790..116V, 2014MNRAS.441.3306H, 2015ApJ...801..126R, 2018ApJ...857..126L, 2019MNRAS.487L..18R}.
However, most of the current cosmological models for dwarf galaxies do not incorporate AGN feedback in them, either due to limitations in resolution or because of assumptions that AGN-powered feedback does not contribute significantly to the evolution of dwarf galaxies
\citep{2015MNRAS.452..575S}.

There have been increasing arguments in favor of incorporating AGN feedback in dwarf galaxies models. \cite{2018MNRAS.473.5698D} found that AGN feedback is more effective than SNe feedback in dwarf galaxies by comparing analytically modeled properties of the outflows from both sources. High-resolution cosmological zoom-in simulations of dwarf galaxies also showed that AGN feedback is more relevant than stellar feedback in dwarf galaxies \citep{2021MNRAS.503.3568K,2022MNRAS.516.2112K}.

Observationally, AGN feedback in the form of AGN-driven outflows has been observed in many dwarf galaxies. \cite{2018MNRAS.476..979P} found evidence of  ionized gas component that is kinematically offset from their stellar component, possibly indicating outflowing gas. \citet{2019ApJ...884...54M} found evidence of strong and powerful outflows in dwarf galaxies that contain AGN. Following that, \citet[][hereafter L20]{2020ApJ...905..166L} did a comprehensive study of the dwarf galaxies with AGN from the \citet{2019ApJ...884...54M} sample using integral field spectroscopy (IFS) from Keck/KCWI.

\begin{deluxetable*}{cccccccccc} 
\tablenum{1}
\tablecaption{\\Properties of the targets and summary of observations}
\tablewidth{0pt}
\tablehead{
\colhead{Name} & \colhead{Short Name} & \colhead{Redshift} & \colhead{log($M_{*}/M_{\odot}$)} & \colhead{$R_{50}$} & \colhead{log($L_{[O\,III]}$/ erg s$^{-1}$)} & \colhead{t$_{exp}$} & \colhead{PSF}  & \colhead{PA} & \colhead{SFR}\\
& & & & kpc & & sec & \arcsec & degrees & $M_{\odot}$ yr$^{-1}$}
\decimalcolnumbers
\startdata
SDSS J083841.97+354350.0 & J0838 & 0.041 & 9.13 & 0.86 & 40.63 & 2 $\times$ 1200 & 0.83 & 4.5 & 0.5\\
SDSS J085024.95+294051.8 & J0850 & 0.026 & 9.35 & 1.10 & 40.05 & 3 $\times$ 900 & 0.85  & 339.6 & 0.1\\
SDSS J101440.21+192448.9 & J1014 & 0.028 & 8.56 & 0.81 & 39.82 & 3 $\times$ 1200 & 0.83 & -57.6 & 0.1\\
SDSS J104733.80+222400.5 & J1047 & 0.048 & 9.59 & 1.14 & 39.84 & 4 $\times$ 1200 & 0.85 & 321 & 0.2\\
SDSS J130240.05+423825.2 & J1302 & 0.043 & 9.42 & 1.50 & 40.26 & 3 $\times$ 1200 & 0.86 & 299.3 & 0.4\\
SDSS J130724.63+523715.2 & J1307 & 0.026 & 9.1 & 1.21 & 40.11 & 4 $\times$ 900 & 0.86  & 74 & 0.3\\
SDSS J132532.34+315333.1 & J1325 & 0.037 & 9.37 & 1.00 & 39.77 & 3 $\times$ 1200 & 0.85 & 0 & 0.1\\
SDSS J141525.26+045602.4 & J1415 & 0.024 & 9.11 & 0.59	& 39.70 & 3 $\times$ 900 & 0.86 & 97.5 & 0.1\\
SDSS J162244.78+323933.0 & J1622 & 0.041 & 9.48 & 0.96 & 40.41 & 2 $\times$ 900 & 0.85 & 44.3 & 0.4\\
SDSS J171759.66+332003.8 & J1717 & 0.015 & 9.85 & 1.00 & 40.26 & 7 $\times$ 600 & 0.86 & 68 & 0.3\\
\enddata
\tablecomments{\footnotesize Column(1): SDSS Name of the target. Column(2):   Short name of the target. Column(3): Redshift of the target taken from SDSS. Column(4): Stellar mass taken from the MPA-JHU catalog \citep{2003MNRAS.341...33K}. Column(5): Half-light radius measured as the Petrosian half-light radius, taken from SDSS. Column(6): Total [O\,III]$\lambda$5007 luminosity based on the observed total [O\,III]$\lambda$5007 fluxes added from spaxel to spaxel in KCWI data without extinction correction, in units of ergs s$^{-1}$. Column(7): Exposure time of the observation in seconds. Column(8): FWHM of the PSF from the spectrophotometric standard star, measured in arcseconds. Column(9): Position Angle of the IFU in degrees measured east of north. Column(10): SFR measured from the extinction-corrected [OII] fluxes using equations given in \citet{2003AAS...20311901K}.\label{table:properties}}
\end{deluxetable*} 
They found that warm ionized outflows were detected in six of the eight targets, with broad, blue-shifted fast outflows (v$_{50}$ $\sim$ -240 km s$^{-1}$ and W$_{80}$ $\sim$ -1200 km s$^{-1}$). They also calculated the energetics of the outflows and determined that a sizable amount of ionized gas mass ($\sim$ 10$^{-3}$ - 10$^{-1}$ M$_{\odot}$ yr$^{-1}$) was present in the outflows. In addition, the outflows had significant kinetic energy outflow rates ($\sim$ 10$^{37}$ - 10$^{40}$ erg s$^{-1}$). These significant values indicate that AGNs can play an important role in clearing out gas and material from dwarf galaxies or otherwise suppressing their star formation.

In order to make robust claims about the relevance of AGN feedback in dwarf galaxies, it is important to directly compare these values to the energetics found in outflows powered by stellar processes. Kinematics of the ionized gas in nearby starburst and irregular dwarf galaxies have been probed using  HI, H$\alpha$, and Na\,D \citep{2004ApJ...610..201S,2009A&A...493..511V,2010MNRAS.407..113V}. They find outflows with velocities around 20 - 60 km s$^{-1}$, which are too slow for the gas in these outflows to escape the gravitational potential of their host galaxies and are therefore less likely to serve as negative feedback in these dwarf galaxies. \citet{2022arXiv220902726M} studied the kinematics of ionized gas in a sample of 19 nearby starburst dwarf galaxies. Based on the modeling of the H$\alpha$ velocity profiles from MUSE data, they found that the velocity fields in these starburst galaxies have speeds of a few tens of km s$^{-1}$, which is much less than the speeds reached by AGN-powered outflows found by L20. They also found that the mass outflow rates caused by winds in the galaxies are around 10$^{-4}$ - 10$^{-1}$ M$_{\odot}$ yr$^{-1}$.
 
In this paper, we attempt to provide a direct comparison between AGN-driven and stellar-driven outflows in dwarf galaxies. We use a similar analysis to that of L20 on a control sample of ten star-forming (SF) dwarf galaxies using IFS. The paper is organized as follows: In Section 2, we describe the sample selection, present details of the targets, and describe the observations and data reduction. In Section 3, we list the details of fitting the emission lines. In Section 4, we present the main characteristics of the outflows that were detected in the sample and the calculations of the energetics. We compare our results to L20 wherever possible and present our conclusions and their implications in Section 5, and in Section 6, we summarize our results. More details about the individual targets can be found in the Appendix. Throughout the paper we assume the same $\Lambda$CDM cosmology as L20, with H$_o$ = 69.3 km s$^{-1}$ Mpc$^{-1}$, $\Omega_{\mathrm{m}}$ = 0.287, and $\Omega_{\Lambda}$ = 0.713 \citep{2013ApJS..208...19H}.

\begin{figure*}[ht!] 
\gridline{\fig{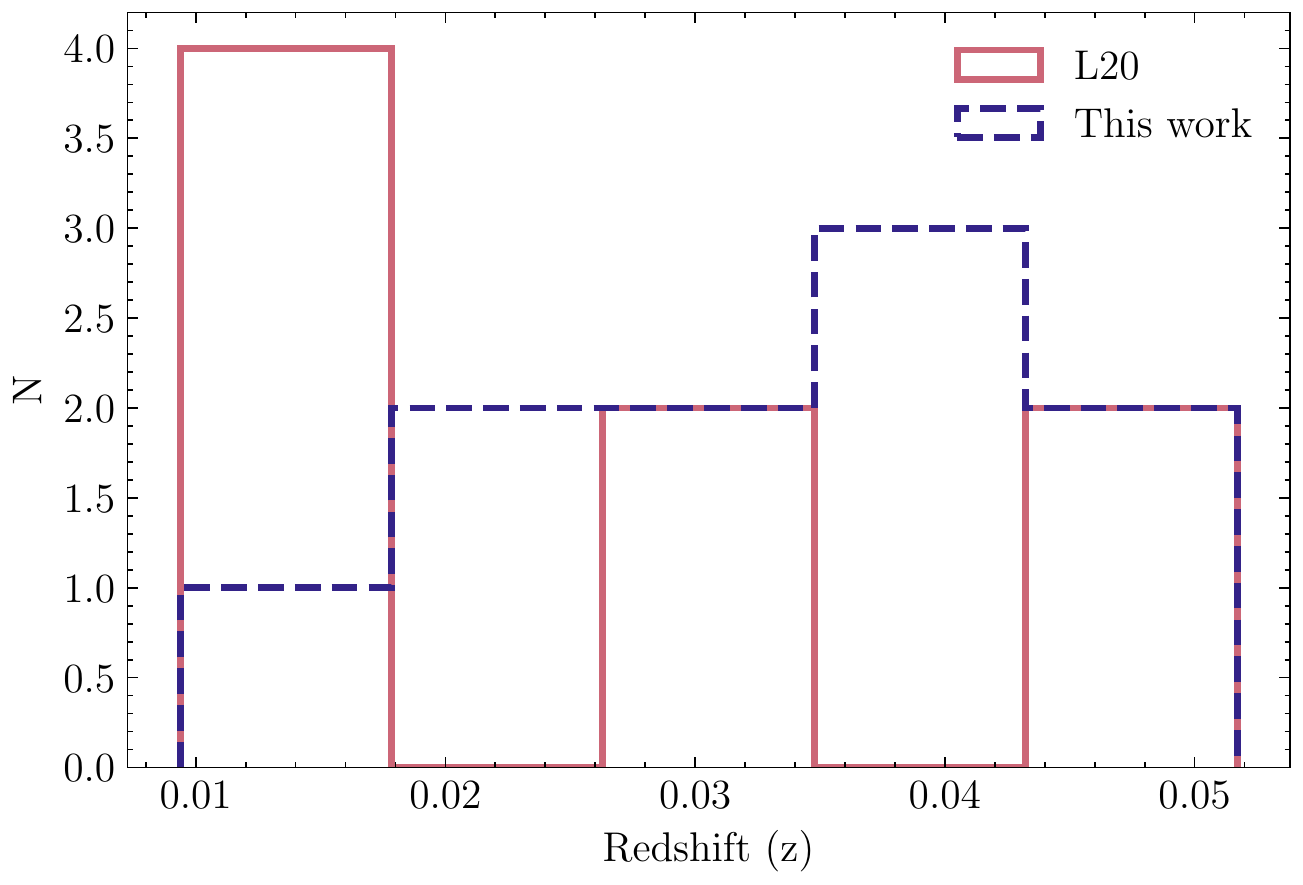}{0.45\textwidth}{(a)}
          \fig{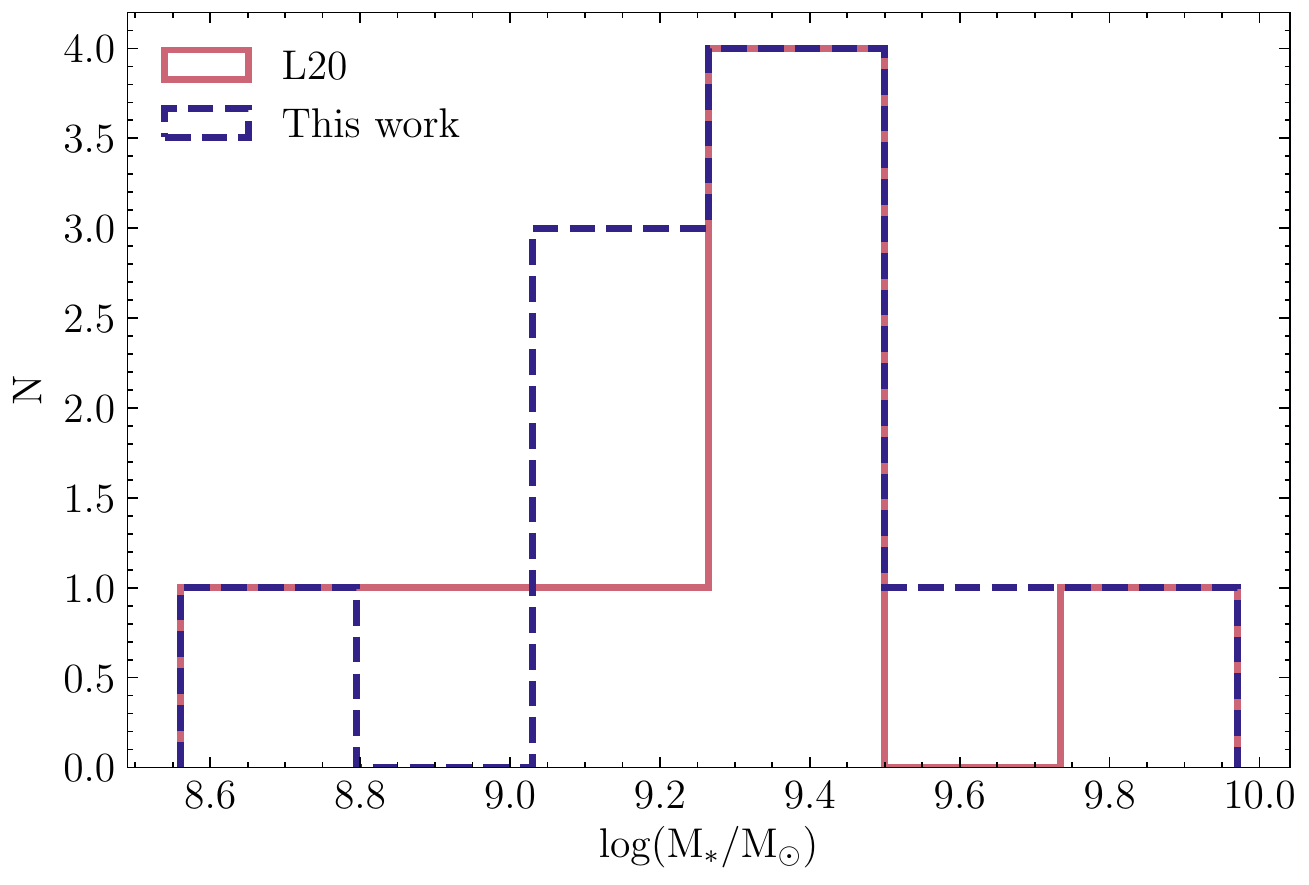}{0.45\textwidth}{(b)}
          }
\caption{ \footnotesize Redshift and stellar mass diagrams comparing the AGN dwarfs from L20 to the SF dwarfs to show that the two samples are matched in stellar mass and redshift. The pink line refers to the sample of dwarfs hosting AGN while the indigo dotted line refers to the SF dwarfs.\label{fig:comp}}
\end{figure*}

\section{Sample, Observations, and Data Reduction} \label{sec:style}
\subsection{Sample}
Our goal was to build a suitable control sample to use for comparison to the study by L20. Thus, we targeted star-forming dwarf galaxies with similar redshift and stellar mass that have prominent outflows but no signs of AGN activity

\begin{figure}[ht!]
\includegraphics[width=0.45\textwidth]{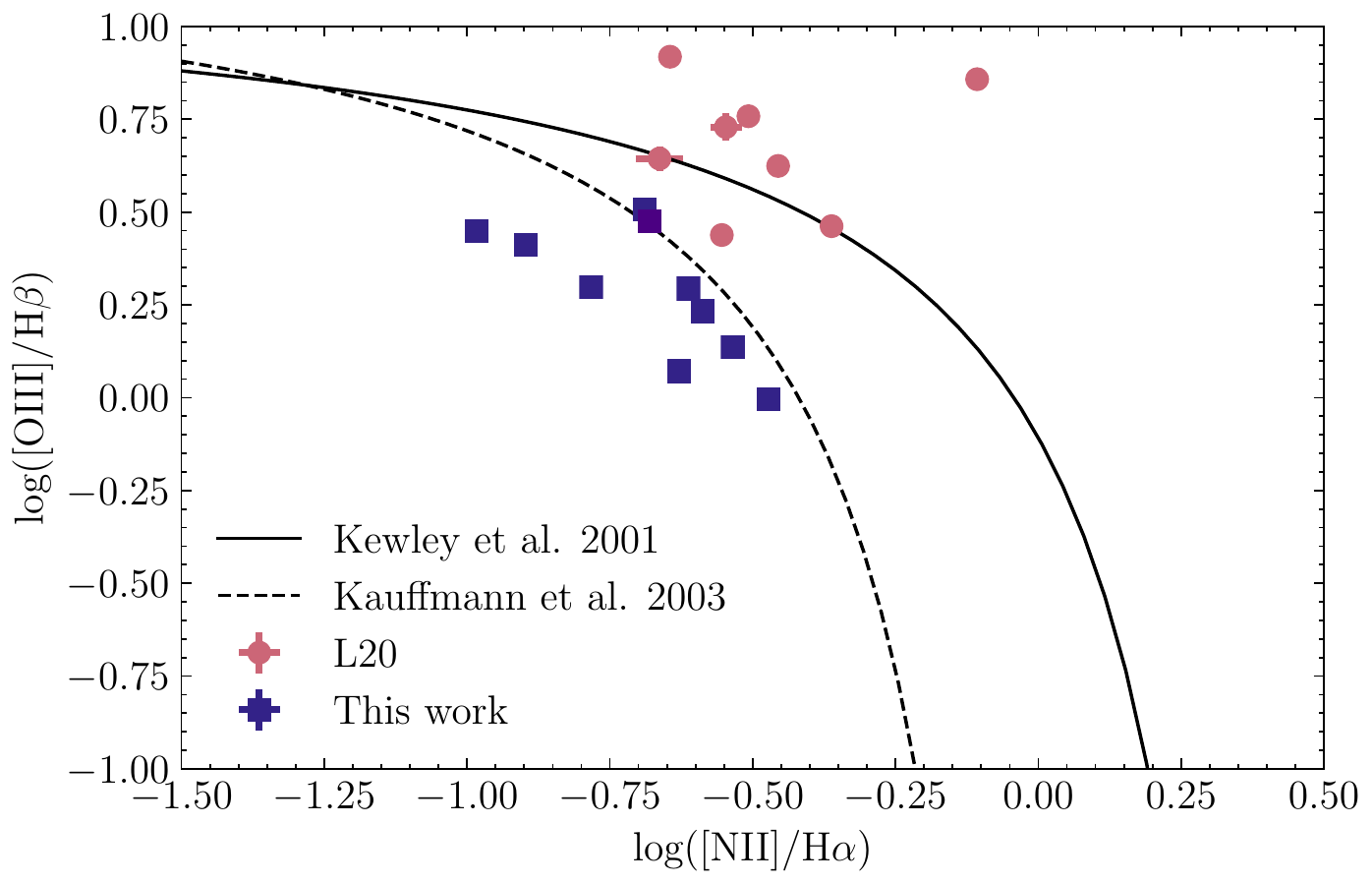}
\caption{ \footnotesize BPT diagram showing the location of the targets in the L20 sample of dwarfs host AGN and our sample of SF dwarfs. The line ratios are measured from SDSS spectra. For most targets, the error bars are smaller than the marker size. The two SF dwarf targets that lie on the boundary with the composite region are J0838 and J0850.\label{fig:bpt}}
\end{figure}

To obtain this sample, we selected all galaxies from the NASA-Sloan Atlas (NSA) with M$_{*}$ $<$ 10$^{10}$ M$_{\odot}$ and z $<$ 0.05 which were optically classified by SDSS as star-forming or starburst. We then used the open-source Python code Bayesian AGN Decomposition Analysis for SDSS Spectra \citep[\textsc{BADASS};][]{10.1093/mnras/staa3278} to fit the SDSS spectra of these galaxies in order to search for broad components to the [O\,III] $\lambda$5007 emission lines which are indicative of outflow activity. Only 72 out of 26,560 galaxies (0.27\%) showed significant broad components, a fraction consistent with that found in samples of more massive galaxies \citep{2022MNRAS.514.4828M}. From this, we selected a subset with the most visible broad [O\,III] $\lambda$5007 components and observed 10 of these targets (shown in Figure~\ref{fig:sdss} and listed in Table~\ref{table:properties}) with the Keck Cosmic Web Imager (KCWI). These included three targets from the  \citet{2019ApJ...884...54M} star-forming dwarf sample. The targets were matched in redshift and stellar mass to those of L20 (their distributions are shown in Figure~\ref{fig:comp}) and were classified as star-forming in the Baldwin, Phillips \& Terlevich (BPT) diagram \citep[][see Figure~\ref{fig:bpt}]{1981PASP...93....5B,1987ApJS...63..295V}, which is the diagnostic diagram used to distinguish between AGNs and star-forming galaxies based on the ratios of optical emission lines. The SF dwarfs were also similar in size to the AGN dwarfs if we compare them based on their half-light radius (average radius of 1.1 kpc for AGN dwarfs vs. 1.0 kpc for SF dwarfs).

\subsection{Observations}\label{observations}
All the targets were observed with KCWI \citep{Morrissey_2018} through the Keck program 2021A-U080 (P.I. G. Canalizo) on 2021 March 19 and 2021 March 20. All targets were observed using the blue grating (BL), which covers a wavelength range from 3500-5500 Å and the small slicer setup (spectral resolution $\sim$ 80 km s$^{-1}$ FWHM at 4550 Å). The observations were obtained under clear sky conditions and a typical seeing of $\sim 0\farcs8$

The PSF of these IFU observations was measured using the observations of the spectrophotometric standard stars that were taken before and after the on-target observations throughout the night. We stacked narrow-band images in the range of 5000-5100 $\text{\AA}$ and fit 2D Moffat profiles \citep{1969A&A.....3..455M} to them. We used a Moffat profile as opposed to the Gaussian profile used by L20, as the Moffat profile fits the radial profile of the star better, as referenced by previous studies \citep{2016RAA....16..139L}. Moreover, there was no significant difference in FWHMs obtained from both profiles, so we adopted the median FWHM obtained from the best-fit Moffat profiles. There would be some discrepancy in the PSF as we could not take observations of the standard stars at the same time as the science target, but the seeing (as measured by the Maunakea Weather Center) ranged from $0\farcs75$ to $1\farcs0$ across both nights, and the FWHM we measured was well within this range.

\subsection{Data Reduction}
All the targets were reduced with the KCWI Data Extraction and Reduction Pipeline (KDERP)\footnote[1]{ \url{https://github.com/Keck-DataReductionPipelines/KcwiDRP/blob/master/AAAREADME}}, following the standard procedures listed in the manual. The pipeline also included both wavelength and flux calibrations. We then used routines from the \textsc{IFSRED} \citep{2014ascl.soft09004R} library to resample the cubes with square spaxels as individual exposures had rectangular spaxels. The data cubes were resampled to 0$\farcs$15 X 0$\farcs$15  using \texttt{IFSR\_KCWIRESAMPLE}, and then data cubes of the same target were median-combined into a single data cube using \texttt{IFSR\_MOSAIC}, which also corrected for the shifts caused due to dithering.

\section{Analysis}
\begin{figure}
\gridline{\fig{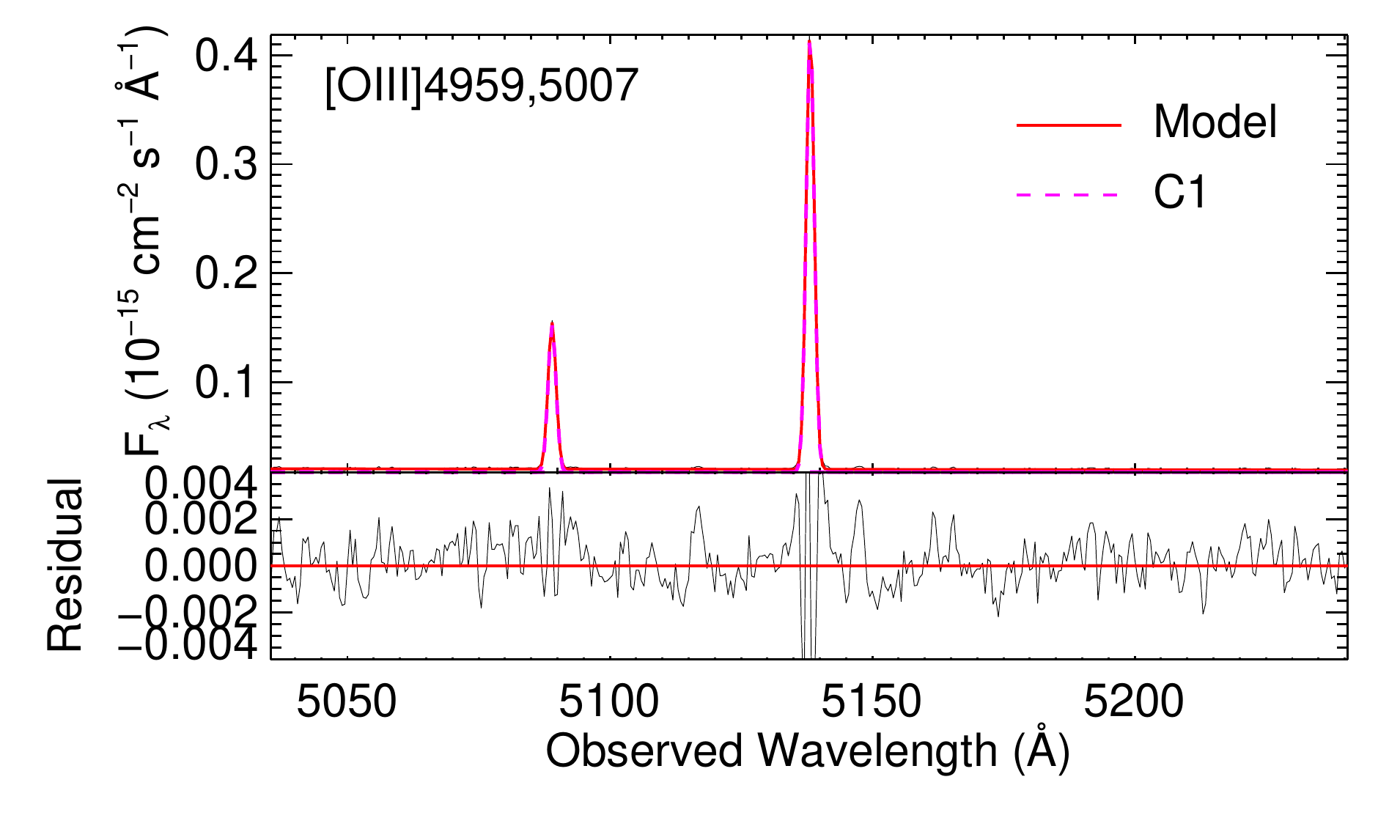}{0.4\textwidth}{(a) 1 component fit}}
\gridline{\fig{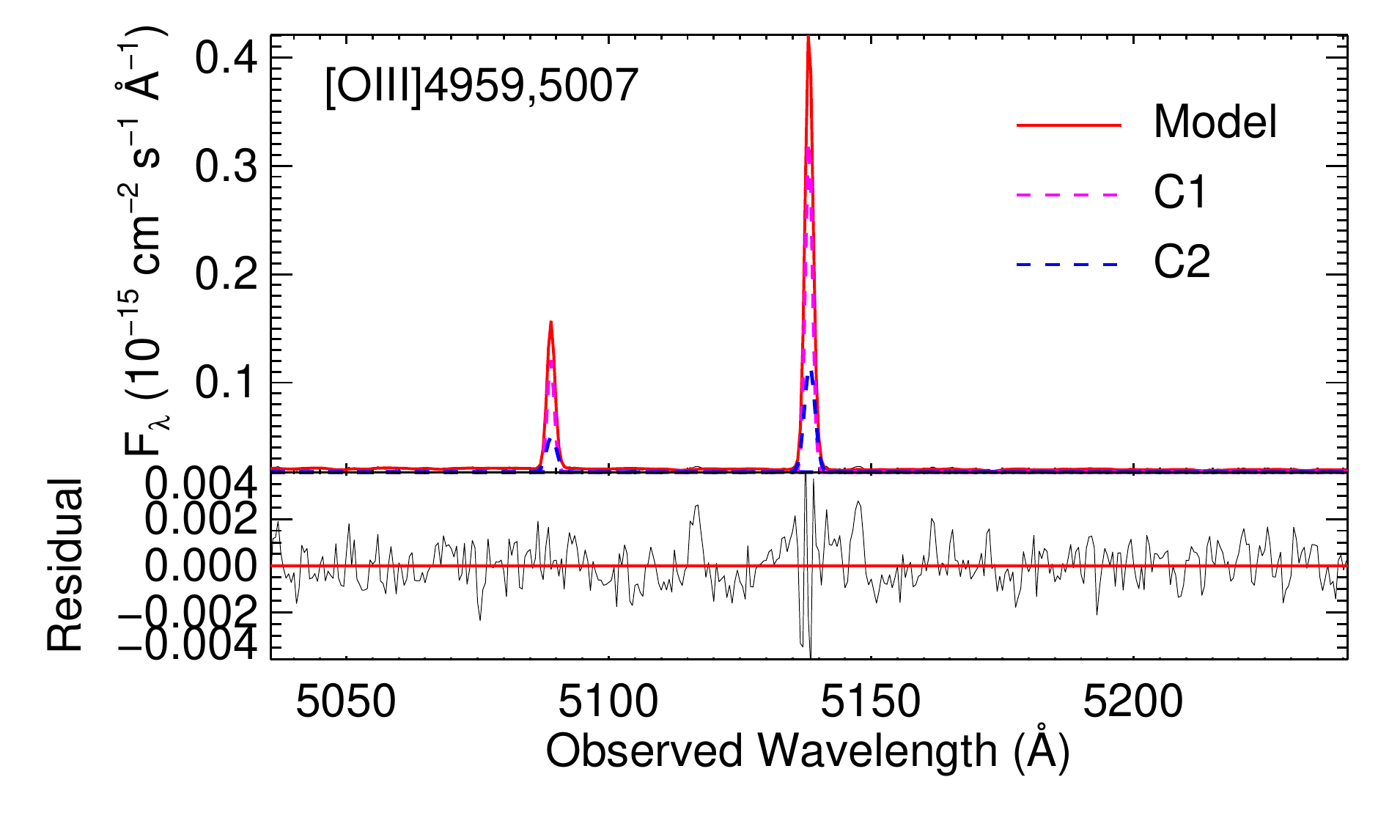}{0.4\textwidth}{(b) 2 component fit}}
\gridline{\fig{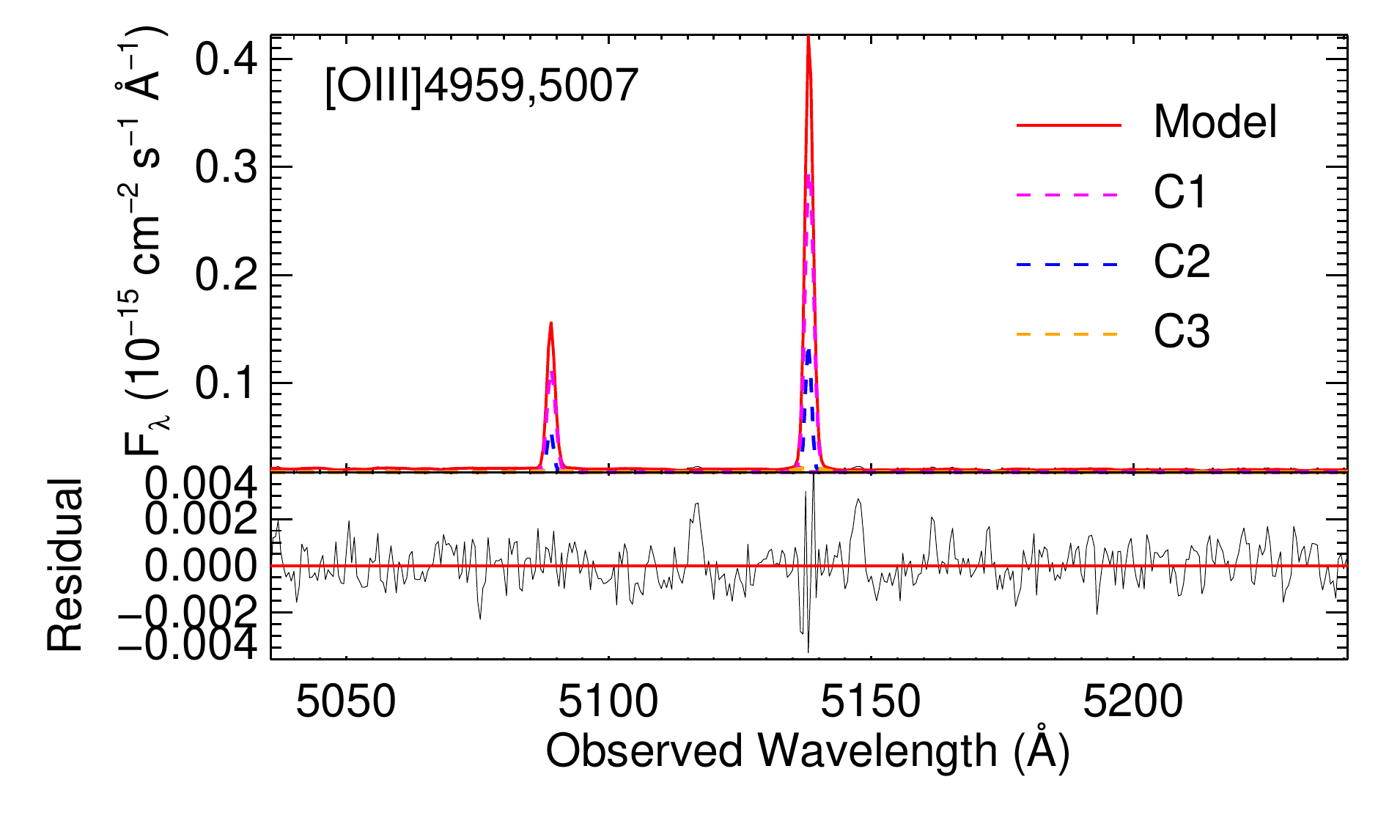}{0.4\textwidth}{(c) 3 component fit}}
\caption{ \footnotesize Examples of (a) one Gaussian component, (b) two Gaussian components and (c) three Gaussian components fit to the [O\,III] doublet line profile for J1307 in the same single spaxel. In each panel, the black spectrum is the observed data, the solid red line is the best fit and the dashed curves represent the Gaussian components. The residuals after subtraction from the best fit are shown at the bottom of each panel. The single component in (a) does not fit the wings accurately, while the third component in (c) does not improve the fit significantly. \label{fig:spectralfits}}
\end{figure}

\subsection{Binning}
We spatially bin the data cubes using the Voronoi binning method \citep{2003MNRAS.342..345C}. Binning is a useful technique that can be used to preserve the spatial resolution of the data while obtaining a higher signal-to-noise ratio (SNR). This would help us to accurately characterize the broad components in the emission lines to determine the outflows, which might otherwise be missed due to a lower SNR. The Voronoi binning method, in particular, bins or divides the data into hexagonal bins, so data from neighboring pixels are grouped as one to achieve the target SNR. Thus, the size of the bins increases with the distance from the photocenter of the target to match the required SNR. We set an SNR of 5 for our targets as we found that was the highest SNR we could obtain before we started losing spatial information from the binning. All subsequent analysis was done on the Voronoi-binned spectra.

\subsection{Spectral fits}
\subsubsection{Fitting the [O\,III] emission lines}
We first masked the emission lines in the range between 4800-5500 $\text{\AA}$ in order to fit the stellar continuum using the public software pPXF \citep{2004PASP..116..138C, 2017MNRAS.466..798C} with 0.5 $\times$ solar metallicity stellar population synthesis models from \cite{2005MNRAS.357..945G}. The emission line profiles were mostly symmetric, with no obvious signatures of blue-shifted wings. In contrast, most of the broad components of the emission line profiles in L20 were blue-shifted.

After subtracting the continuum, the [O\,III] $\lambda$4959, $\lambda$5007 emission lines were fitted with Gaussian components using the IDL library MPFIT \citep{2012ascl.soft08019M}. The line centers of the velocities and widths of both lines were tied together, and only the amplitudes were allowed to vary. We also did not fix the doublet ratios in [O\,III] in order to allow fits even when a Gaussian component was only detected in one of the two emission lines. We initially tried three different sets of fits. First, we allowed only a maximum of one component, then a maximum of two, and the final set of fits comprised a maximum of three components. We then used an $F$-test (described in Section \ref{ftest}) to choose the number of components to fit the [O\,III] emission lines.

\subsubsection{F-tests to determine the optimum number of components to fit the [O\,III] emission lines}\label{ftest}
The $F$-test determines the significance of the fit between a complex (higher-order) and simple (lower-order) model in order to justify the number of Gaussian components we include in our fit. To do this, we calculated the standard deviation of the residuals of the fit (given by $\sigma$) for each spaxel. We did the $F$-test, given as $F=(\sigma_{\rm{lower-order}})^2/(\sigma_{\rm{higher-order}})^2$, for one vs. two components as well as two vs. three components for the Gaussian fits to the [O\,III] $\lambda$4959, $\lambda$5007 lines. The F-test was calculated over a wavelength range of 4800 - 5500 $\text{\AA}$ which includes the [O\,III] $\lambda$4959, $\lambda$5007 lines. An $F$ value greater than 3 indicates that the higher order fit (i.e., the fit using more components) is justifiable.
We performed the $F$-test for the fits to the [O\,III] $\lambda$4959, $\lambda$5007 lines in every spaxel of our targets and mapped the results of the $F$-test. All targets failed the $F$-test for one component, so more than one component was required to fit the profile. For some of the targets (J1325, J1622, J1047, J1014), it was evident that we needed only two components.
The rest of the targets were more ambiguous since they contained a fraction of spaxels with $F$ values close to, but below 3. For those targets, we inspected the fits visually to assess the effect that including an additional component had on the residuals and to determine whether the additional component was fitting a real feature in the emission lines. From Figure \ref{fig:spectralfits} (a), we see that fitting a single component is not sufficient, and it misses a crucial part of the emission line. In \ref{fig:spectralfits} (c), we see that the third component is not fitting any real feature. 
We, therefore, chose to use a maximum of two components (Figure~\ref{fig:spectralfits} (b)) for all the targets to keep the fits as simple as possible, with the caveat that our flux measurements may miss a small fraction in the wings of the emission lines.
This is also different from the analysis of L20, in which three of the eight targets needed three components to get the best fit, while two needed just one component.

\subsubsection{Fitting the full spectrum} 
After obtaining robust fits to the [O\,III] $\lambda$4959, $\lambda$5007 emission lines, we fit the emission lines to the entire spectral range and simultaneously fit strong emission lines such as H$\beta$, H$\gamma$, H$\delta$, [O\,II] $\lambda\lambda$3726, 3729 and [Ne\,III] $\lambda$3869. Following the results of the $F$-Test for the [O\,III] $\lambda$4959, $\lambda$5007 lines, a maximum of two components were fit for these lines.

\subsection{Star formation rates} \label{section:sfr}
The star formation rates (SFR) of the star-forming dwarfs were calculated from the [O\,II] lines using the following equation \citep{2003AAS...20311901K}:

\begin{equation}
    \mathrm{SFR [M_{\odot}\hspace{0.1cm} yr^{-1}] = \frac{7.9 \times 10^{-42} L[O\,II] (erg\hspace{0.1cm}s^{-1})}{(-1.75)[log(O/H) + 12] + 16.73}}
\end{equation}

where [log(O/H) + 12] was taken to be 8.9 for solar metallicities (as adopted by L20). If we consider 0.5 $\times$ solar metallicity, the SFR would be $\sim$ 30 \% lower.

These values are listed in column 10 in Table \ref{table:properties}. We find that for most of our targets, the SFR is higher than the values listed for the AGN dwarfs from L20, which have an average upper limit of 0.20 M$_{\odot}$ yr$^{-1}$. In comparison, the average for the SF dwarfs is 0.25 M$_{\odot}$ yr$^{-1}$. We note that the SFR values for the SF dwarfs are likely to be higher than those of the AGN dwarfs, as there could be a significant contribution from the AGN to the [O\,II] emission lines, which in turn could lead to an overestimate in the value of SFR.

The SFR for 9 out of 10 objects in our sample could also be obtained from the MPA-JHU catalog which are calculated based on the techniques used in \cite{2004MNRAS.351.1151B}. These values are much higher than our calculated values, with an average of 1.38 M$_{\odot}$ yr$^{-1}$. This could be because the MPA-JHU estimates use multiple lines as SF indicators, while we only used the [O\,II] lines for our estimates. The biggest difference was for J0838, which had an MPA-JHU estimate for SFR of 3.94 M$_{\odot}$ yr$^{-1}$ as compared to the SFR of 0.45 M$_{\odot}$ yr$^{-1}$ we measure from [O\,II].

\begin{deluxetable*}{cccccccc}
\tablenum{2}
\tablecaption{\\Kinematic Properties of the Targets}
\tablewidth{0pt}
\tablehead{
\colhead{Name} & \colhead{$N_{comp}$} & \colhead{Component} & \colhead{Median $v_{50}$} & \colhead{Min $v_{50}$} & \colhead{Max $v_{50}$} & \colhead{Median $W_{80}$} & \colhead{Max $W_{80}$}\\
\colhead{} & \colhead{} & \colhead{} & \colhead{(km s$^{-1}$)} & \colhead{(km s$^{-1}$)} & \colhead{(km s$^{-1}$)} & \colhead{(km s$^{-1}$)} & \colhead{(km s$^{-1}$)}
}
\decimalcolnumbers
\startdata
J0838 & 2 & C2 & -5 & -53 & 61 & 397 & 625 \\
	  &   & C1 & 15 & -2 & 69 & 113 & 180 \\
	  &   & Total & 16 & -7 & 54 & 315 & 474 \\
\hline
J0850 & 2 & C2 & 21 & -53 & 58 & 593 & 832 \\
	  &   & C1 & 30 & -40 & 86 & 124 & 168 \\
	  &   & Total & 38 & -19 & 82 & 278 & 648 \\
\hline
J1014 & 2 & C2 & 7 & -80 & 62 & 378 & 855 \\
	  &   & C1  & -7 & -50 & 20 & 87 & 139 \\
	  &   & Total &  19 & -42 & 43 & 127 & 701\\
\hline
J1047 & 2 & C2 & -28 & -82 & 22 & 651 & 830\\
	  &   & C1 & 13 & -10 & 32 & 118 & 171\\
	  &   & Total & 19 & -1 & 43 & 522 & 778\\
\hline
J1302 & 2 & C2 & 23 & -21 & 122 & 721 & 1203\\
	  &   & C1 & 27 & -40 & 22 & 130 & 238\\
	  &   & Total &	31 & -11 & 82 & 551 & 1042\\
\hline
J1307 & 2 &	C2 & 9 & -20 & 50 & 385 & 385\\
	  &   & C1 & 15 & -54 & 60 & 67 & 98\\
	  &   & Total &	8 & -32 & 71 & 131 & 369\\
\hline
J1325 & 2 & C2 & -12 & -66 & 72 & 586 & 760\\
	  &   & C1 & 3 & -8 & 26 & 117 & 154\\
	  &   & Total & 12 & -11 & 35 & 453 & 649\\
\hline
J1415 & 2 & C2 & -4 & -41 & 41 & 488 & 641\\
	  &   & C1 & 5 & -12 & 42 & 122 & 152\\
	  &   & Total & 12 & -61 & 42 & 318 & 586\\
\hline
J1622 & 2 & C2 & -20 & -93 & 58 & 462 & 671\\
	  &   & C1 &  -12 & -58  & 99 & 122 & 162\\
	  &   & Total & -14 & -57 & 62 & 246 & 395\\
\hline
J1717 & 2 & C2 & 10 & -77 & 100 & 178 & 725\\
	  &   & C1 & 6 & -67  & 171 & 79 & 159\\
	  &   & Total & 8 & -48 & 76 & 139 & 244\\
\enddata
\tablecomments{\footnotesize Column(1): Short name of the target. Column(2): Number of components used for the fit. Column(3): Component number Column(4)-(6): Median, maximum and minimum values of v$_{50}$ measured across the cube. Column(7)-(8): Median and maximum values of W$_{80}$ measured across the cube. The spaxels with the highest and the lowest 5\% of v$_{50}$ were ignored from the calculations. Measurement errors for individual values of v$_{50}$ range between 1 and 5 km s$^{-1}$. \label{table:kinem}}
\end{deluxetable*}

\begin{figure*}[]
\gridline{\fig{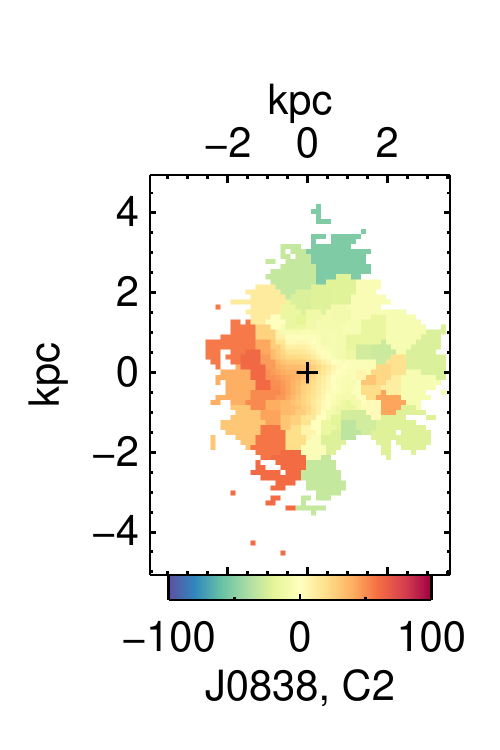}{0.14\textwidth}{}
\fig{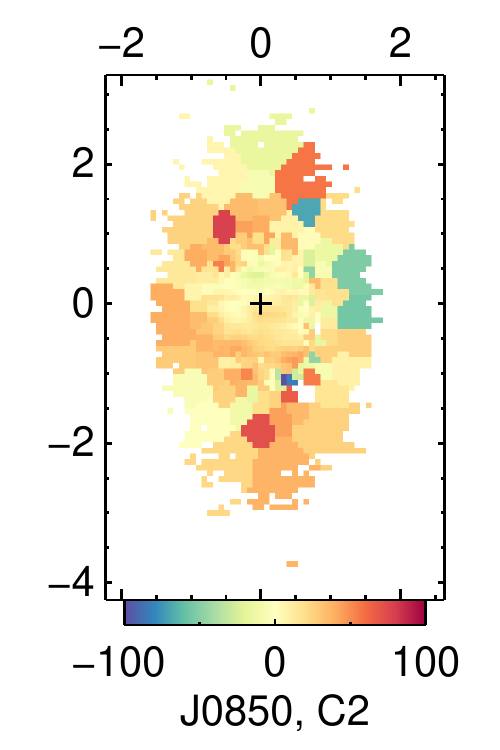}{0.14\textwidth}{}
\fig{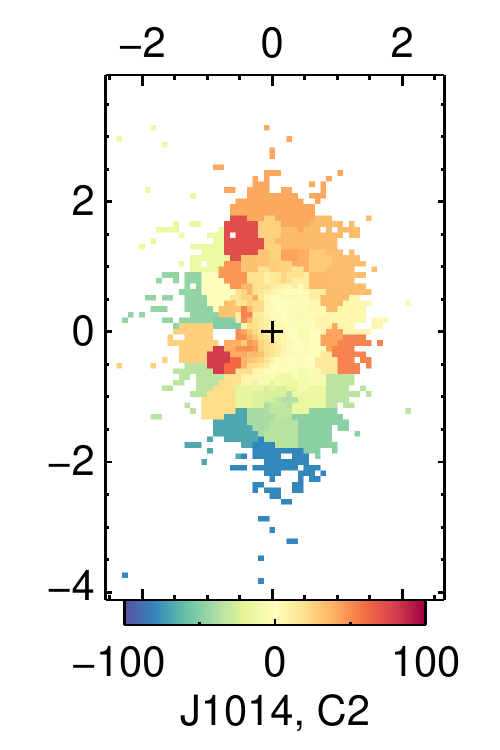}{0.14\textwidth}{}
\fig{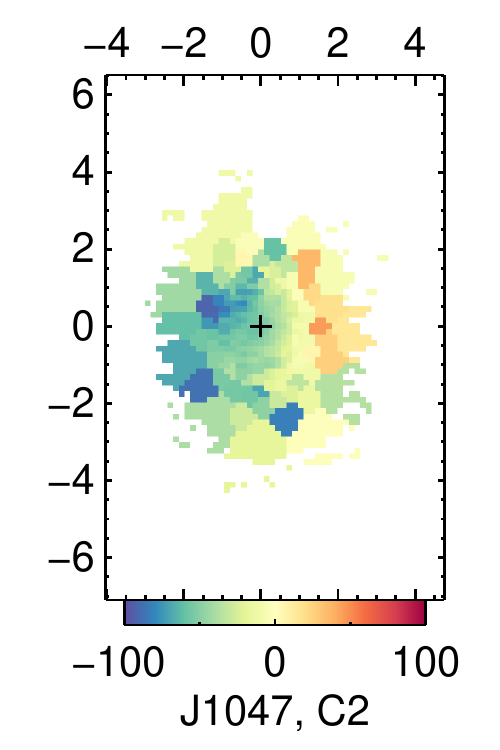}{0.14\textwidth}{}
\fig{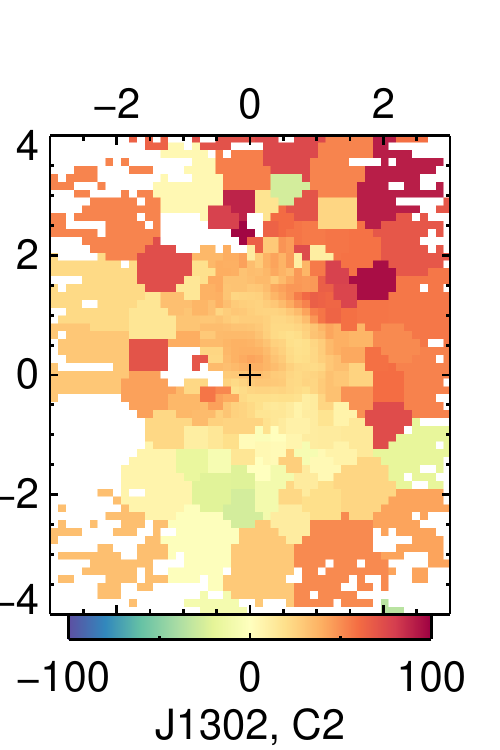}{0.14\textwidth}{}}
\gridline{\fig{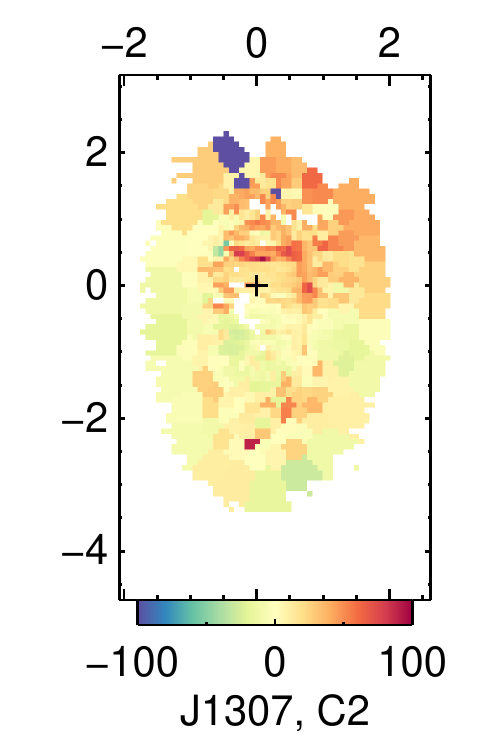}{0.14\textwidth}{}
\fig{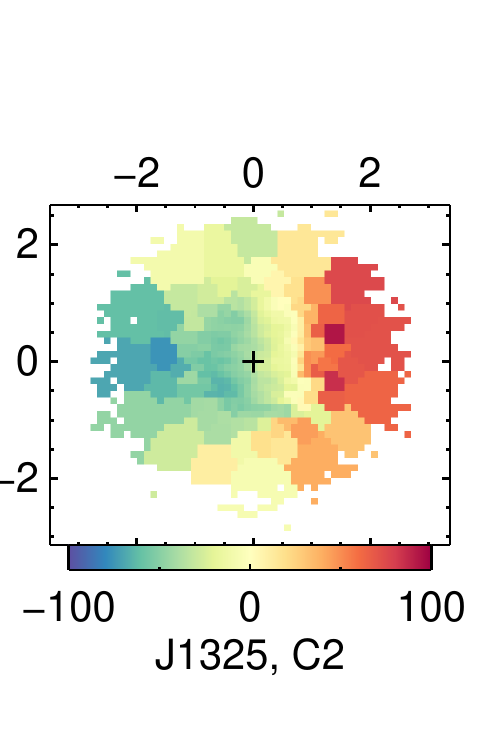}{0.14\textwidth}{}
\fig{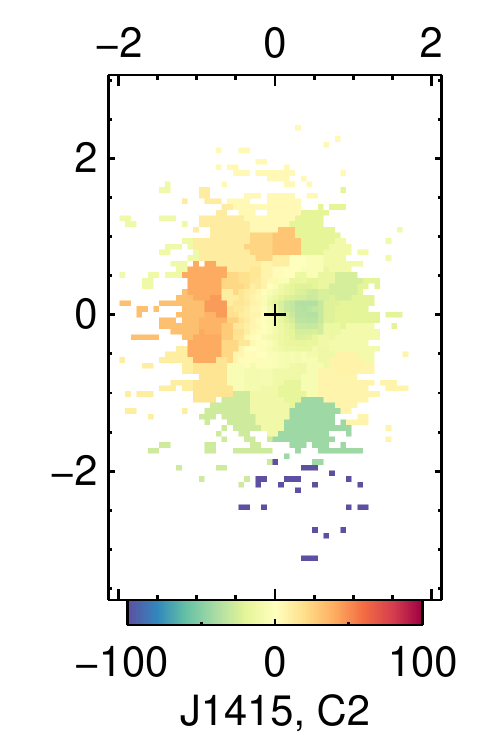}{0.14\textwidth}{}
\fig{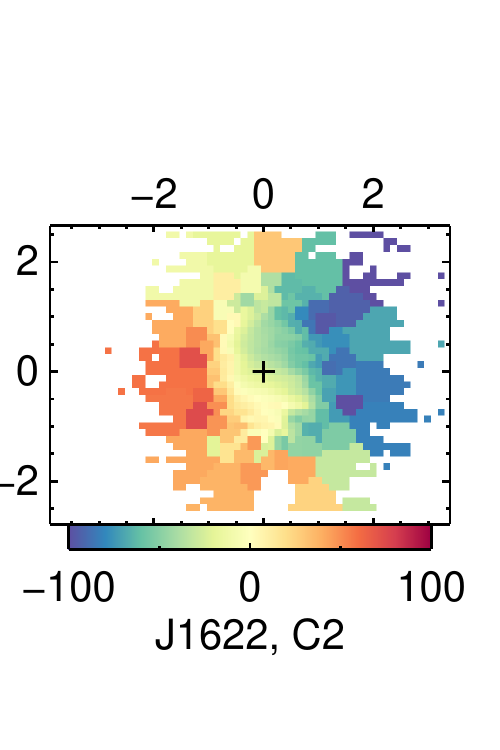}{0.14\textwidth}{}
\fig{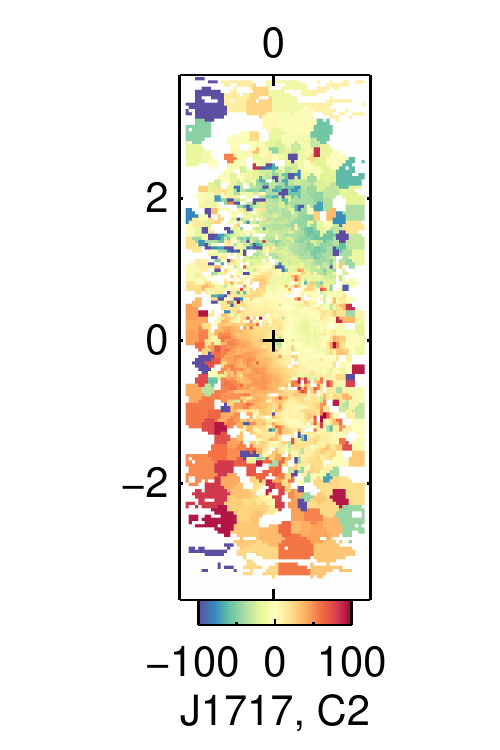}{0.14\textwidth}{}}
\caption{ \footnotesize Velocity maps with the v$_{50}$ of the C2 component measured in km s$^{-1}$ for all the SF dwarfs. The extent of the outflows in kpc is indicated by the x-axis and y-axis. The black cross indicates the center of the galaxy. Measurement errors for individual values of v$_{50}$ range between 1 and 5 km s$^{-1}$. However, velocity variations from bin to bin indicate that the true uncertainty is closer to 20 km s$^{-1}$.}
\label{fig:v50}
\end{figure*}

\begin{figure*}[t!]
\gridline{\fig{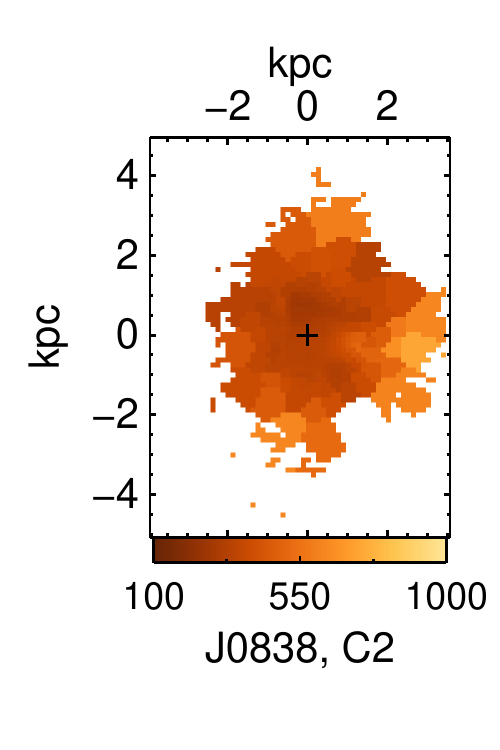}{0.14\textwidth}{}
\fig{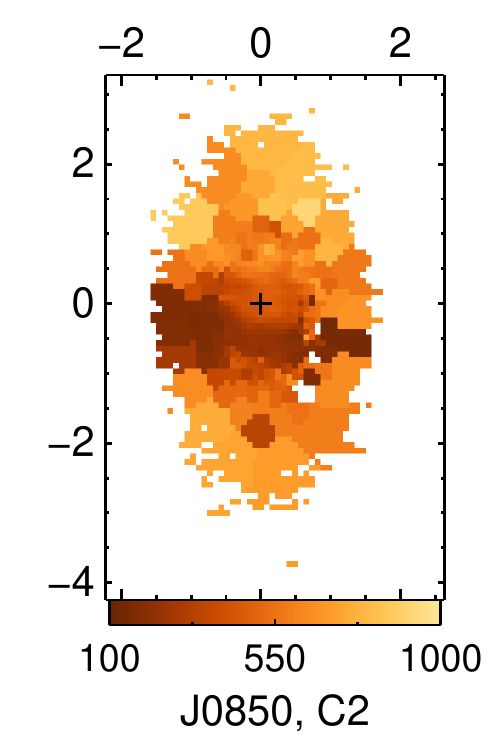}{0.14\textwidth}{}
\fig{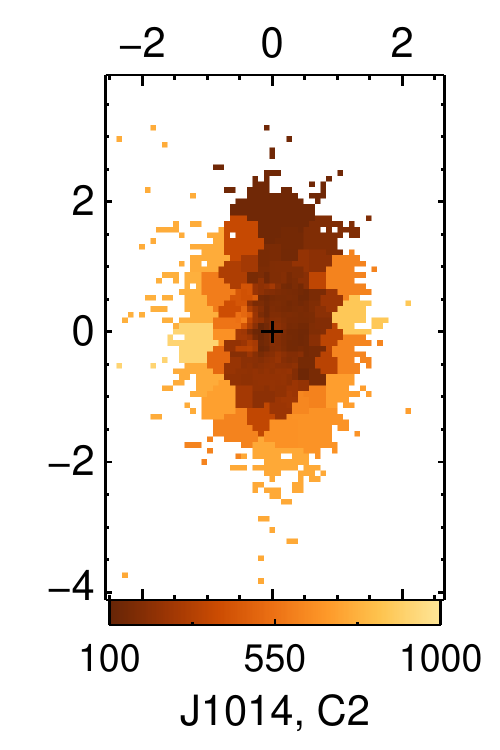}{0.14\textwidth}{}
\fig{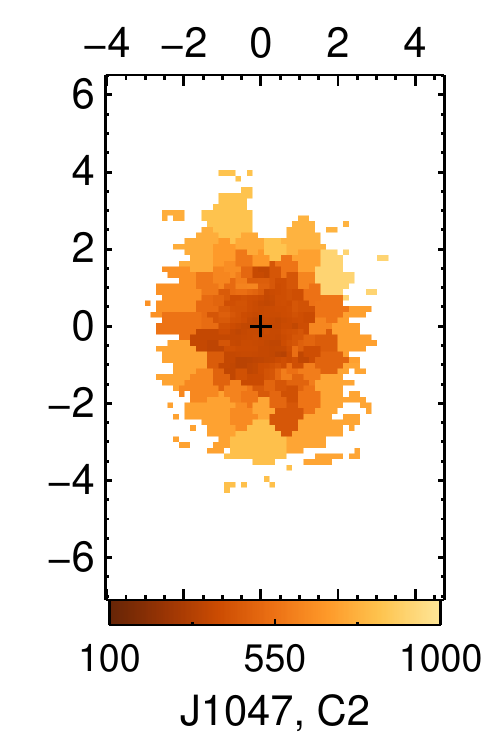}{0.14\textwidth}{}
\fig{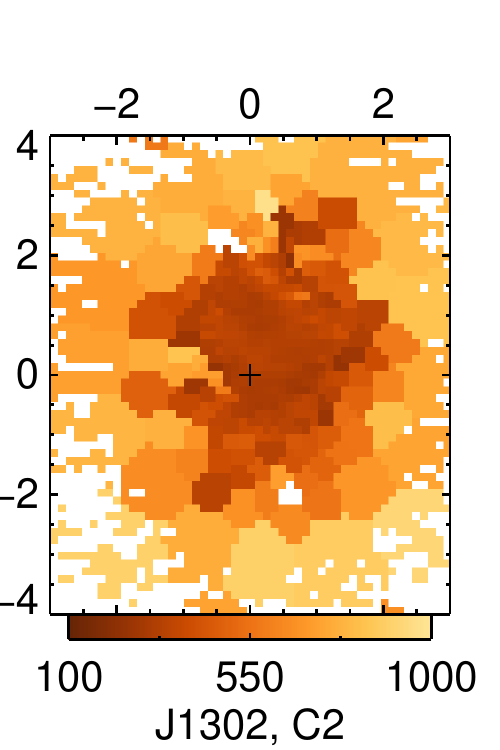}{0.14\textwidth}{}}
\gridline{\fig{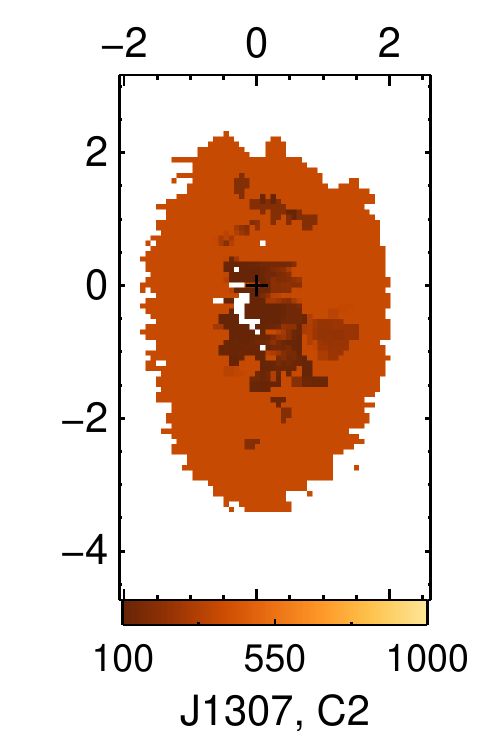}{0.14\textwidth}{}
\fig{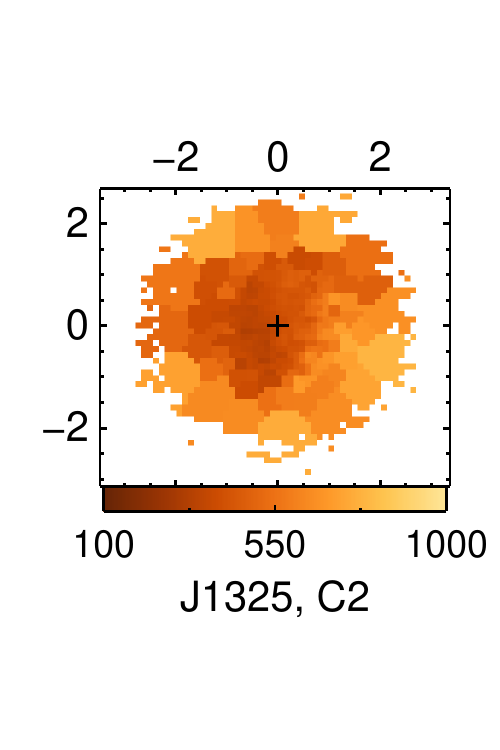}{0.14\textwidth}{}
\fig{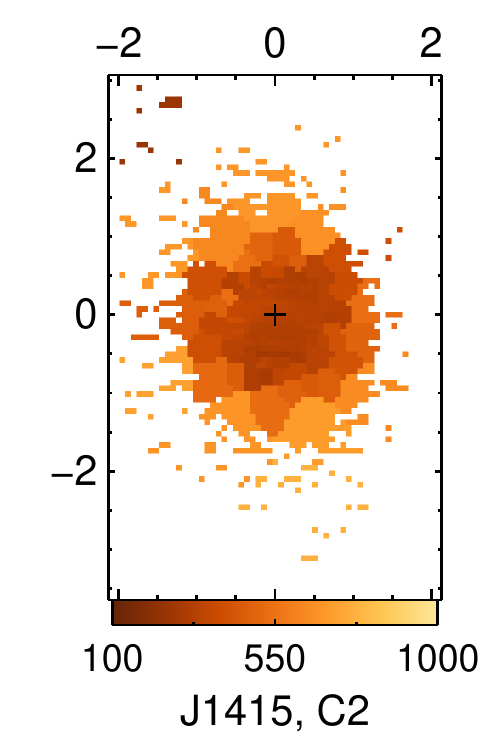}{0.14\textwidth}{}
\fig{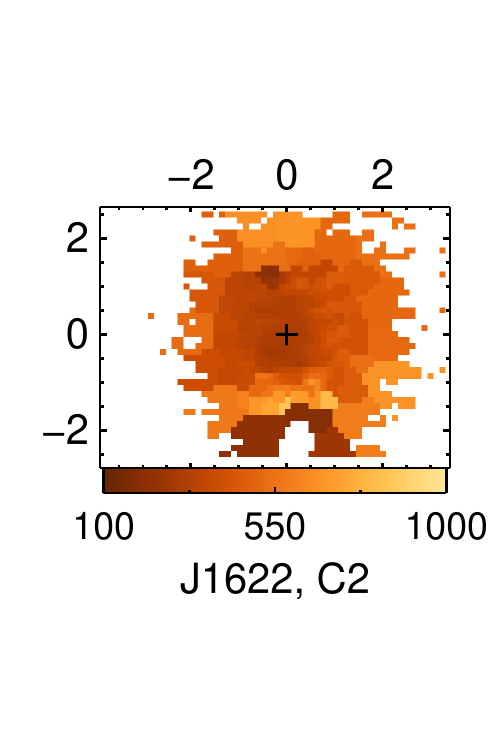}{0.14\textwidth}{}
\fig{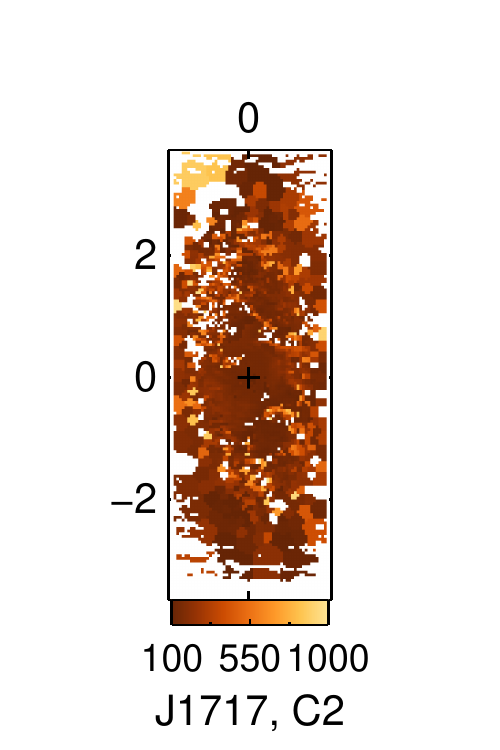}{0.14\textwidth}{}}
\caption{ \footnotesize W$_{80}$ maps showing the width of the velocity profile in km s$^{-1}$ of the C2 component in all the SF dwarfs. Measurement errors for W$_{80}$ range from 2 to 10 km s$^{-1}$. There is a large range in dispersion values, with a median W$_{80}$ of $\sim$ 480 km s$^{-1}$.}\label{fig:w80}
\end{figure*}

\begin{figure*}[]
\gridline{\fig{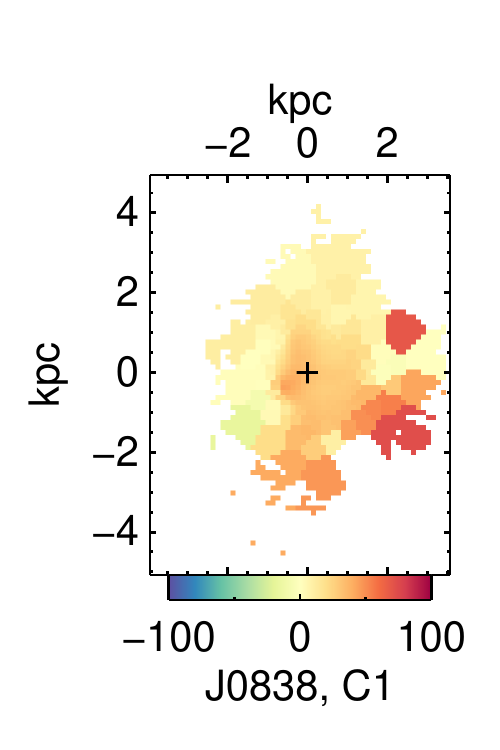}{0.15\textwidth}{}
\fig{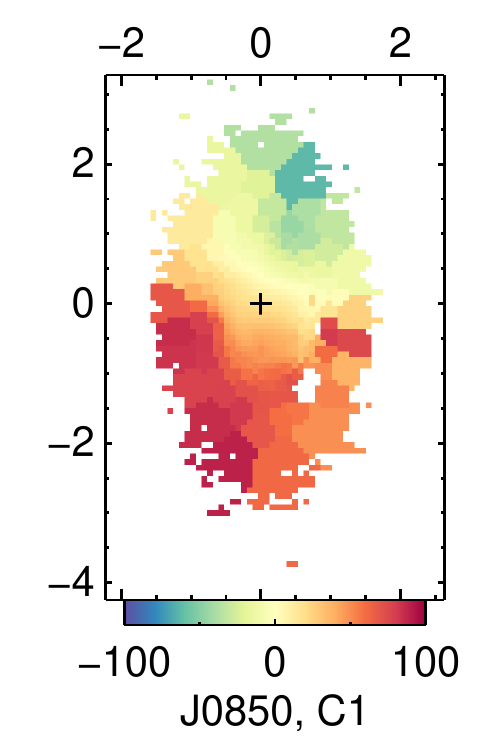}{0.15\textwidth}{}
\fig{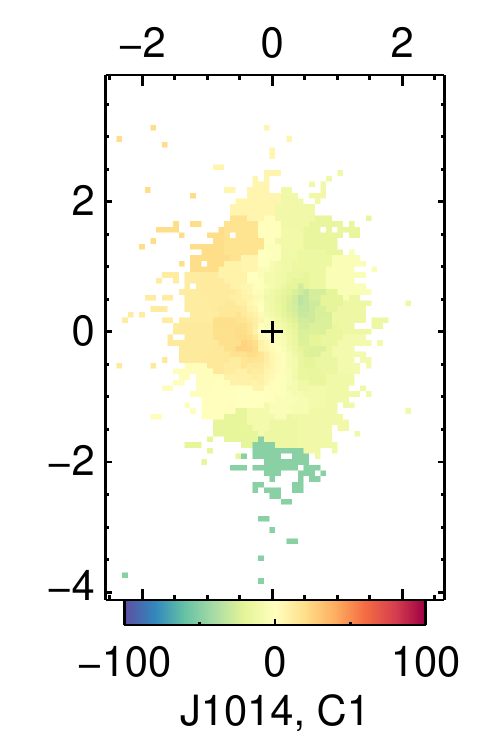}{0.15\textwidth}{}
\fig{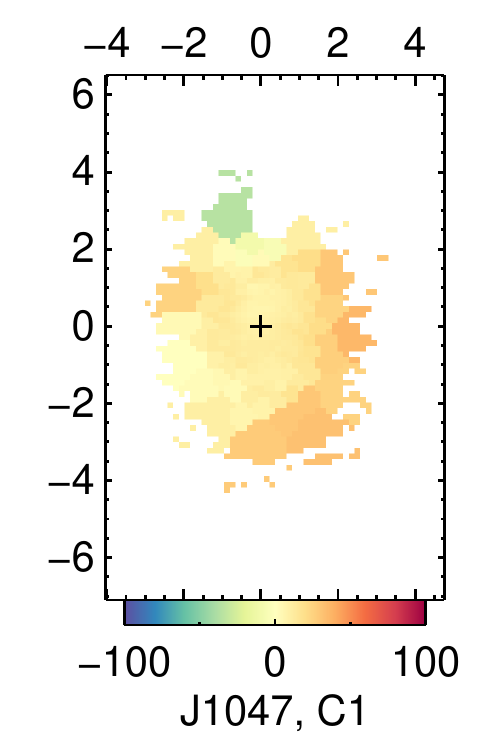}{0.15\textwidth}{}
\fig{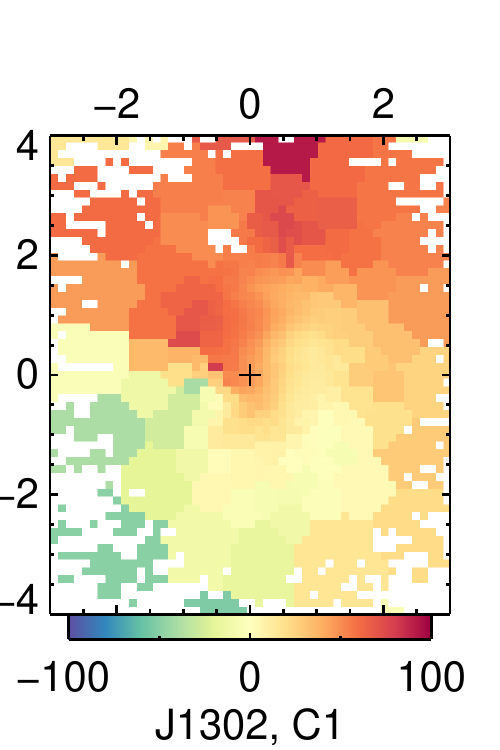}{0.15\textwidth}{}}
\gridline{\fig{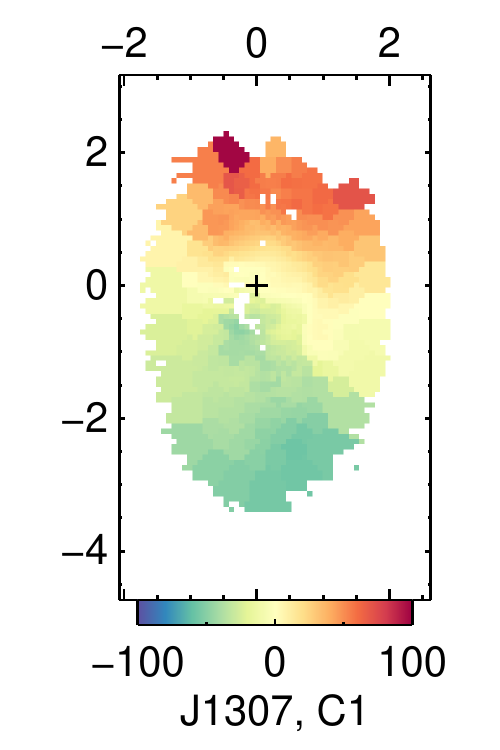}{0.15\textwidth}{}
\fig{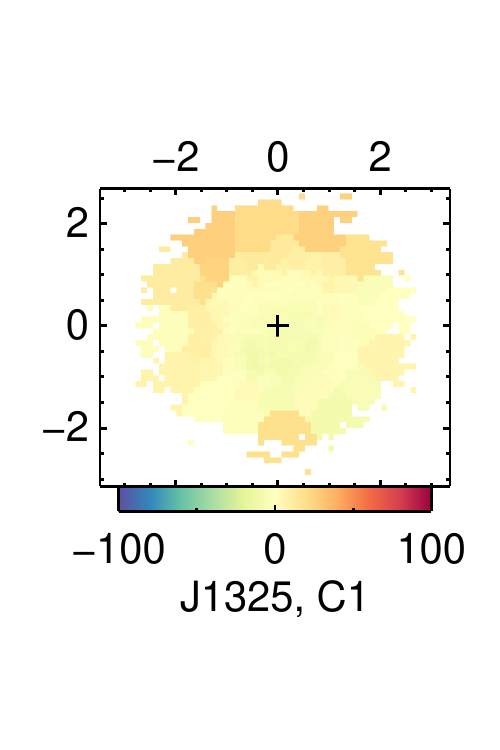}{0.15\textwidth}{}
\fig{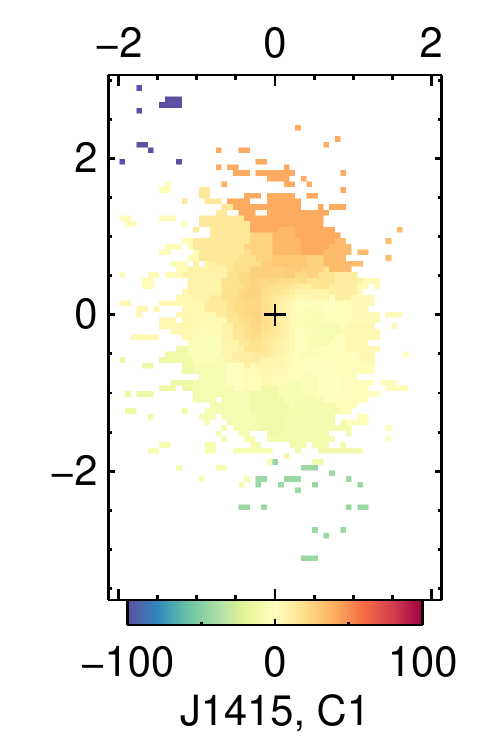}{0.15\textwidth}{}
\fig{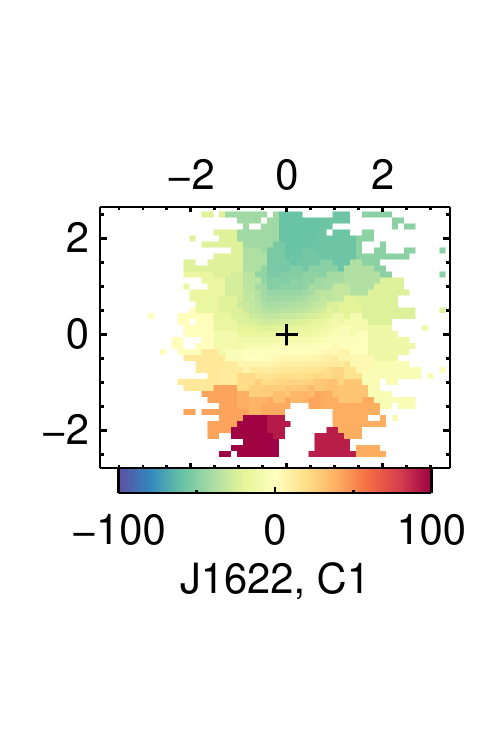}{0.15\textwidth}{}
\fig{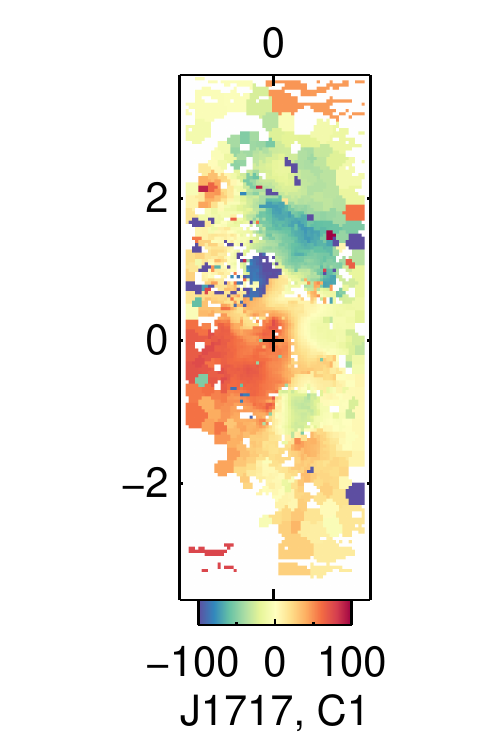}{0.15\textwidth}{}}
\caption{ \footnotesize Velocity maps with the v$_{50}$ of the C1 component measured in km s$^{-1}$ for all the SF dwarfs.}
\label{fig:v50_c1}
\end{figure*}

\begin{figure*}[t!]
\gridline{\fig{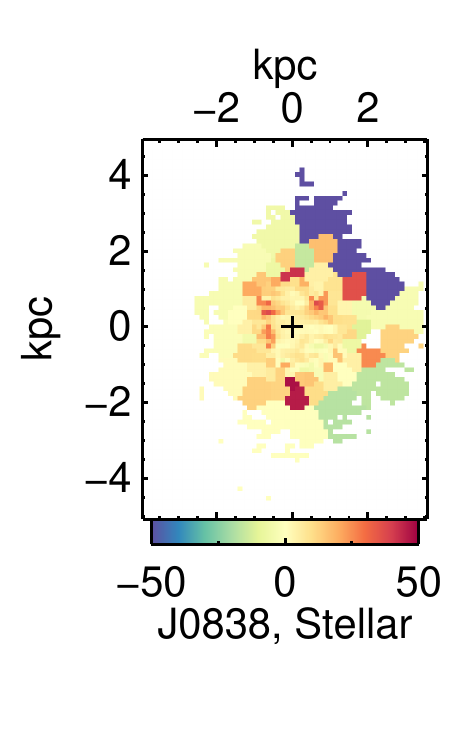}{0.16\textwidth}{}
\fig{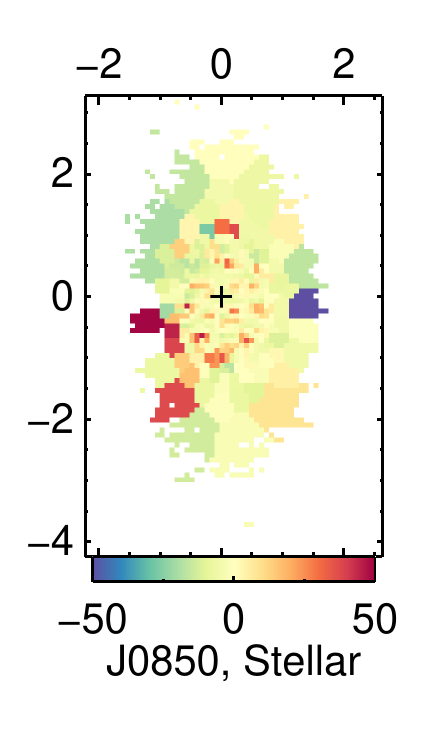}{0.16\textwidth}{}
\fig{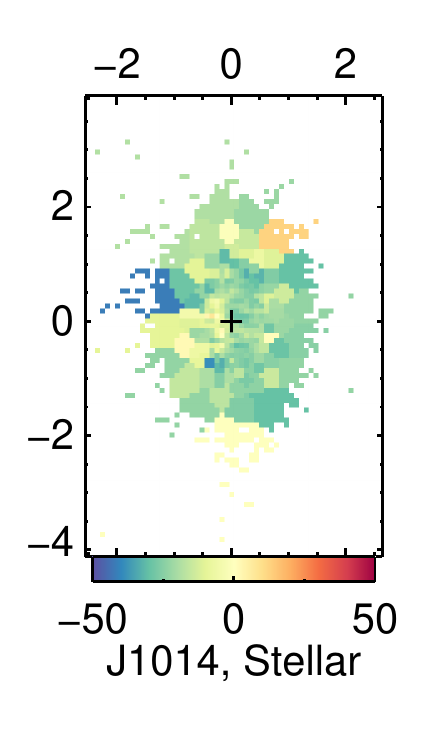}{0.16\textwidth}{}
\fig{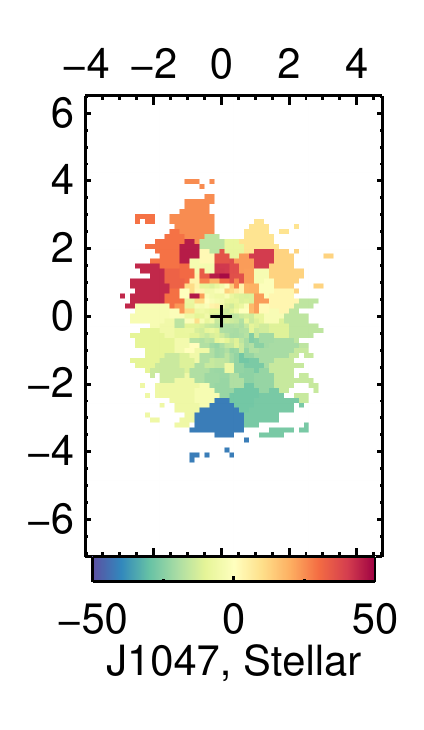}{0.16\textwidth}{}
\fig{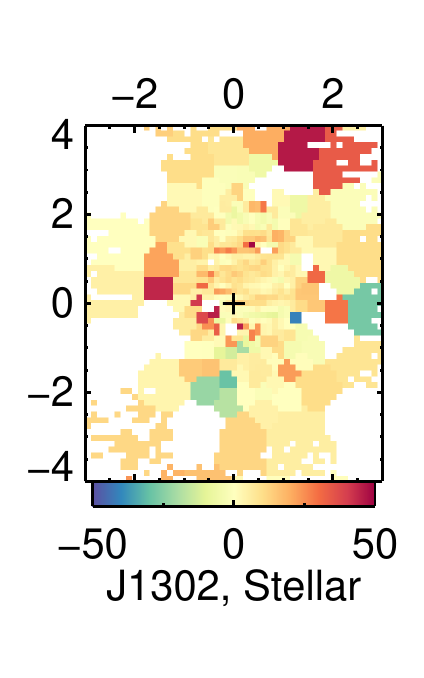}{0.16\textwidth}{}}
\gridline{\fig{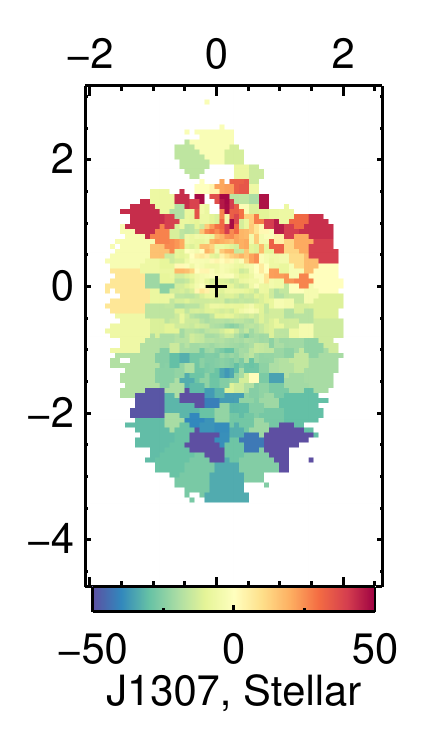}{0.16\textwidth}{}
\fig{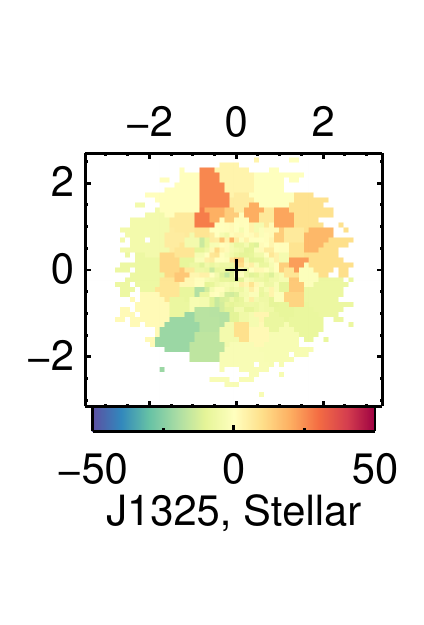}{0.16\textwidth}{}
\fig{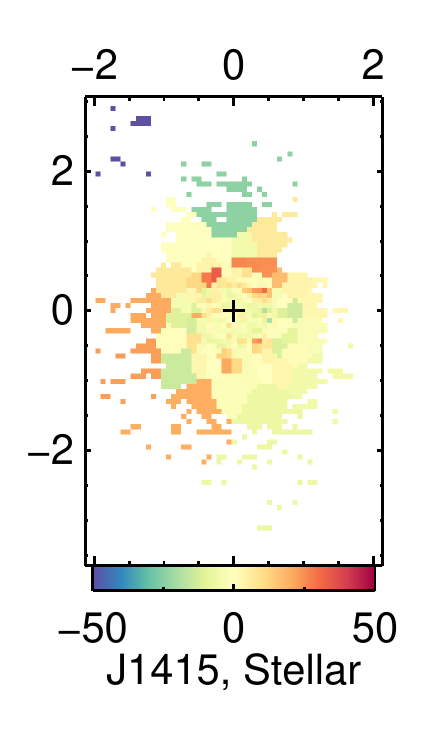}{0.16\textwidth}{}
\fig{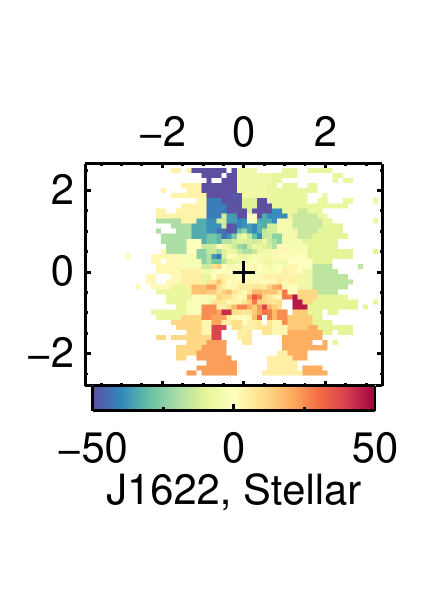}{0.16\textwidth}{}
\fig{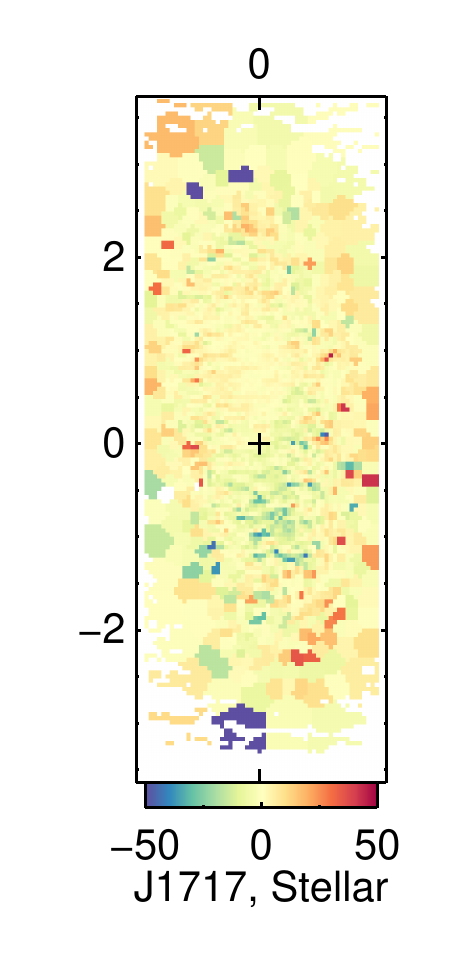}{0.16\textwidth}{}}
\caption{ \footnotesize Velocity maps with the v$_{50}$ of the stellar component measured in km s$^{-1}$ for all the SF dwarfs.}
\label{fig:v50_stellar}
\end{figure*}

\section{Outflows detected in the Sample} \label{sec:Outflows}
\subsection{Gas Kinematics} \label{kin}
We calculate the median, minimum, and maximum values of v$_{50}$ and W$_{80}$ for the emission lines in our targets, similar to the methodology adopted by L20 for the AGN dwarfs. v$_{50}$ is defined as the median velocity at the 50th percentile of the total flux, while W$_{80}$ is the line width that includes 80\% of the total flux. W$_{80}$ is calculated as v$_{10}$-v$_{90}$, where v$_{10}$ and v$_{90}$ are the velocities at the 10th and 90th percentile of the flux, calculated from the red side of the line. For a schematic diagram depicting the different quantities mentioned, refer to Figure 3 in L20.

These values were determined relative to the systemic velocity of the galaxy, which we determined by fitting stellar absorption features in the integrated spectrum of the data cube. Uncertainties in the velocity measurements were estimated by Monte Carlo methods over 100 iterations.

Measurement uncertainties in v$_{50}$ estimated in this way are typically below 5 km s$^{-1}$. However, velocity maps (see Figure \ref{fig:v50}) show bin-to-bin variations in the order of 20 km s$^{-1}$ in some cases. These fluctuations are more likely to be due to error measurements than true changes in velocity. Therefore, we adopt 20 km s$^{-1}$ as a conservative error in our velocity measurements. Our measurements for the whole sample are summarized in Table \ref{table:kinem}. 

A maximum of two components were fit to the emission lines for each of the targets. The C1 component refers to the narrow component, and C2 refers to the broad component. L20 assumed that C2, or in some cases C3, refers to the outflow component since nearly all of their C2/C3 v$_{50}$ values were blue-shifted, which is a clear indicator of outflows. In our case, the broad components were at approximately systemic velocity or even red-shifted,  with only certain regions showing significant blue-shifted velocity (Figure \ref{fig:v50}).
We only consider spaxels with a blue-shifted velocity relative to systemic to constitute an outflow. The red-shifted values of the C2 v$_{50}$ could indicate disturbed or stirred gas in the galaxy. If we choose the red-shifted spaxels to constitute the outflows as well, then that would indicate a larger fraction of the spaxels contributes to outflowing material, which could change the values of the energetics carried out by the flow. This could lead to either over-estimating or under-estimating the effect of the outflow. Since we do not have a good indicator to determine whether the red-shifted spaxels truly constitute an outflow, we tend to agree with the generally used assumption that only the blue-shifted spaxels are likely to depict an outflow. However, for completeness, we consider both cases (using blue-shifted vs using both blue- and red-shifted spaxels) in our analysis of the energetics of the outflows (see Section \ref{sec:discussion}).

On average, the median values of v$_{50}$ for the C2 component was close to 0 km s$^{-1}$. This could be due to the fact that most of the spaxels did not have a significant velocity offset. However, since we considered only the blue-shifted velocity values to constitute the outflows, the average of the minimum (maximally blue-shifted) values of v$_{50}$ was found to be $\sim -60$ km s$^{-1}$. From the velocity maps in Figure \ref{fig:v50}, we see that blue-shifted outflows constitute a small fraction of the kinematics in the galaxy. The C2 W$_{80}$ velocity maps are plotted in Figure \ref{fig:w80} and show a significant spread in the dispersion values. The average of the median values of W$_{80}$ were $\sim$ 480 km s$^{-1}$. 

As mentioned earlier, the shape of the line profile of the broad component in our targets is not preferentially blue-shifted and is symmetric, in contrast to the clear blue-shifted line profile in L20's AGN dwarfs. This is consistent with previous observations of line profiles associated with star-formation-driven outflows \citep{2017A&A...606A..36C,2019ApJ...873..122D, 2022MNRAS.514.4828M}. The symmetric shape of the line profiles in the star-forming dwarfs could indicate lower extinction along the line of sight in at least some of these outflows. AGN-powered outflows originate from the dusty centers of galaxies, and so they are affected by extinction which would lead to their profiles being more blue-shifted \citep{1981ApJ...247..403H}. However, stellar-driven outflows are not restricted to originate from the centers of galaxies, as energetic stellar processes can occur anywhere in the galaxy. Any outflows near the edge of the galaxies, i.e., those that are least affected by extinction, are more likely to dominate the flux of the final spectrum, leading to a profile that appears to be more symmetric, with velocity values centered around 0 km s$^{-1}$ (S.\ Kadir et al., in prep.)

\subsection{Gas and Stellar kinematics}

\begin{figure}[h]
\gridline{\fig{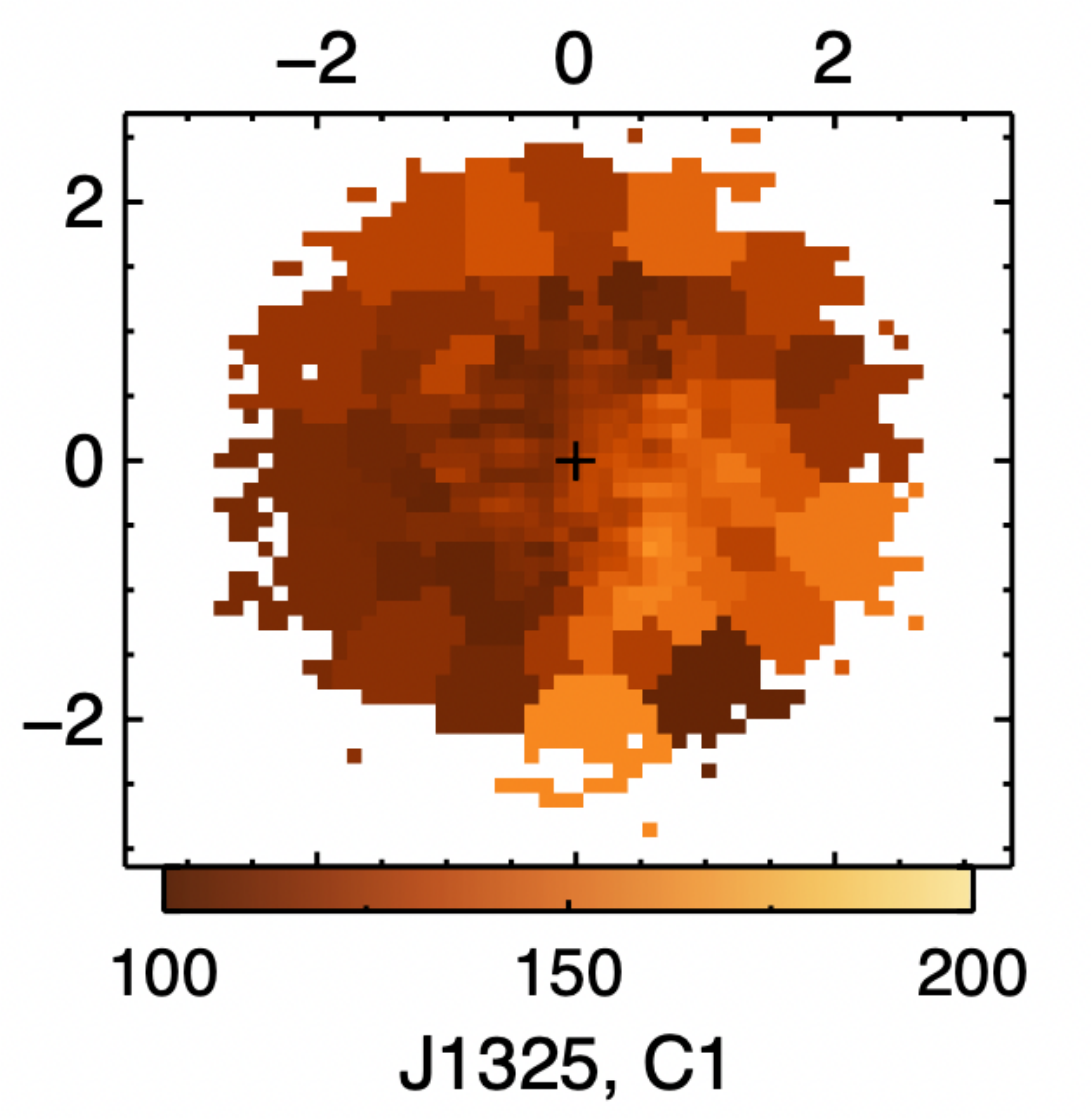}{0.255\textwidth}{(a)}
\fig{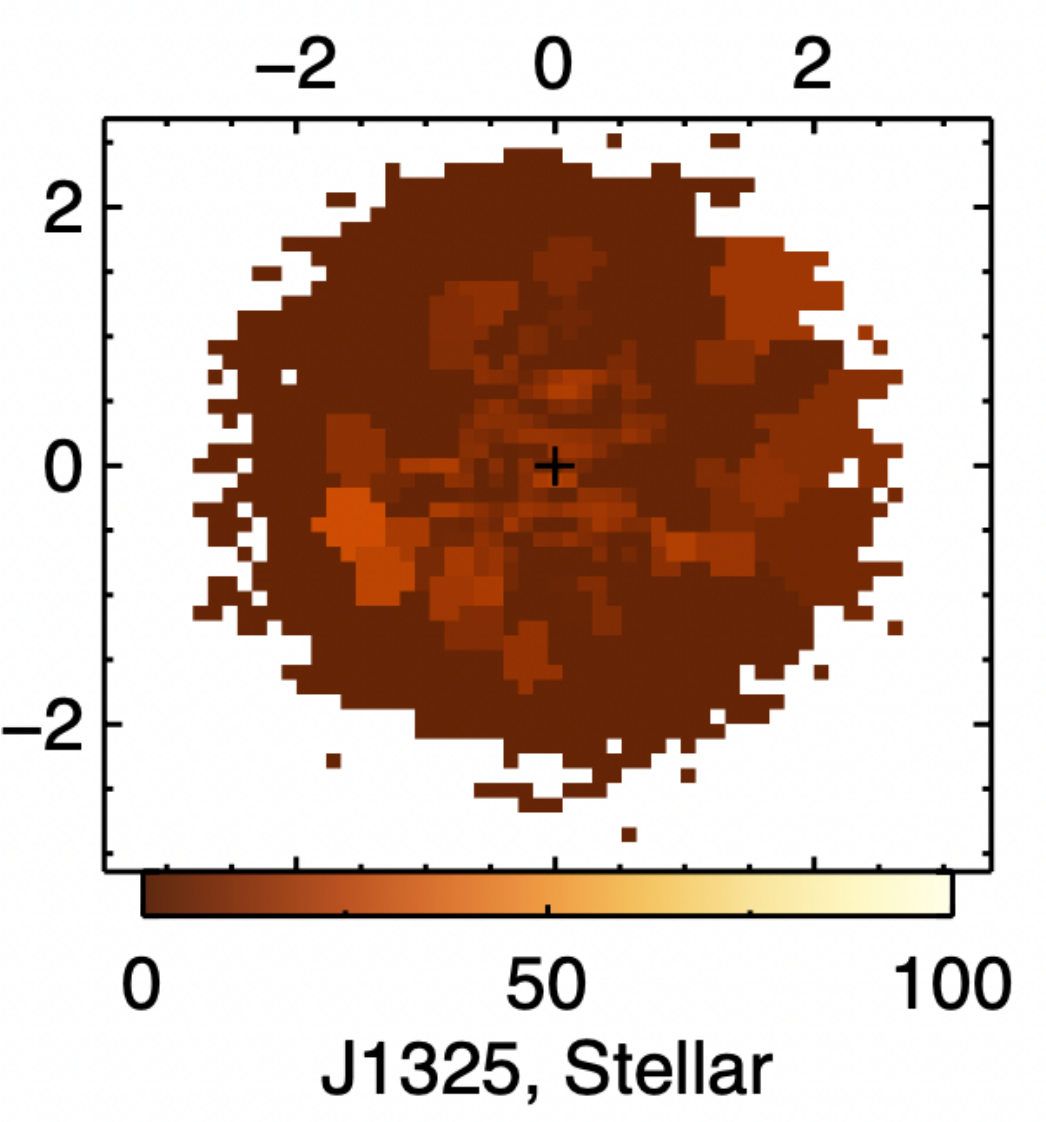}{0.245\textwidth}{(b)}}
\caption{ \footnotesize Example of the distribution of (a) W$_{80}$ for the C1 component and (b) the stellar velocity dispersion ($\sigma$) for J1325. The rest of the targets also show similar W$_{80}$ and $\sigma$ maps indicating minimal dispersion, in contrast to the W$_{80}$ maps of the C2 component.\label{fig:c1_stel_w80}}
\end{figure}

We plot the v$_{50}$ velocity maps for the C1 component and the stellar fields in Figures \ref{fig:v50_c1} and \ref{fig:v50_stellar}, respectively on, page 9. The C1 component represents the gas that follows the stellar component in the galaxy. Comparing Figures \ref{fig:v50} and \ref{fig:v50_c1}, we see that for nearly all the targets, the C1 and C2 components have distinct kinematics, implying that the two components are likely tracing different physical processes. For instance, the velocity fields for the C1 and C2 components for J1622 in Figures \ref{fig:v50} and \ref{fig:v50_c1} are perpendicular to each other, possibly indicating that the outflow is oriented perpendicular to the rotation of the galaxy. As mentioned earlier, the gas in the C1 component follows the stellar component in the galaxy as can be readily seen by comparing Figures \ref{fig:v50_stellar} and \ref{fig:v50_c1}. We see signs of rotation in the stellar field of some targets, such as J1047, J1307, J1622, and J1717. For the rest of the targets, there does not seem to be a clear rotation in the stellar velocity, indicating that either we are seeing the galaxies face-on in some cases or the galaxies may not be supported by rotation in other cases.

The W$_{80}$ velocity maps of the C1 component of J1325 in Figure \ref{fig:c1_stel_w80} (a) show narrower widths of the line profile as compared to those of the C2 component. This trend was prevalent among all objects in the sample, possibly indicating that the C2 component is likely tracing processes like outflows that can cause the broadening of the line profiles, while the C1 component is relatively quiescent. We refer the reader to the appendix for a more detailed analysis of the v$_{50}$ and W$_{80}$ maps of the C1 and C2 components of the individual targets. The stellar velocity dispersion of all the targets also have low values as indicated by Figure \ref{fig:c1_stel_w80} (b).

\subsection{Spatial extent and radial profiles of the broad component}
Figure \ref{fig:v50} also shows the spatial distribution of the C2 component for each galaxy, with the axes in units of kpc. The outflows appear to
extend to significantly larger radii than the half-light radius of these targets as given by SDSS (see  Table \ref{table:properties}). We also map out the radial profile of the outflows and compare it to the PSF (see appendix \ref{appendix:radial_profiles}). We fit a Moffat profile to a standard star that was observed close to the target to reduce uncertainties caused by changes in atmospheric turbulence and to determine the physical extent of the outflows in our targets. The deviation from the profile of the star for all the targets shows that the outflows are spatially extended.

\subsection{Multiple HII regions} \label{ms}
[O\,III] $\lambda$5007 flux maps of our targets show that multiple HII regions are present in some of the targets (See appendix figures \ref{fig:[OIII5007]_1307}, \ref{fig:[OIII5007]_1717}). These are most apparent in J1307 and J1717, but they are also likely present in J0838, J0850, J1014, and J1302. This makes it challenging to set a single radius for the outflow while determining the energetics (see Section \ref{sec:discussion}), since different regions may have outflows of different sizes. L20 used a single spherical shell model for their sample of AGN dwarfs as AGN-powered outflows originate from a central source and the morphology of the outflows are spherical on the 2D sky-plane. For comparison purposes, we also considered a similar spherical shell model for the SF dwarfs. The radius of the outflow is taken to be the extent to which the outflow component could be measured from the radial profiles and velocity maps. Thus, the results need to be analyzed with caution since the radius of the outflows could be much smaller.

\subsection{Outflow ionization: [O\,III]/H$\mathrm{\beta}$ ionization maps} \label{sec:ionization}

\begin{figure}[h!]
\gridline{\fig{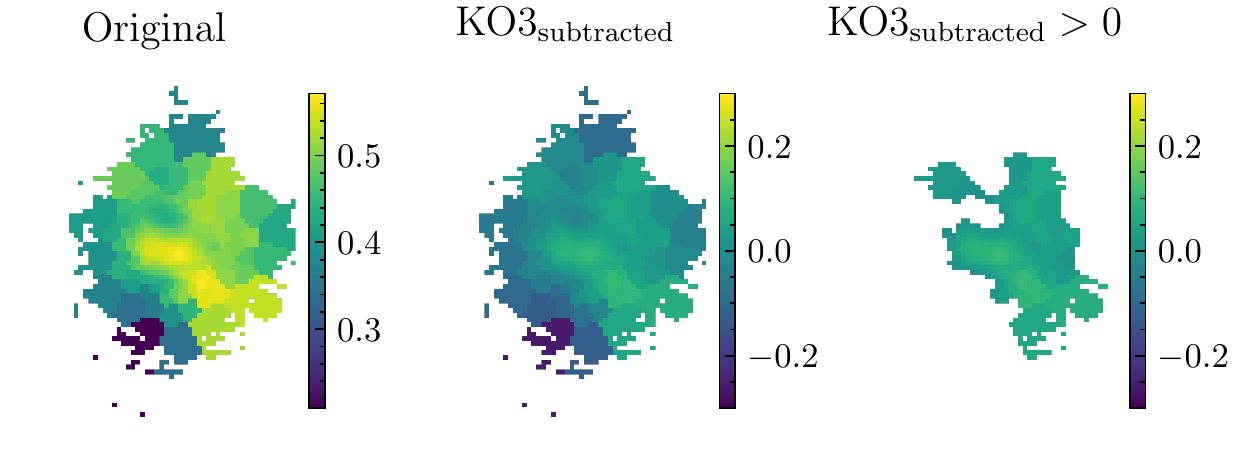}{0.5\textwidth}{(a) J0838}}
\gridline{\fig{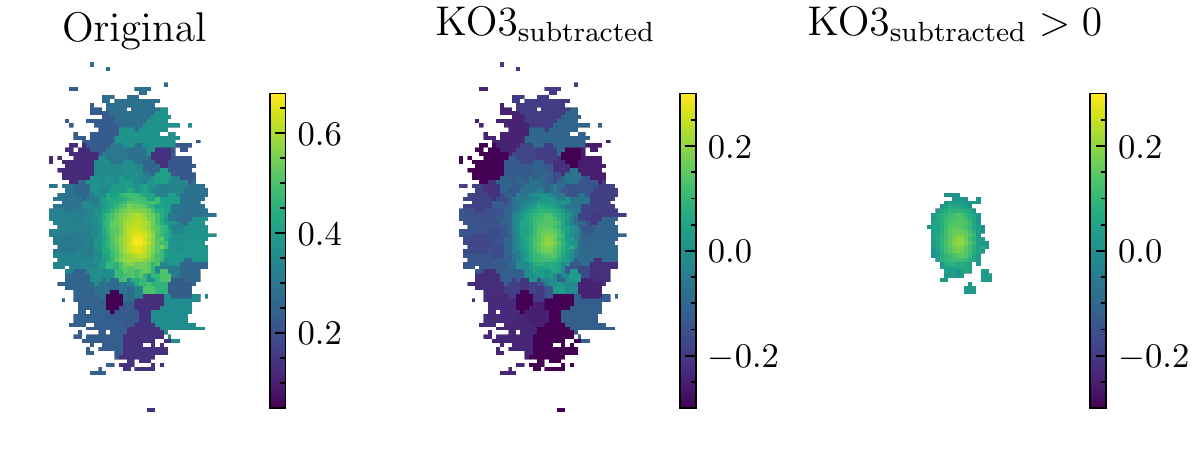}{0.5\textwidth}{(b) J0850}}
\caption{\footnotesize Ionization test maps for (a) J0838, (b) J0850. The first panel in each figure is the actual [O\,III]/H$\beta$ flux map, and the second panel is the flux map after subtracting the expected [O\,III]/H$\beta$ value. The third panel shows if any part of the galaxy has a [O\,III]/H$\beta$ line ratio in the composite region. For (a) and (b), some parts have a higher line ratio than expected for starforming regions from the BPT diagram. These are the only two targets in the sample with signatures of potential AGN contamination. \label{fig:ionization}}
\end{figure}

Since our main goal is to compare outflows powered by star formation to those powered by AGN, it is important to make sure the galaxies in our sample do not host any AGN activity. The targets in our sample are classified as star-forming based on integrated SDSS spectra (Figure \ref{fig:bpt}). However, a weak AGN signal could be diluted by the strong emission lines associated with star-forming regions. Using MANGA IFU data, \cite{2020ApJ...898L..30M} find AGN activity in individual spaxels based on ionized line ratios of the lines used in the BPT diagram that is not apparent in the integrated spectra of the same targets. We take a similar approach for our data cubes to ensure that all of the ionization is likely caused by SF and to rule out any possibility of AGN activity that can manifest itself in the various comparison ratios.

To do this, we obtained the [O\,III]/H$\beta$ ratio for each spaxel and compared it to the value of the Kauffmann demarcation line \citep[K03;][]{2003MNRAS.341...33K} between AGN and SF. We chose the Kauffman demarcation line because it has lower limits than the Kewley line \citep{2001ApJ...556..121K}. Since the KCWI wavelength coverage does not extend beyond 5500 $\text{\AA}$, we were unable to measure the [N\,II]/H$\alpha$ ratio directly. So for each target, we obtained the value of [N\,II]/H$\alpha$ from SDSS and determined the corresponding [O\,III]/H$\beta$ ratio on the Kauffmann line. Note that this value of [N\,II]/H$\alpha$ is obtained using an integrated aperture and thus may not be indicative of the actual spaxel-to-spaxel variation of the line ratio. This could lead to the biasing of the ratio towards the centers of the galaxy and lead to significant errors while determining values in the outer regions. Being mindful of this caveat, we subtracted the obtained value from our calculated value and plotted those spaxels where the difference was greater than 0 (see Figure \ref{fig:ionization}). If the difference is greater than 0, this indicates that the ratio is larger than the accepted value for possible star-forming origins.

For 8 of the 10 targets, no part of the galaxy indicated [O\,III]/H$\beta$ ratios above the star-forming line. In two targets, J0838 and J0850, the ratios in some spaxels were 0.2 dex greater than the expected value; Figure~\ref{fig:ionization} shows the specific location of these spaxels. While the larger [O\,III]/H$\beta$ ratios in these spaxels are likely due to shocks, we cannot rule out the possibility that there may be some AGN contamination in these two targets. However, all the outflow properties in these two targets appear to be consistent with the rest of the sample, so their outflows are most likely powered by star formation.
\subsection{Electron densities}
The electron densities of the targets can be obtained from the [S\,II] line ratios or the [O\,II] line ratios from the relation given by \cite{2016ApJ...816...23S}. We took the [S\,II] line ratios from SDSS for all our targets (similar to L20). The values of electron densities (average value of $\sim$ 165 cm$^{-3}$) were lower compared to the values for the AGN dwarfs (average value of $\sim$ 400 cm$^{-3}$). This difference between the two samples could be because AGNs are capable of producing strong ionization and thus lead to higher values of the ionized electron densities than the SF dwarf galaxies.

\subsection{Dust extinction} \label{dust}
Dust extinction in the targets was measured using the H$\beta$/H$\gamma$ emission line ratios of the integrated spectrum of the targets. The extinction was calculated using the extinction curve from \citet{1989ApJ...345..245C} with an $R_v$=3.1 and an intrinsic H$\beta$/H$\gamma$ ratio of 2.13 using  Case B from \citet{2006agna.book.....O} at a temperature of $10^4$ K. We also compared these values to the extinction calculated from the H$\alpha$/H$\beta$ lines ratios obtained from SDSS, and they were comparable within error limits. So we used the values of extinction $A_v$ determined from the H$\beta$/H$\gamma$ ratios for further calculations. The values of the electron densities and extinction for the SF dwarfs are given in columns (3) and (4) in Table \ref{table:energetic}. 

\subsection{Energetics of the Outflows} \label{sec:discussion}
Following L20, the ionized gas masses were calculated using the H$\beta$ luminosity of the outflowing gas. We converted the H$\beta$ luminosity to the H$\alpha$ luminosity by a correction factor $L_{H\alpha,corr}$ = 2.86 $L_{H\beta,corr}$. The correction factor is the intrinsic H$\alpha$/H$\beta$ ratio for HII regions, which is appropriate for the case B recombination assumed from \cite{2006agna.book.....O} with T= 10$^4$ K. We also corrected for extinction based on an intrinsic H$\beta$/H$\gamma$ ratio of 2.13 (for more details, refer to Section \ref{dust}). The equation for the ionized mass is adapted from Equation (29) in \citet{2020A&ARv..28....2V} by assuming a solar metallicity and is given as:
\begin{equation}
    \mathrm{M_{out}= 4.48\hspace{0.1cm}M_{\odot}\Big(\frac{L_{H\alpha,corr}}{10^{35}\hspace{0.1cm} erg s^{-1}} \Big) \Big(\frac{\langle n_e \rangle}{100\hspace{0.1cm}cm^{-3} } \Big)^{-1}} 
\end{equation}\label{eq1}

We also calculate the mass loss rates (dM/dt), the momentum rates in units of energy (cdp/dt), and the kinetic energy rates (dE/dt). We assumed a time-averaged, thin-shell free wind model, which has been previously used by \cite{2010ApJ...724.1430S}, \cite{2013ApJ...768...75R} and L20, with a spherically symmetric outflow.
\begin{longrotatetable}
\begin{deluxetable*}{cccccccccccc} \label{energetics}
\tablenum{3}
\tablecaption{Energetics of the Outflows}
\tablewidth{0pt}
\tablehead{
\colhead{Name} & \colhead{Comp} & \colhead{$\mathrm{n_e}$(cm$^{-3}$)}& \colhead{Extinction(E(B-V))} & \colhead{log(M/M$_{\odot}$)} & \colhead{$\mathrm{R_{out}}$(kpc)} & \multicolumn{2}{c}{log[(dM/dt)/(M$_{\odot}$ yr$^{-1}$)]} & \multicolumn{2}{c}{log[(dE/dt)/(ergs s$^{-1}$)]} & \multicolumn{2}{c}{log[(cdp/dt)/(L$_{\odot}$)]}\\
\cline{7-8}\cline{9-10}\cline{11-12}
& & & & & & Blue & Red+Blue & Blue & Red+Blue & Blue & Red+Blue}
\decimalcolnumbers
\startdata
J0838 & C1 & 121$\pm$0 & 0.2 & 4.4$_{-0.2}^{+0.2}$ & 2.9 & -4.1$_{-0.2}^{+0.2}$ & -1.9$_{-0.2}^{+0.2}$ & 37.0$_{-0.2}^{+0.2}$ &
37.0$_{-0.2}^{+0.2}$ & 4.8$_{-0.2}^{+0.2}$ & 7.4$_{-0.2}^{+0.2}$\\	 & C2 &  &  & 5.9$_{-0.2}^{+0.2}$ & 2.9 & -2.3$_{-0.2}^{+0.2}$ &  -2.0$_{-0.2}^{+0.2}$ & 36.7$_{-0.2}^{+0.2}$ & 34.0$_{-0.2}^{+0.2}$  & 7.0$_{-0.2}^{+0.2}$ & 7.4$_{-0.2}^{+0.2}$\\
J0850 &	C1 & 150$\pm$0 & 0.2 & 5.0$_{-0.2}^{+0.2}$ & 1.2 & -2.8$_{-0.2}^{+0.2}$ & -2.0$_{-0.2}^{+0.2}$ & 36.2$_{-0.2}^{+0.2}$ & 37.03$_{-0.2}^{+0.2}$ & 6.5$_{-0.2}^{+0.2}$ & 7.6$_{-0.2}^{+0.2}$ \\
 & C2 &  &  & 4.4$_{-0.2}^{+0.2}$ & 1.2 & -3.5$_{-0.2}^{+0.2}$ & -2.5$_{-0.2}^{+0.2}$ &	35.5$_{-0.2}^{+0.2}$ & 36.7$_{-0.2}^{+0.2}$ &	5.8$_{-0.2}^{+0.2}$ & 6.9$_{-0.2}^{+0.2}$\\	
J1014 & C1 & 63$\pm$24 & 0.1 & 5.7$_{-0.2}^{+0.3}$ & 2.7 & -1.9$_{-0.2}^{+0.3}$ & -2.3 $_{-0.2}^{+0.3}$ & 37.3$_{-0.2}^{+0.3}$ & 36.2$_{-0.2}^{+0.3}$ &	7.6$_{-0.2}^{+0.3}$ & 6.8$_{-0.2}^{+0.3}$ \\
 & C2 & & & 4.6$_{-0.2}^{+0.3}$ & 2.7  & -1.9$_{-0.2}^{+0.3}$ & -2.5$_{-0.2}^{+0.3}$ & 37.3$_{-0.2}^{+0.3}$ & 36.2$_{-0.2}^{0.3}$ &	7.6$_{-0.2}^{+0.3}$ & 6.7$_{-0.2}^{+0.3}$ \\
J1047 & C1 & 265$\pm$2 & 0.2 & 3.0$_{-0.3}^{+0.4}$ & 2.0 & -5.1$_{-0.3}^{+0.4}$ & -2.4$_{-0.3}^{+0.4}$ & 33.2$_{-0.3}^{+0.3}$ & 35.8$_{-0.3}^{+0.3}$ & 3.9$_{-0.3}^{+0.4}$ & 7.0$_{-0.3}^{+0.4}$ \\	
 & C2 & & & 2.0$_{-0.3}^{+0.4}$ & 2.0  & -1.8$_{-0.3}^{+0.4}$ & -2.7$_{-0.3}^{+0.4}$ & 37.3$_{-0.3}^{+0.4}$ & 36.2$_{-0.3}^{+0.4}$ & 7.6$_{-0.3}^{+0.4}$ & 6.7$_{-0.3}^{+0.4}$ \\
J1302 & C1 & 146$\pm$2 & 0.2 & 4.6$_{-0.2}^{+0.2}$ & 2.1 & -3.3$_{-0.2}^{+0.2}$ &
-2.4$_{-0.2}^{+0.2}$ & 35.5$_{-0.2}^{+0.2}$ & 37.1$_{-0.2}^{+0.2}$ & 6.0$_{-0.2}^{+0.2}$ & 7.0$_{-0.2}^{+0.2}$ \\	
 &	 C2 & & & 3.9$_{-0.2}^{+0.2}$ & 2.5    & -4.2$_{-0.2}^{+0.2}$ & -2.6$_{-0.2}^{+0.2}$ & 37.2$_{-0.2}^{+0.2}$ & 37.2$_{-0.2}^{+0.2}$ & 5.1$_{-0.2}^{+0.2}$ & 7.0$_{-0.2}^{+0.2}$ \\
J1307 & C1 & 113$\pm$15 & 0.2 &	6.0$_{-0.2}^{+0.2}$ & 1.9 & -1.8$_{-0.2}^{+0.2}$ & -2.1$_{-0.2}^{+0.2}$ & 34.6$_{-0.2}^{+0.2}$ & 36.8$_{-0.2}^{+0.2}$ & 7.6$_{-0.2}^{+0.2}$ & 7.2$_{-0.2}^{+0.2}$ \\	
 &	C2 & &  & 5.2$_{-0.2}^{+0.2}$ & 1.4 & -2.7$_{-0.2}^{+0.2}$ & -2.5$_{-0.2}^{+0.2}$ & 35.7$_{-0.2}^{+0.2}$ & 37.3$_{-0.2}^{+0.2}$ & 6.4$_{-0.2}^{+0.2}$ & 7.2$_{-0.2}^{+0.2}$ \\
J1325 &	C1 & 233$\pm$40 & 0.1 &	5.4$_{-0.2}^{+0.2}$ & 1.6 & -3.0$_{-0.2}^{+0.2}$ & -3.2$_{-0.2}^{+0.2}$ & 34.1$_{-0.2}^{+0.2}$ & 35.0$_{-0.2}^{+0.2}$ & 5.5$_{-0.2}^{+0.2}$ & 5.7$_{-0.2}^{+0.2}$ \\
  & C2 & &  & 5.3$_{-0.2}^{+0.2}$ &	1.5   & -2.2$_{-0.2}^{+0.2}$ & -2.8$_{-0.2}^{+0.2}$ & 35.8$_{-0.2}^{+0.2}$ & 36.3$_{-0.2}^{+0.2}$ & 7.1$_{-0.2}^{+0.2}$ & 6.6$_{-0.2}^{+0.2}$ \\	
J1415 &	C1 & 271$\pm$18 & 0.2 & 4.7$_{-0.2}^{+0.2}$ & 1.5 & -3.7$_{-0.2}^{+0.2}$ & -3.0$_{-0.2}^{+0.2}$ & 34.1$_{-0.2}^{+0.2}$ & 35.9$_{-0.2}^{+0.2}$ & 5.0$_{-0.2}^{+0.2}$ & 6.1$_{-0.2}^{+0.2}$ \\	
 &	C2 & & & 5.3$_{-0.2}^{+0.2}$ &	1.4   & -2.5$_{-0.2}^{+0.2}$ & -3.1$_{-0.2}^{+0.2}$ & 35.2$_{-0.2}^{+0.2}$ & 35.3$_{-0.2}^{+0.2}$ & 6.5$_{-0.2}^{+0.2}$ & 6.0$_{-0.2}^{+0.2}$ \\	
J1622 &	C1 & 193$\pm$5 & 0.1 & 6.0$_{-0.2}^{+0.2}$ & 3.2 & -2.0$_{-0.2}^{+0.2}$ &
-2.2$_{-0.2}^{+0.2}$ & 36.7$_{-0.2}^{+0.2}$ & 36.5$_{-0.2}^{+0.2}$ & 7.0$_{-0.2}^{+0.2}$ & 7.0$_{-0.2}^{+0.2}$ \\	
 & C2 & & & 5.8$_{-0.2}^{+0.2}$ & 2.9	& -2.1$_{-0.2}^{+0.2}$ & -2.3$_{-0.2}^{+0.2}$ & 37.2$_{-0.2}^{+0.2}$ & 37.0$_{-0.2}^{+0.2}$ & 7.4$_{-0.2}^{+0.2}$ & 7.2$_{-0.2}^{+0.2}$ \\	
J1717 &	C1 & 85$\pm$11 & 0.2 & 5.7$_{-0.3}^{+0.3}$ & 3.4 & -2.4$_{-0.3}^{+0.3}$ & -2.0$_{-0.3}^{+0.3}$ & 37.0$_{-0.3}^{+0.3}$ & 37.4$_{-0.3}^{+0.3}$ & 7.3$_{-0.3}^{+0.3}$ & 7.4$_{-0.3}^{+0.3}$ \\	
 & C2 & & & 5.7$_{-0.3}^{+0.3}$ &  3.0  & -2.4$_{-0.3}^{+0.3}$ & -2.4$_{-0.3}^{+0.3}$ & 38.0$_{-0.3}^{+0.3}$ & 37.5$_{-0.3}^{+0.3}$ & 7.4$_{-0.3}^{+0.3}$ & 7.2$_{-0.3}^{+0.3}$  \\	
\enddata
\tablecomments{ \footnotesize Column(1): Short name of the target. Column(2): Component number. Column(3): Electron densities measured from the [SII] line ratios from SDSS. Column(4): The extinction values calculated from the H$\beta$/H$\gamma$ measured from the integrated spectrum of the KCWI data. Column(5): Ionized gas mass of the outflow. Column(6): Outflow radius used in the calculation of the energetics. The mass loss rate calculated using only the blue-shifted spaxels (Column(7)) and both blue-shifted and the red-shifted spaxels (Column(8)) to constitute the outflow. The  energy  rate calculated using only the blue-shifted spaxels (Column(9)) and both the blue-shifted and the red-shifted spaxels (Column(10)). The momentum rate calculated using only the blue-shifted spaxels (Column(11)) and both the blue-shifted and the red-shifted spaxels (Column(12)). The errors in the energetics are dominated by extinction. \label{table:energetic}}
\end{deluxetable*}    
\end{longrotatetable}
The energetics are calculated by summing up quantities over individual spaxels:
\begin{equation}
    \mathrm{dM/dt} = \sum dm/dt = \sum \frac{m_{out} v_{50,out} \mathrm{sec} \theta}{R_{out}}
\end{equation}
\begin{equation}
    \mathrm{dp/dt} = \sum(v_{50,out}\mathrm{sec} \theta)dm/dt
\end{equation}
\begin{equation}
    \mathrm{dE/dt} = \frac{1}{2} \sum[(v_{50,out}\mathrm{sec} \theta)^2 + 3 \sigma_{out}^2]dm/dt,
\end{equation}
where $\mathrm{m_{out}}$ is the outflowing mass in each spaxel, $v_{50,out}$ is the value of v$_{50}$ in each spaxel, and $\sigma_{out}$ is the velocity dispersion calculated as W$_{80}$/2.73. $\mathrm{R_{out}}$ is the radial extent of the spherical outflow and was determined from the maximum radial extents indicated in the C2 velocity maps of the individual targets (Figure \ref{fig:v50} and \ref{fig:w80}). The angle between the velocity vector of the outflow in 3D space and the line of sight, is defined as  $\theta$ = sin$^{-1}$(r$_{\mathrm{spaxel}}$/$\mathrm{R_{out}}$).

The values of the energetics are shown in Table \ref{table:energetic}. The amount of ionized gas mass in Column (5) was calculated using Equation \ref{eq1}, and it gives a measure of the gas moved by the outflows for C2. The values of the energetics in columns 7, 9, and 11 indicate the mass loss rates, energy rates, and momentum rates of the ionized gas, respectively.

As indicated earlier, we only considered the blue-shifted values of v$_{50, out}$ as being part of the outflow. This was added over the entire radial extent to get the total values of the energetics. If we also include the red-shifted spaxels to constitute the outflow, we find that the value of $dE/dt$ is on average 10\% larger than the values calculated for only the blue-shifted spaxels, implying that there would be a small increase in the values of the energetics. The values of the energetics calculated by considering both red-shifted and blue-shifted spaxels to constitute the outflow are given in columns 8, 10, and 12 in Table \ref{table:energetic} for comparison.

Comparing the values given in Tables \ref{table:kinem} and \ref{table:energetic} to the SFRs in Table \ref{table:properties} shows that there is no correlation between SFR (or sSFR) and outflow properties for SF dwarfs.  However, this could be due to the poor constraints we have on SFRs, as noted in Section \ref{section:sfr}.

\subsection{Does the outflowing gas escape the galaxy?}
Similar to the analysis done by L20, we compare the outflow velocities of our sample of star-forming dwarfs to the escape velocities of the galaxies in order to determine whether a significant fraction of the outflowing gas escapes the galaxy. 

While there are several methods to calculate the escape velocities of galaxies, we use abundance matching in order to obtain results that can be compared directly to those of L20. We calculate the escape velocities by first obtaining the halo masses of the galaxies using abundance matching \citep{2013MNRAS.428.3121M}. We assume an NFW dark matter density profile \citep{1996ApJ...462..563N}. Using the integrated spectrum of the broad component of [O\,III] $\lambda$5007 for each galaxy, we find the ratio of the flux that has velocities larger than the escape velocity to the total flux of the line, and we define this as the escape fraction. The escape fraction in percentage for the SF dwarfs is given in Table \ref{table:escapevel}. Note that we define escape fraction based on the flux ratios, not the mass ratios, so the caveats listed in L20 still apply here. Our escape velocities are also measured from the center of the galaxy and should be considered conservative upper limits, as the escape velocities are the largest at the center. 

The escape fractions listed in Table \ref{table:escapevel} demonstrate that only a small fraction of the gas (if any) is able to escape the galaxies. Since escape velocities derived via abundance matching are susceptible to significant systematic biases, we also estimated them using stellar velocities and dispersions. We determine the circular velocity, v$_{\mathrm{circ}}$ = $\sqrt{\mathrm{(v_{*}^2 + \sigma_{*}^2)}}$, where v$_{*}$ and $\sigma_{*}$ is the maximum stellar velocity and stellar dispersion respectively. Note that this is a lower limit, as many of our rotation curves do not extend to the flat part of the curve.  We then used the relation v$_{\mathrm{esc}}$ $\approx$ 3v$_{\mathrm{circ}}$ \citep{2020A&ARv..28....2V} to determine the escape velocity. We find that the escape velocities determined in this way are on average $\sim$30\% higher than the escape velocities given in Table \ref{table:escapevel}, indicating that an even lower fraction of the gas would be likely to escape.

\begin{deluxetable}{ccc}
\tablenum{4}
\tablecaption{\\Escape Fractions of the Targets} 
\tablewidth{0pt}
\tablehead{
\colhead{Name} & \colhead{$v_{esc}$ [km s$^{-1}$]} & \colhead{$f_{esc}$ [\%]}
}
\decimalcolnumbers
\startdata
J0838 & 260 & 2 \\
J0850 & 279 & 0 \\
J1014 & 213 & 0 \\
J1047 & 308 & 1 \\
J1302 & 288 & 1 \\
J1307 & 255 & 0 \\
J1325 & 282 & 1 \\
J1415 & 256 & 1 \\
J1622 & 294 & 1 \\
J1717 & 337 & 0 \\
\enddata
\tablecomments{\footnotesize Column (1): Short name of the target. Column (2): Escape velocities rounded to the nearest whole number. Column (3): Escape fractions calculated, rounded to the nearest whole number.  \label{table:escapevel}}
\end{deluxetable}

\section{Discussion}
\subsection{Comparison with AGN dwarfs}
There is a difference in the kinematics, extents, and energetics between the SF dwarfs and AGN dwarfs. The average of the median values of v$_{50}$ for SF dwarfs, found in Section \ref{kin}, is lower than the average of the median values of the AGN dwarfs ($-$64 km s$^{-1}$). Comparing the maximally blue-shifted velocities also show that SF-driven outflows are slower than AGN-driven outflows (average of $\sim$ $-$65 km s$^{-1}$ for SF dwarfs vs $-$145 km s$^{-1}$ for AGN dwarfs). The median values are red-shifted or centered around 0, compared to the highly blue-shifted values for AGN dwarfs. If we consider just the blue-shifted v$_{50}$ velocities to contribute significantly to the outflow, we see that there are very few blue-shifted spaxels in the broad component of the SF dwarfs, whereas nearly all the spaxels in the broad component of the L20 AGN dwarfs are blue-shifted. This could possibly indicate that strong blue-shifted outflows are not as prominent in SF dwarfs as in AGN dwarfs, and most of the effect of stellar activity may be to stir the gas rather than to expel it from the hosts. 

\begin{figure}[ht!]
\gridline{\fig{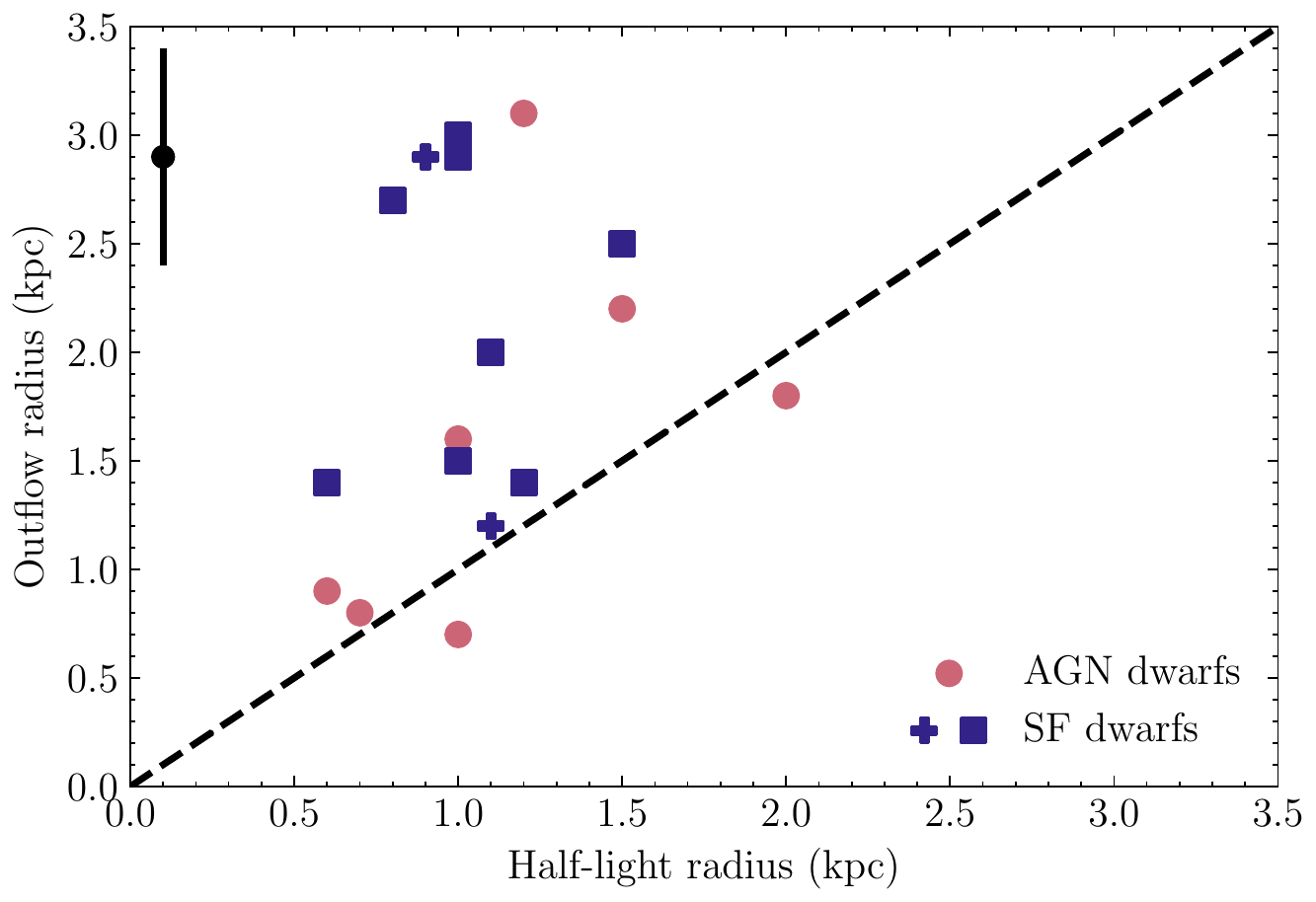}{0.5\textwidth}{}}
\caption{\footnotesize A comparison between the radius of the C2 component (outflow radius) and the half-light radius of the targets. The values for the half-light radius for SF dwarfs are indicated in Table \ref{table:properties}, and the values for the half-light radius for the AGN dwarfs are taken from L20. A cross symbol is used to indicate the two SF targets that could potentially have AGN contamination based on their BPT line ratios; see Section \ref{sec:ionization}. The dotted line represents equal values for the outflow radius and half-light radius. The average value for the errors in the data points for the SF dwarfs is indicated by the black sign.\label{fig:radial}}
\end{figure}

The overall extent of the stellar-driven outflows is greater than that of the AGN-driven outflows as can be clearly seen in Figure \ref{fig:radial}. Note that the detectability limits were comparable in the measurement of outflow extents in both the SF and AGN samples. Therefore the difference in sizes is real and not the result of different sensitivities in the two samples. The AGN-driven outflows extend to distances between 0.3 kpc to 3.1 kpc, with an average of 1.2 kpc. On the other hand, the stellar-powered outflows extend to further distances, between 1.2 kpc and 3.2 kpc, with an average of 2.2 kpc. This could be because the stellar-powered outflows may originate at multiple places within the galaxy (see Section \ref{ms}), and are not necessarily confined to the center as AGN-powered outflows are. They can thus reach to larger scales, even though they might not be as fast as AGN-powered outflows.

\begin{figure*}
\gridline{\fig{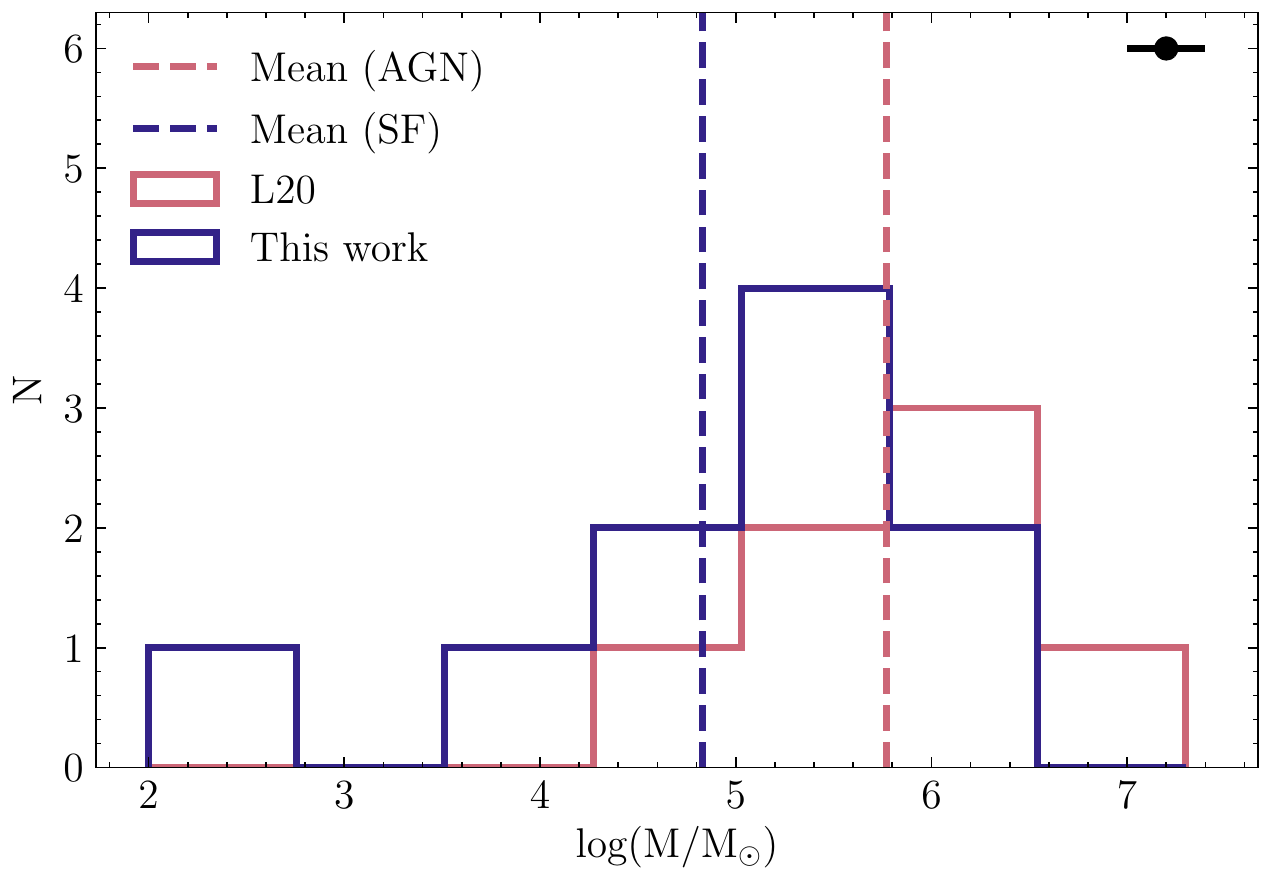}{0.5\textwidth}{(a)}
          \fig{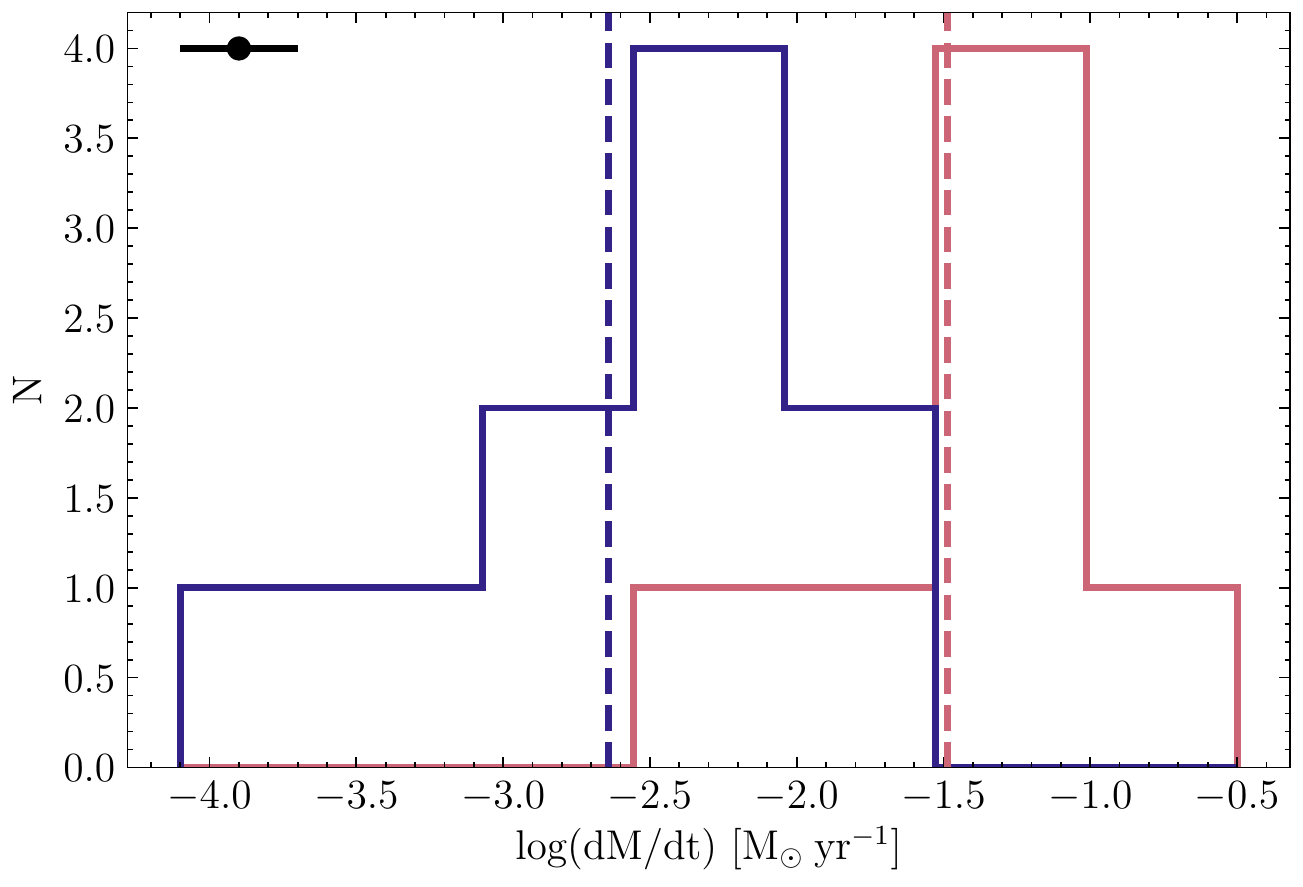}{0.5\textwidth}{(b)}
          }
\gridline{\fig{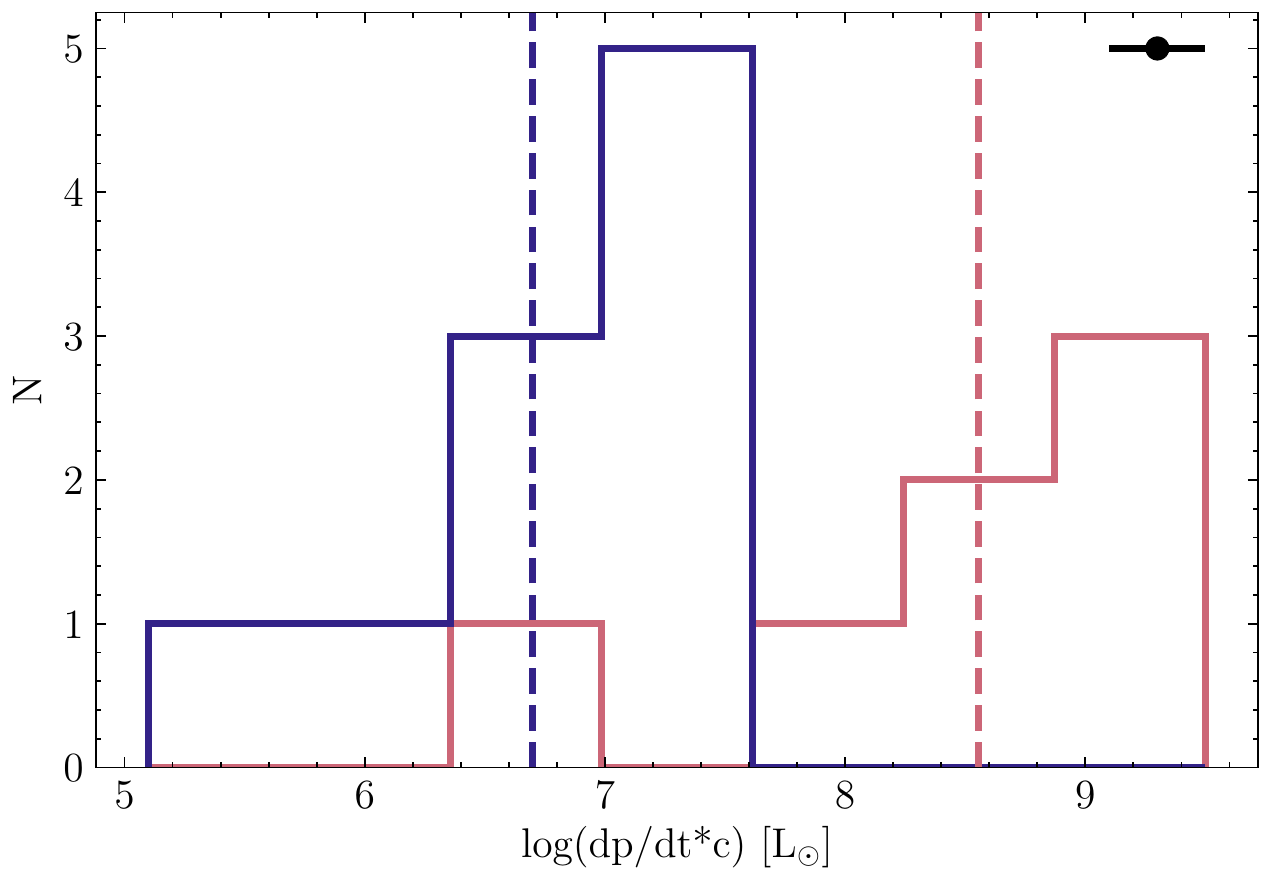}{0.5\textwidth}{(c)}
          \fig{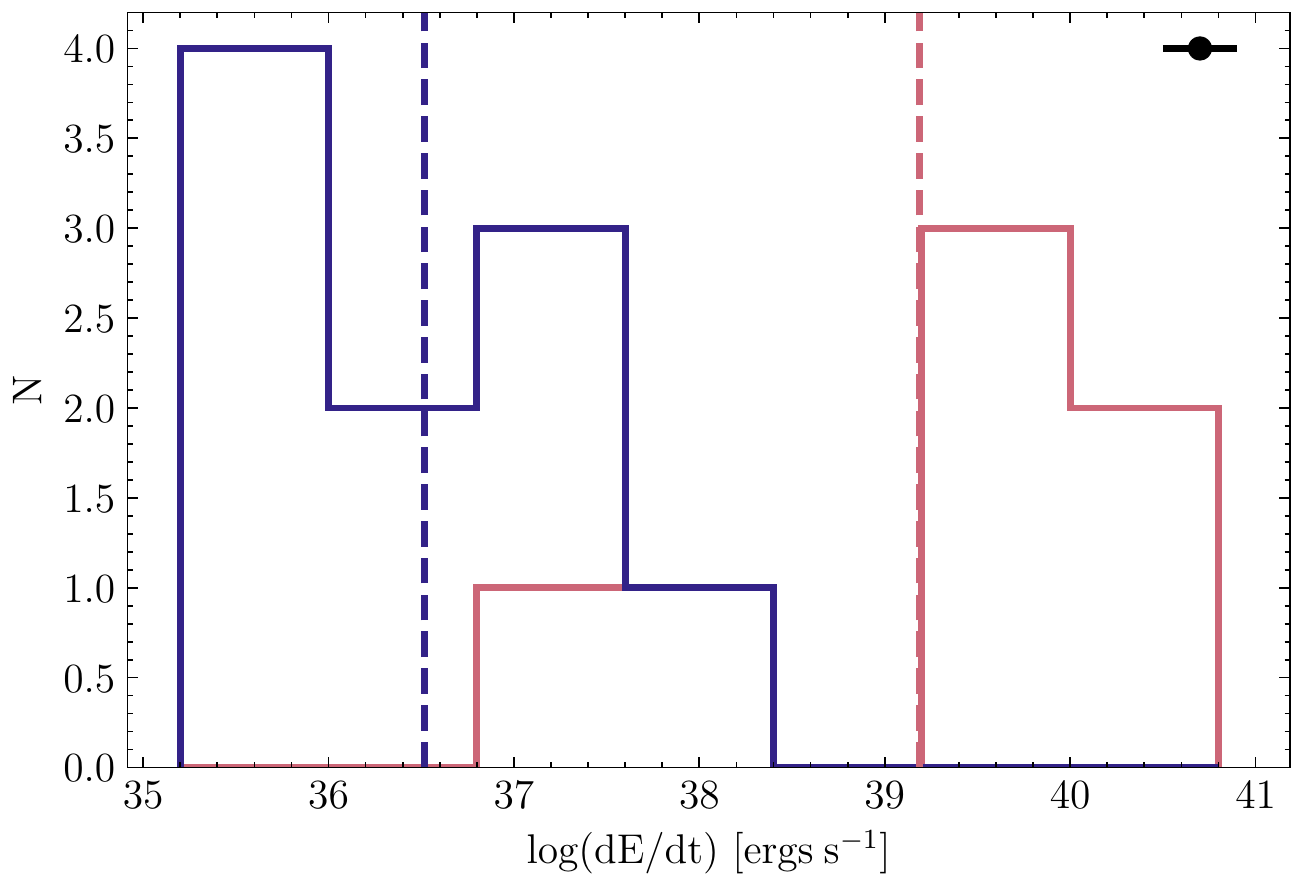}{0.5\textwidth}{(d)}
          }
\caption{\footnotesize Histograms comparing the energetics of the outflow component calculated in AGN dwarfs and SF dwarfs. Figure 12(a) compares the amount of mass contained in the outflow, and Figures 12(b), 12(c), and 12(d) compare the outflow rates of mass, momentum, and kinetic energy, respectively. The dotted lines represent the median values of either sample. The values for the SF dwarfs across all the energetics are lower than the AGN dwarfs. The average value for the errors in the data points for the AGN and SF dwarfs is indicated by the black sign.
\label{fig:energetics}}
\end{figure*}

The mass, momentum, and kinetic energy rates are all much lower for the SF dwarfs than for the AGN dwarfs, as indicated in Figure \ref{fig:energetics}. As mentioned earlier, assuming both the red-shifted and blue-shifted spaxels contribute to the outflows, we find that the value of the energetics increases by a small fraction, but overall, they are still lower by two orders of magnitude when compared to the AGN dwarfs.

The differences between the quantities in the two samples increase from the mass loss rates to the momentum rates and are maximum for the energy rates. These calculations depend on both line luminosities and velocity terms. While the [O\,III] $\lambda$5007 luminosities of the AGN dwarfs are comparable to those of the star-forming dwarfs, the median $\mathrm{v_{50}}$ velocities of the AGN-driven outflows are much higher than those of the SF-driven outflows.

The radial extent of the outflows could also play a role in lowering the energetics for the SF dwarfs, as the outflows are more radially extended than those of the AGN dwarfs.

The escape fractions from Table \ref{table:escapevel} for the SF galaxies are comparable to the escape fractions of the AGN galaxies. However, the AGN dwarfs have a maximum of 6\% for two galaxies. Although these fractions are not significant enough to translate to an effective portion of the gas being ejected from the host galaxies by the outflows, it could still show that the AGN dwarfs could play a more effective role in stirring up gas in the dwarf galaxies, which can still considerably affect its environment. The escape velocities are also upper limits as they are calculated from the center of the galaxy, and we would also need to factor in the frequencies of these outflows.

\subsection{Impact on the galaxy}

There is a significant difference between the outflows powered by AGN and those powered by stellar processes. The lower speeds of the SF-driven outflows as compared to the AGN-driven outflows indicate that the gas is moving more slowly, and the lower energetic rates indicate that they are possibly not transporting large amounts of mass or energy from the galaxy. The greater values of the kinematics and energetics from the AGN-driven outflows suggest that AGNs could play a more significant role in stripping the galaxy of gas and halting or slowing down star formation. Thus it is important to consider AGN contribution to feedback in dwarf galaxies in the study of galaxy evolution.

\begin{figure}[ht!]
\gridline{\fig{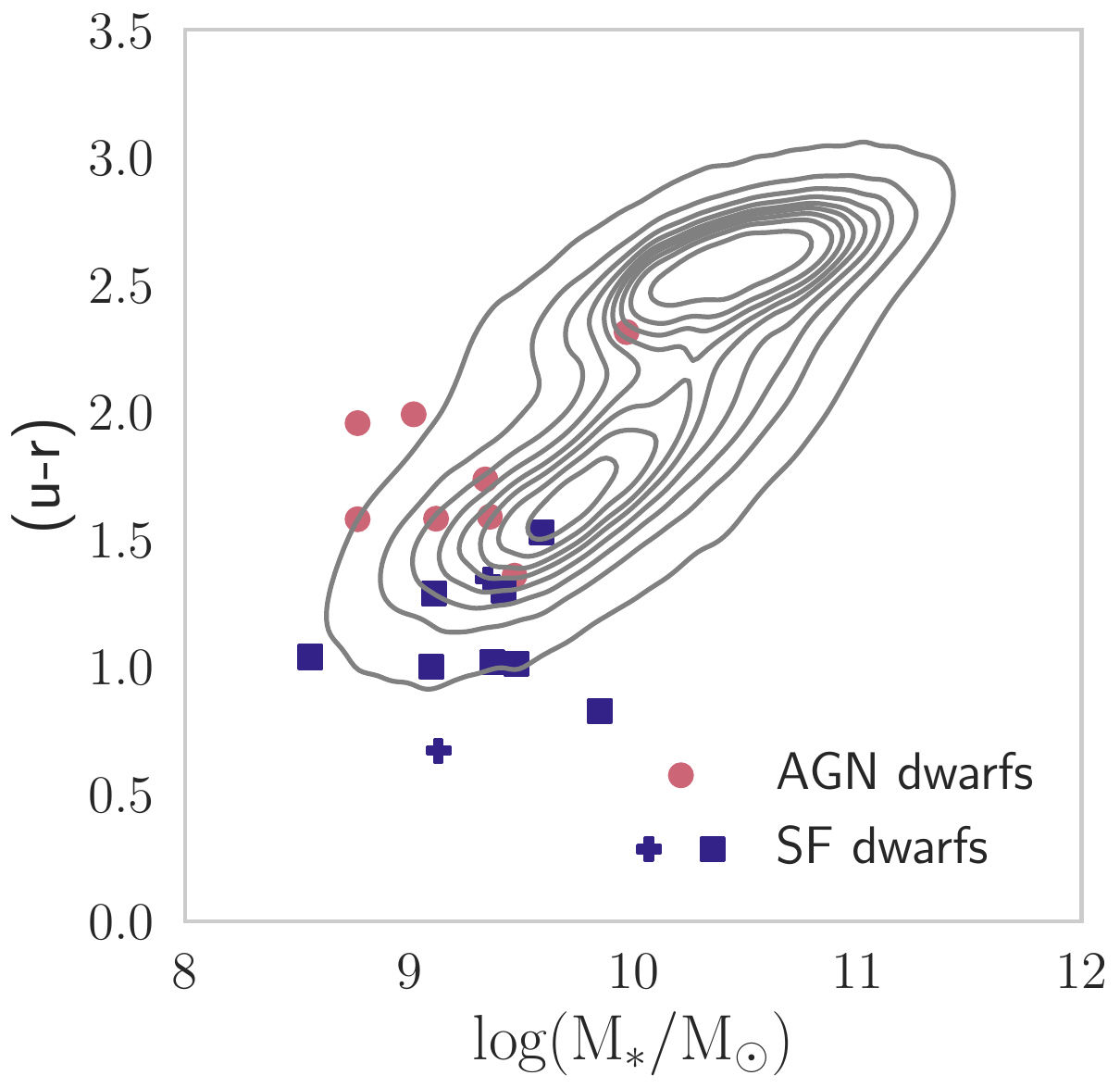}{0.45\textwidth}{}}
\caption{\footnotesize The u-r model magnitude colors from SDSS DR 16 are plotted against the MPA-JHU stellar mass. The colors are corrected for galactic extinction following \citet{1998ApJ...500..525S}, and the contours are from \citet{2014MNRAS.440..889S}. The cross symbols in the SF dwarfs are again used to specify J0838 and J0850 as the two galaxies with some signs of AGN ionization.\label{fig:cmd}}
\end{figure}

From the color-magnitude diagram for the AGN dwarfs and SF dwarfs indicated in Figure \ref{fig:cmd}, we see that the AGN dwarfs are redder than the SF dwarfs. The correlation between galaxy colors and the nature of the galaxy has been extensively studied \citep{2007ApJS..173..342M,2007MNRAS.382.1415S,2010ApJ...711..284S}. The bluer colors of the SF dwarfs could indicate that their outflows have not played a significant role in effectively quenching star formation. On the other hand, the relatively more energetic AGN-powered outflows may have hampered star formation. We note, however, that the positions of the galaxies in the color-magnitude diagram could be a selection effect, as it is generally easier to detect AGNs in galaxies with lower star formation rates. 

The extent of the SF outflows is found to be greater than those of AGN. This could indicate that the SF outflows can reach greater scales, and thus their overall time-averaged kinetic energy outflows rates appear smaller. SF outflows could be comparable to AGN outflows in stirring up the gas at smaller scales in the host galaxy. The escape velocities could also be lower at these scales, so it would be easier for the material in the SF outflows to escape the host galaxy.

While star formation could be instantaneously less energetic compared to AGN-driven outflows, it is important to consider the long-term effects of the two processes. Multiple bursts from stellar activities could lead to stellar processes dominating over AGN-driven outflows in regulating the environment of the galaxy. For AGN-driven outflows to be significant over a long time, it would be important to consider the duty cycles of AGN activity or how frequently the AGN-powered outflows occur as compared to stellar-driven outflows.

\subsection{Comparison with previous studies}
A number of studies have measured the kinematics of ionized gas in dwarf galaxies, and we compare our results with some of the work found in the literature. \citet{2004ApJ...610..201S} used high-resolution Echelle spectrographs to study the NaD absorption lines in 6 nearby dwarf starburst galaxies. They discovered that 3 of them have outflows, with an average outflow velocity of -27 km s$^{-1}$. This is lower (more blue-shifted) than our median v$_{50}$ values. However, our maximum blue-shifted v$_{50}$ values for the broad component are comparable.

\citet{2009A&A...493..511V} looked at emission lines in HI and H$\alpha$ to estimate outflow velocities in the nearby starforming irregular dwarf galaxy NGC 2366. They characterize the galaxy to have two major outflows, one with blue-shifted velocities of 30 km s$^{-1}$ and the other has red-shifted velocities up to 50 km s$^{-1}$ with a radial extent of 1.4 kpc. It is interesting to note that they considered red-shifted velocities to indicate an outflowing component, in contrast to our analysis. We find that if we consider the red-shifted velocities in our calculations of the energetics, the values do not change significantly. However, it would imply that it is possible that the red-shifted velocities in our C2 maps are indicative of actual outflowing gas. 

Our results are also in line with \citet{2022arXiv220902726M}, who studied ionized gas kinematics in a sample of 19 nearby starburst galaxies. The line profile they observe, as depicted in Figure (6) in \citet{2022arXiv220902726M}, also appears to be symmetric, and there is no apparent blue-shift. They find ionized gas outflow rates in the range of 10$^{-4}$ - 10$^{-1}$ M$_{\odot}$ yr$^{-1}$, which is an order of magnitude higher than the values calculated in this work for SF dwarfs. However, they still conclude that stellar feedback only stimulates a gentle gas cycle and is less likely to cause major blowouts that affect the evolution of the dwarf galaxy.

\citet{2019ApJ...884...54M} observed 3 of the 10 SF dwarfs in this work and found wind speeds that are much higher than what we measured. This could be due to the fact that they defined outflow velocity differently from this work. It could also be attributed to the slit spectroscopy used as compared with the higher resolution power of KCWI and the larger field of view, which helped us make more accurate measurements. Some differences in profiles were also observed by L20 in their comparison of KCWI data with LRIS data for the AGN dwarfs. While the results generally agreed, they found that for two of the targets, the KCWI data showed narrower profiles with smaller blue-shifts as compared to the LRIS data (For a more in-depth analysis, refer to Section 4.3 in L20).

Our analysis shows that stellar feedback might be less significant than previously expected. With the spatial information provided by IFS, we are able to see the effects of stellar-powered outflows in the galaxies. There appear to be some regions of strong outflow activity, but even then, they are not as powerful as AGN activity. It is possible that, at least in many cases, stellar feedback might only disturb the gas locally and not cause major effects on the host galaxy as a whole.

\section{Summary and Conclusions}
In this paper, we report the results from a Keck/KCWI integral field spectroscopic study of a sample of 10 low redshift star-forming dwarf galaxies with ionized gas outflows. This study is a companion study to \citet{2020ApJ...905..166L}, so we also compare our results to their sample of dwarf galaxies with AGN-powered outflows.
\begin{enumerate}
    \item The [O\,III] $\lambda$5007 emission line shows a clear broad component in all our targets, indicating either outflowing or disturbed gas in the galaxy. These broad components have more symmetric line profiles and are less blue-shifted than those found in the sample of AGN-dwarfs, with an average v$_{50}$ close to zero km s$^{-1}$, and an average value W$_{80}$ of 484 km s$^{-1}$ across 10 targets. In contrast, the broad component of the [O\,III] $\lambda$5007 lines in the AGN dwarfs shows strong blue-shifts with an average v$_{50}$ $\sim$ -64 km s$^{-1}$.
    Nevertheless, the average value of W$_{80}$ in both samples is comparable ($\sim$ 456 km s$^{-1}$ for the AGN dwarfs).
    \item All of the outflows detected in SF dwarfs are spatially resolved with sizes up to two to three times the half-light radius of the galaxies. These outflows are significantly more extended than those of AGN dwarfs. While the latter show a biconical structure in some cases, the stellar-driven outflows tend to be more spherically symmetric.
    \item We calculate the amount of ionized mass in the outflows, mass loss rates (dM/dt), kinetic energy rates (dE/dt), and momentum rates (cdp/dt). These energetics help us determine how much of the gas mass is being carried out by the outflows and, in turn, can let us know how potentially effective the outflows can be in influencing the evolution of the galaxy. We find that the values of ionized gas mass in the outflows range from $\sim$ 3 $\times$ 10$^{-5}$ to 2 $\times$ 10$^{-2}$ M$_{\odot}$ yr$^{-1}$, and the kinetic energy rates for the outflows range from $\sim$ 2 $\times$ 10$^{35}$ to 1 $\times$ 10$^{38}$ erg s$^{-1}$. These values for the SF outflows are all lower than the corresponding values for the AGN-driven outflows, with the energy outflow values being 2-3 orders of magnitude lower.
    \item A small fraction ($<$ 2\%) of the outflowing ionized gas in the targets have velocities larger than the escape velocities of the host galaxy, calculated from the center of the galaxy. These values are similar to the ones found by L20 for the AGN-driven outflows. 
\end{enumerate}
Thus we find that SF dwarfs clearly have significantly lower values of energetics in the outflows as compared to AGN dwarfs. This is despite the fact that our sample of SF dwarfs is biased toward stronger outflows, since we selected the targets that showed the clearest and broadest second components to the [O\,III] $\lambda$5007 emission lines. Determining the relative importance between AGN feedback and SF feedback in the evolution of dwarf galaxies will require detailed models that take into consideration factors such as duty cycles and BH occupation fractions, as well as the energetics presented in this study and that of L20. What is certainly clear is that AGN feedback is an important ingredient that cannot be ignored in the study of dwarf galaxy evolution. 

\begin{acknowledgments}
We thank the anonymous referee for their thoughtful feedback and constructive comments that greatly helped to improve this paper. The authors would also like to thank Laura Sales for her useful comments on the paper and Marie Wingyee Lau for her support with IFSFIT. Partial support for this project was provided by the National Science Foundation under Grant No. AST 1817233. VU acknowledges funding support from NASA Astrophysics Data Analysis Program (ADAP) grant 80NSSC20K0450. The data presented herein were obtained at the W. M. Keck Observatory, which is operated as a scientific partnership among the California Institute of Technology, the University of California, and the National Aeronautics and Space Administration. The Observatory was made possible by the generous financial support of the W. M. Keck Foundation. The authors wish to recognize and acknowledge the very significant cultural role and reverence that the summit of Maunakea has always had within the indigenous Hawaiian community. We are most fortunate to have the opportunity to conduct observations from this mountain. 
\end{acknowledgments}
\facilities{SDSS, KCWI/Keck}

\software{astropy \citep{2013A&A...558A..33A,2018AJ....156..123A},
          BADASS \citep{10.1093/mnras/staa3278},
          IFSFIT \citep{2014ascl.soft09005R},
          IFSRED \citep{2014ascl.soft09004R},
          MPFIT \citep{2012ascl.soft08019M},
          pPXF \citep{2017MNRAS.466..798C},
          SciencePlots \citep{SciencePlots},
          Source Extractor \citep{1996A&AS..117..393B}}
\appendix

The detailed results from our analysis are presented in this appendix. We show the [O\,III] $\lambda$5007 flux maps, the $\mathrm{v_{50}}$ velocity maps, as well as the $\mathrm{W_{80}}$ velocity maps for the narrow velocity component (C1), the broad velocity component (C2) and the total. We also show the radial profiles of each target. In all cases, the systemic velocities were determined from the stellar velocities measured from the stellar absorption lines in the spectra integrated over the whole cube.
\\

\section{[O\,III] $\lambda$5007 Flux, velocity, and stellar kinematic maps}
\subsection{J0838}
\begin{figure}[h!]
\gridline{\fig{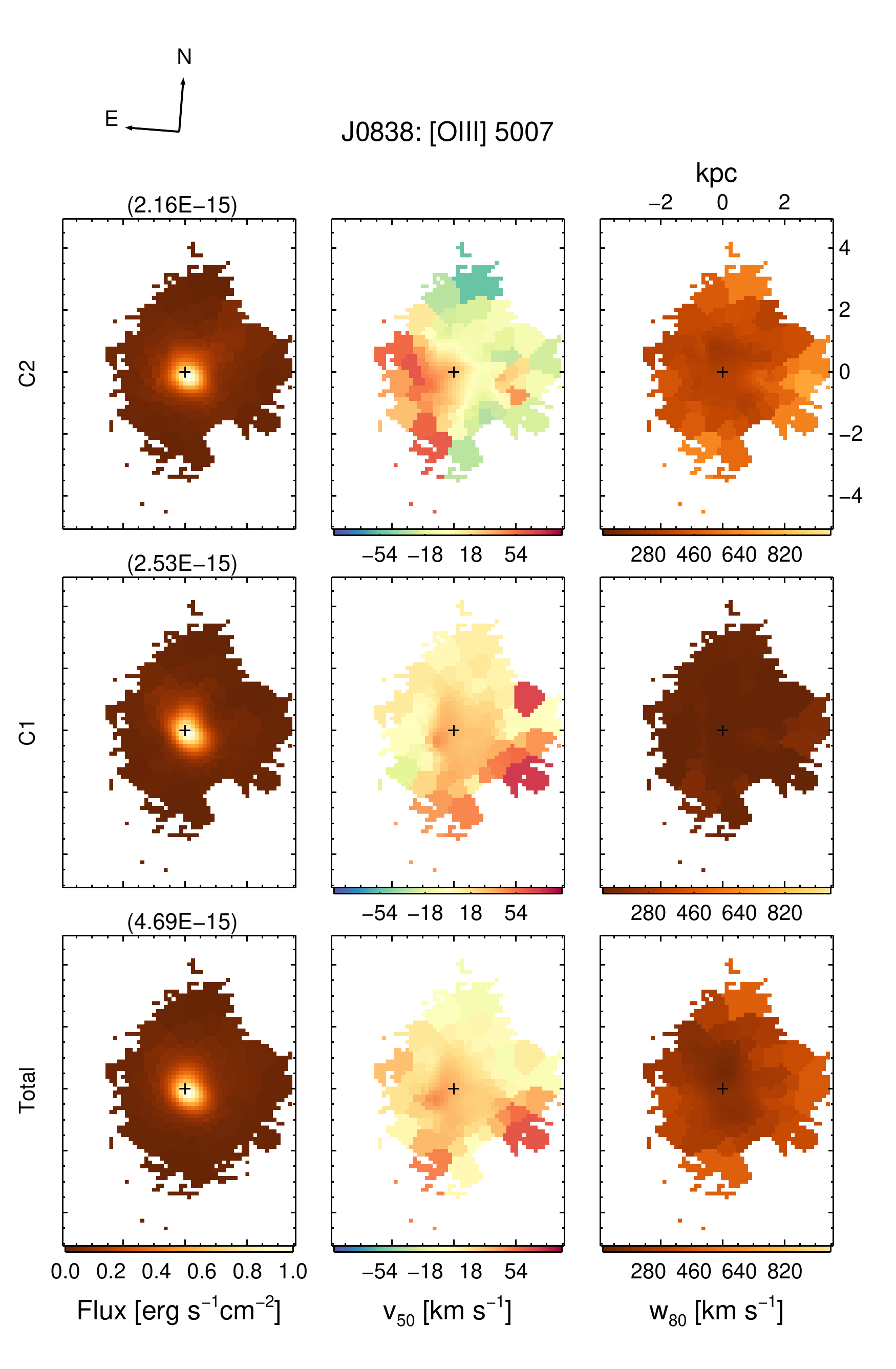}{0.5\textwidth}{}}\label{fig:[OIII5007]}
\caption{\footnotesize Voronoi-binned [O\,III] $\lambda$5007 flux maps for J0838 in the left panel, $\mathrm{v_{50}}$ velocity maps in the middle panel and the $\mathrm{W_{80}}$ velocity maps in the right panel. The first row is C2, the second row is C1, and the third row (Total) represents the overall emission profile. The maximum value of the flux in cgs units is indicated above the flux maps in the left panel. The axis represents the extent in kpc, and the orientation of the maps is indicated by the black compass at the top-left of the figure. \label{fig:[OIII5007]_0838}}
\end{figure}
Figure \ref{fig:[OIII5007]_0838} shows the maps of the [O\,III] $\lambda$5007 flux, velocity maps, stellar kinematics maps, and the radial flux profiles for J0838. The first row represents the broad component, the second row represents the narrow component, and the third panels represent the total flux map. The flux is normalized to the value indicated by the black cross to mark the approximate center of the galaxy. This is slightly different from L20, where they marked the black cross with the highest flux value, which also coincided with the center of the galaxy due to the nature of the AGN dwarfs. In our samples, the maximum value of the flux need not necessarily be close to the center as there is no preferential location for strong stellar activities. To maintain uniformity with the L20 analysis, we consider our normalizing flux value to come from a spaxel in the apparent center of the galaxy. The maximum values of the flux are indicated above the panels. The plot of the C2 flux shows some extended emissions distinct from that traced by C1.

The C2 $\mathrm{v_{50}}$ values have a significant gradient compared to the C1 values. There does not seem to be a strong blue-shifted component, but there is evidence of a red-shifted component. In contrast, the C1 velocities are relatively quiescent with velocities very close to 0 km s$^{-1}$. The strong red-shifted blobs on the right are likely caused due to noise and do not have any physical meaning. The $\mathrm{v_{50}}$ velocity maps of the broad component is clearly distinct from the narrow component to indicate that the broad component is not tracing the gas in the galaxy. There are also no apparent correlations between the C2 flux and the strength of the velocity maps.

The relation between W$_{80}$ and $\sigma$ indicating dispersion in the line profile is given as  W$_{80}$ = 2.63$\sigma$. The W$_{80}$ maps indicate that the narrow component does not have a wide dispersion which is expected as the narrow component traces the quiescent gas in the galaxy. In contrast, the broad component clearly has greater values of W$_{80}$ to show that the component it is tracing is significantly disturbed. They range between 300 - 700 km s$^{-1}$, with a median value of 397 km s$^{-1}$.

\subsection{J0850}

\begin{figure}
\gridline{\fig{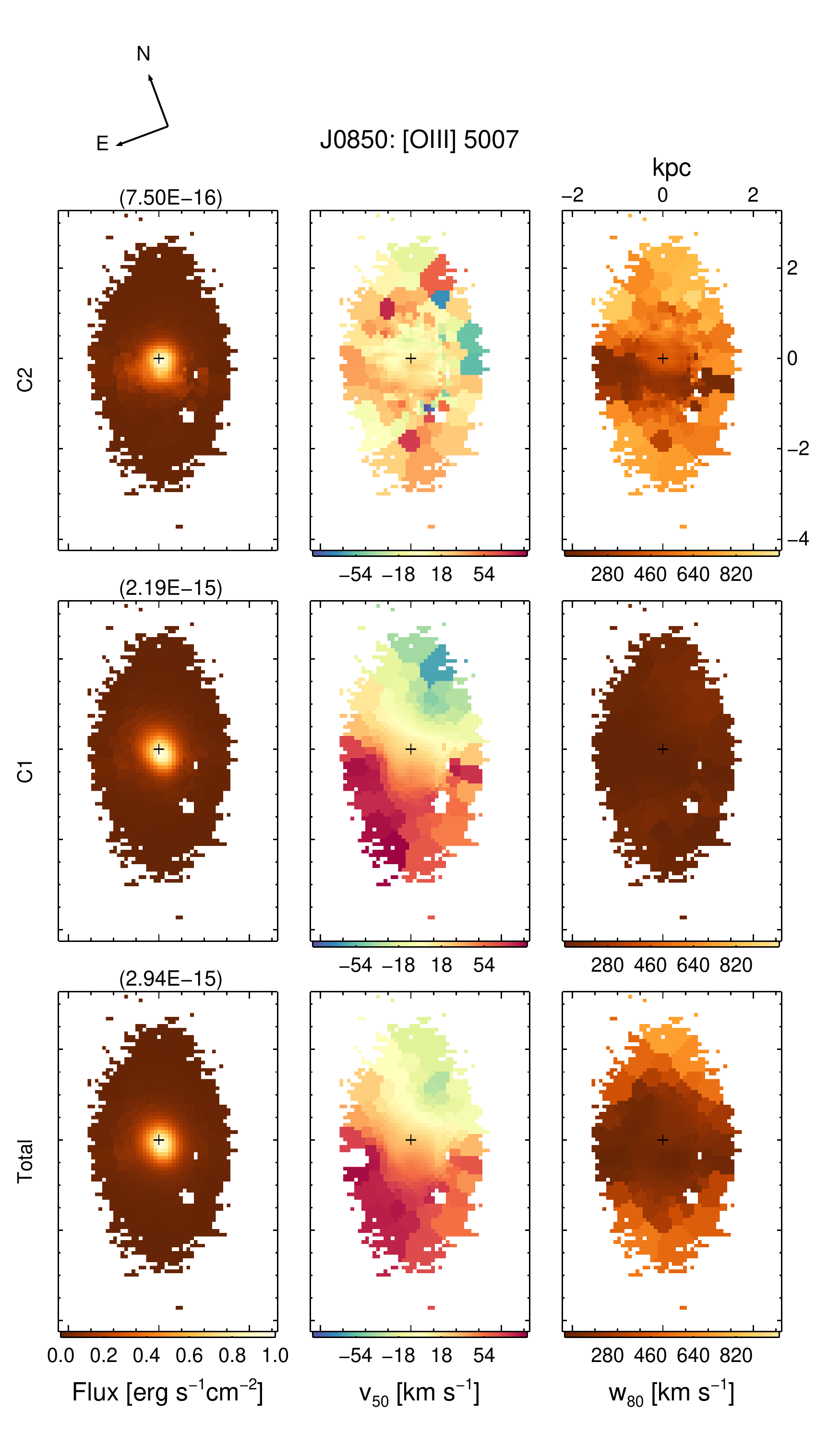}{0.5\textwidth}{}}
\caption{ \footnotesize Same as Figure \ref{fig:[OIII5007]_0838}, for J0850 \label{fig:[OIII5007]_0850}}
\end{figure}

\begin{figure}[ht!]
\gridline{\fig{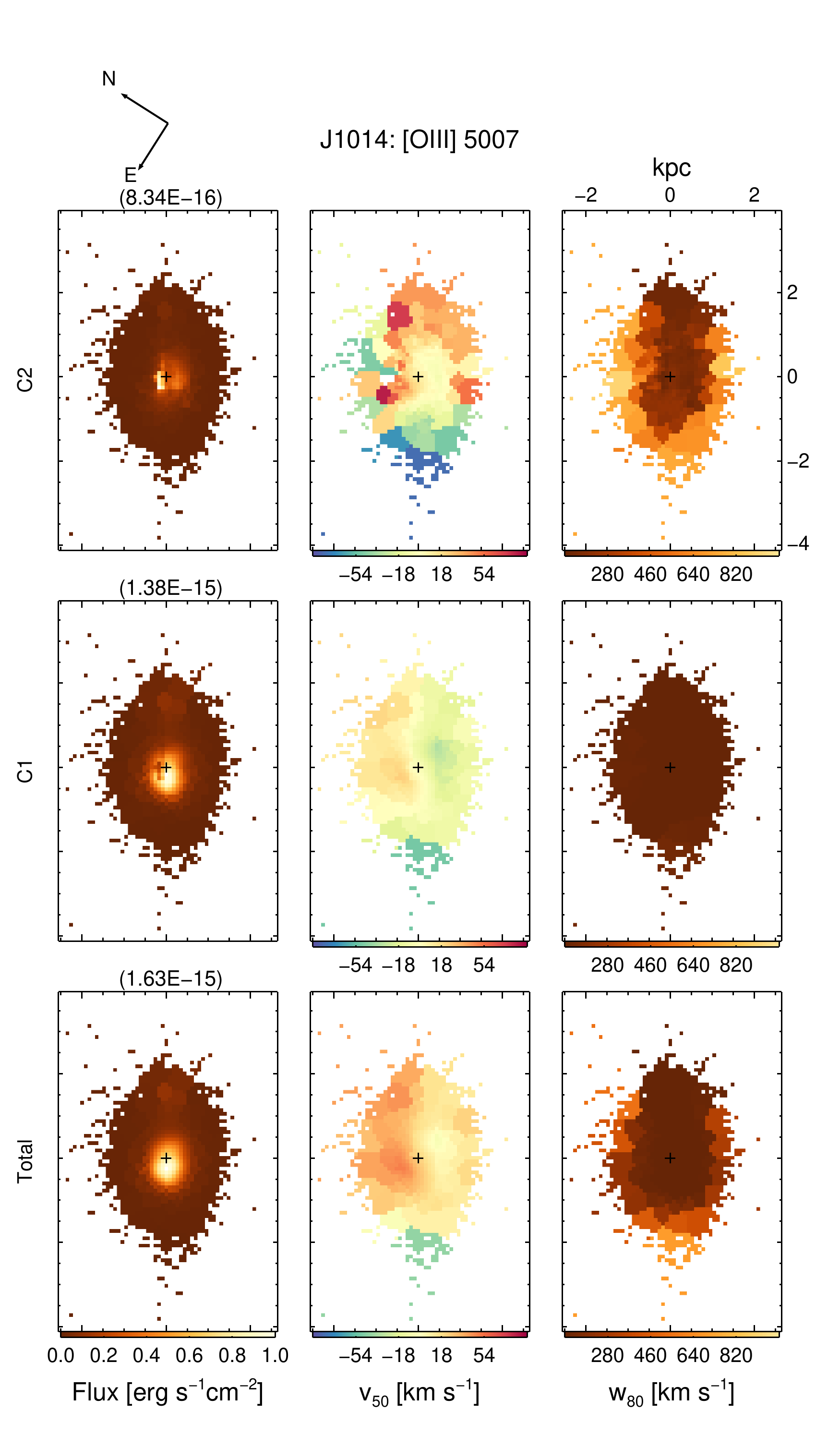}{0.5\textwidth}{}}
\caption{ \footnotesize Same as Figure \ref{fig:[OIII5007]_0838}, for J1014 \label{fig:[OIII5007]_1014}}
\end{figure}

In the [O\,III] $\lambda$5007 flux maps for J0850 indicated in Figure \ref{fig:[OIII5007]_0850}, the C2 flux component shows an extended flux making it distinct from the C1 flux component. The velocity maps of C1 show that the systemic gas in the galaxy is likely rotating. However, we see that the C2 component once again does not trace the gas in the galaxy. There does not seem to be a pattern in the C2 velocity maps, but that could just indicate that the broad component is tracing stirred and disturbed gas in the galaxy, which could be caused by stellar processes. The average value of the C2 v$_{50}$ is roughly 0 km $s^{-1}$  when averaged over the entire galaxy. The W$_{80}$ maps in C2 seem to trace certain regions of low dispersion, and other regions show higher dispersion, which could possibly indicate turbulence caused by the C2 component.

\begin{figure}[ht!]
\gridline{\fig{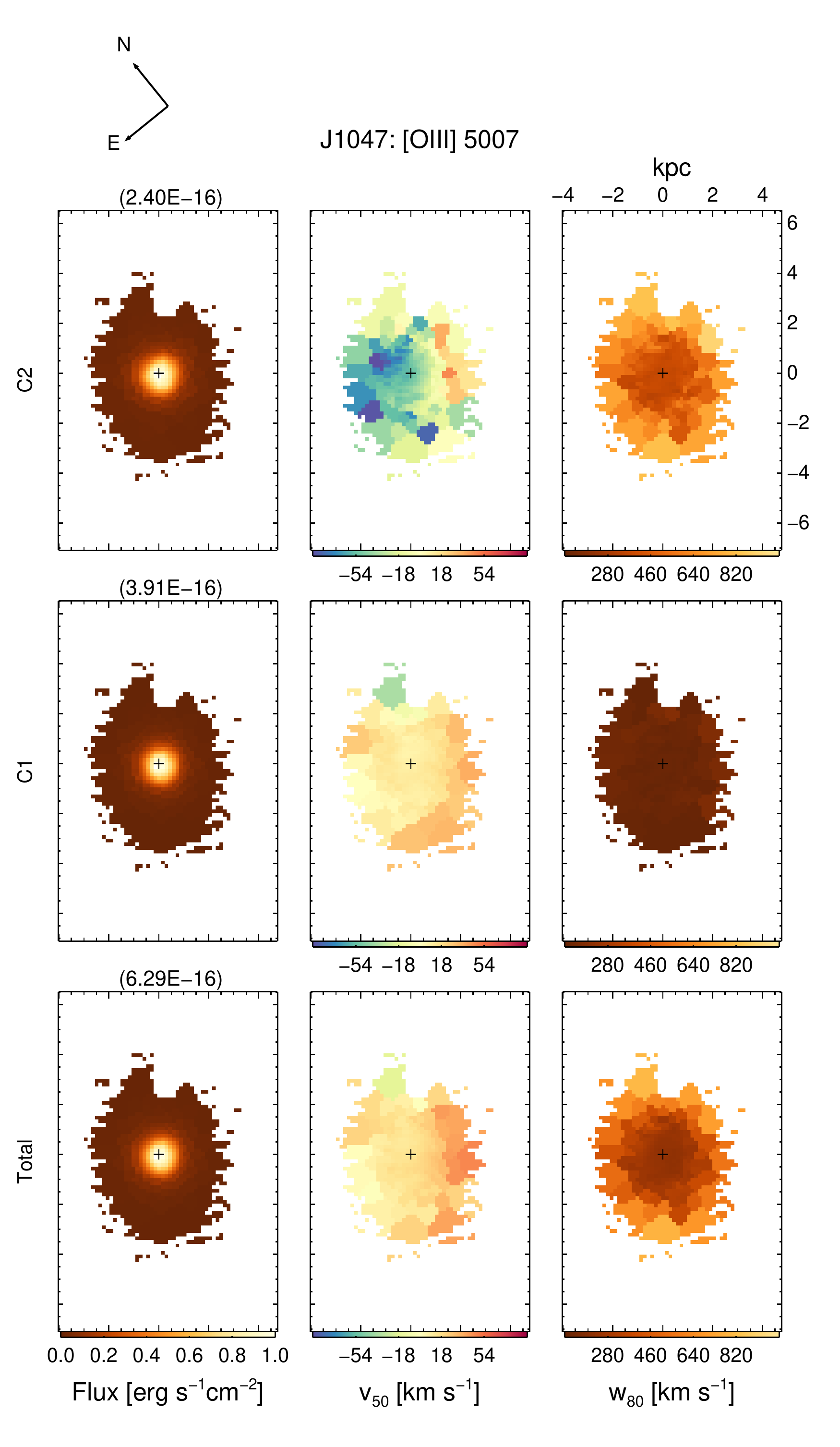}{0.5\textwidth}{}}
\caption{ \footnotesize Same as Figure \ref{fig:[OIII5007]_0838}, for J1047 \label{fig:[OIII5007]_1047}}
\end{figure}

\subsection{J1014}

Figure \ref{fig:[OIII5007]_1014} traces the [O\,III] $\lambda$5007 flux and velocity maps. The C2 flux component seems diminished, indicating that there is not much contribution from the broad component to the flux. There seems to be a faint blob near the top, but it is more prominent in the C1 flux map as compared to C2. It also does not translate over to the v$_{50}$ velocity maps, as we do not see any corresponding distinct feature near the top of the velocity maps. The C1 v$_{50}$ component appears quiescent, and the bottom blue component is out of place and is likely caused by noise. The C2 v$_{50}$ does not have any significant features, although it does appear to be more strongly red-shifted as compared to the narrow gas component. The velocity dispersion indicated by the W$_{80}$ maps also show higher dispersion in the outer regions for C2, and there seems to be a region of lower dispersion being traced out as well.

\begin{figure}[ht!]
\gridline{\fig{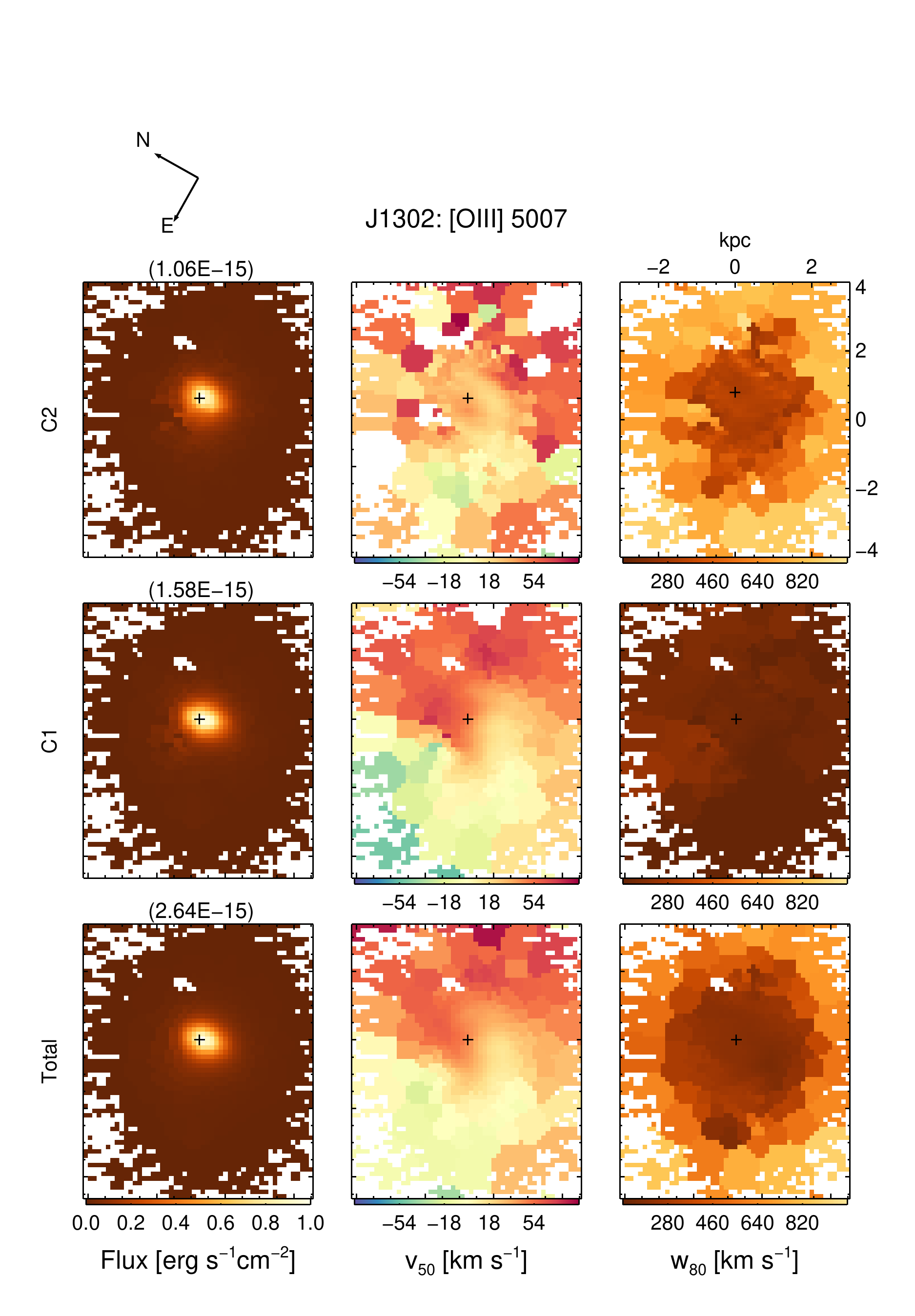}{0.5\textwidth}{}}
\caption{ \footnotesize Same as Figure \ref{fig:[OIII5007]_0838}, for J1302 \label{fig:[OIII5007]_1302} }
\end{figure}

\begin{figure}[ht!]
\gridline{\fig{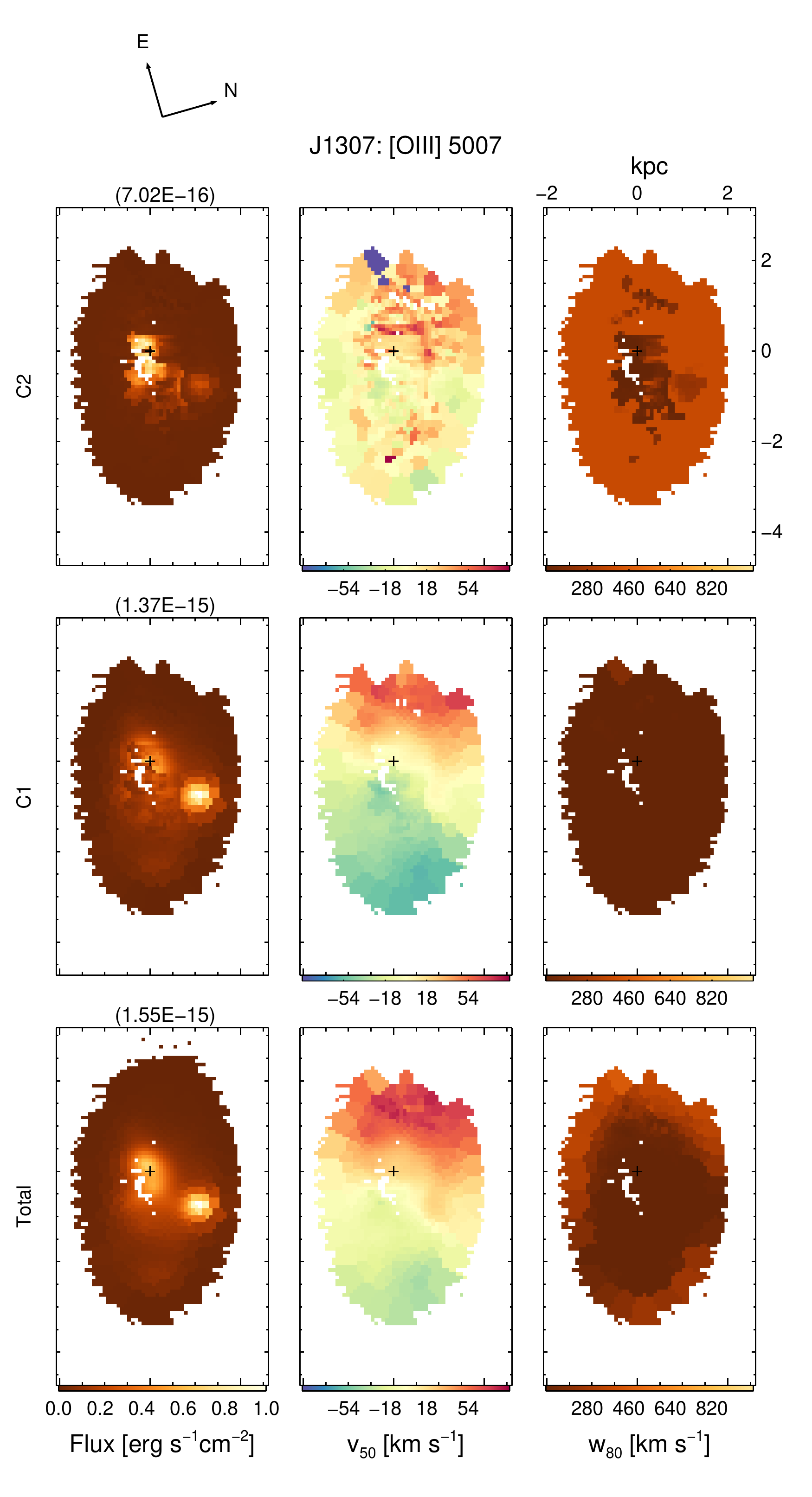}{0.5\textwidth}{}}
\caption{ \footnotesize Same as Figure \ref{fig:[OIII5007]_0838}, for J1307 \label{fig:[OIII5007]_1307}}
\end{figure}

\begin{figure}[ht!]
\begin{center}
    \includegraphics[scale=0.5]{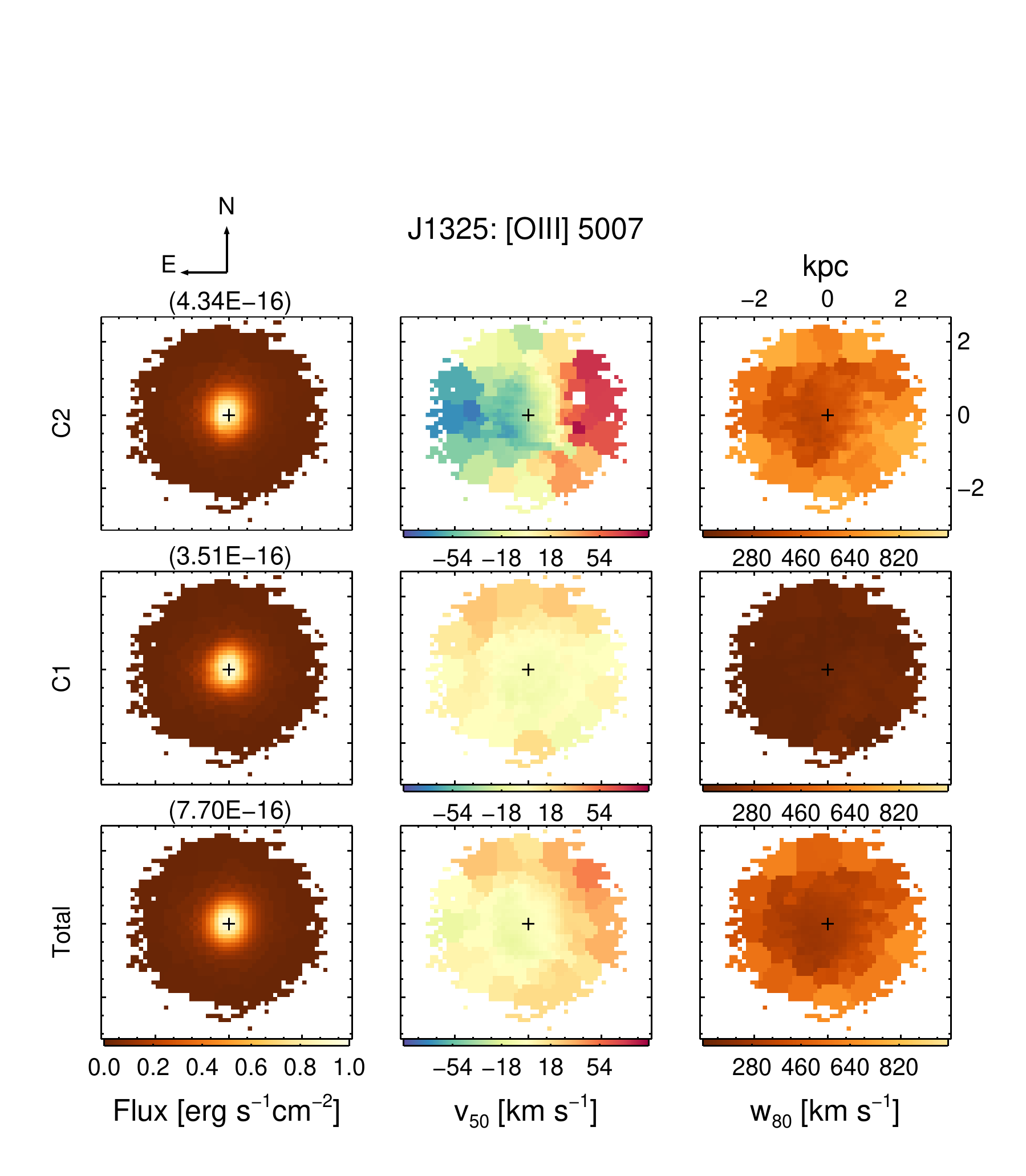}
\end{center}
\caption{\footnotesize Same as Figure \ref{fig:[OIII5007]_0838} but for J1325. \label{fig:[OIII5007]_1325}}
\end{figure}

\subsection{J1047}

From Figure \ref{fig:[OIII5007]_1047}, we see that the v$_{50}$ velocity maps for C2 is clearly distinct from C1. They appear to be more blue-shifted, indicating that this target likely has some prominent stellar outflows, also indicated by the high values of dispersion from the W$_{80}$ velocity maps for C2. The C1 is once again quiescent with little dispersion. 

\subsection{J1302}

Figure \ref{fig:[OIII5007]_1302} shows the [O\,III] $\lambda$5007 flux maps and velocity maps for J1302. C1 traces the gas in the galaxy and shows that the gas in the galaxy is likely to be rotating. The C2 velocity maps do not follow the rotation but they do not have any distinct features. They mostly seem to have low-velocity values, although they do show significant dispersion in their W$_{80}$ values as compared to the C1.

\subsection{J1307}
From Figure \ref{fig:[OIII5007]_1307}, it is evident that J1307 has multiple sources visible in the flux maps. They appear to be fainter in the C2 maps as compared to C1 maps, suggesting that these sources likely do not contribute to the broad component. The v$_{50}$ velocity maps of C1 show some rotation in the galaxy, but there does not seem to be a strong correlation between the bright flux region in the flux map and the velocity maps. There is a slight blue-shifted velocity observable in the C2 velocity map in the regions around the bright sources, but they are not that significantly high. It is possible that the bright flux region corresponds to another stellar source producing high ionization activity, but it still does not produce fast outflows. This is supported by the absence of significant dispersion in the W$_{80}$ velocity maps in the same region.

\subsection{J1325}
The C2 v$_{50}$ map shows that the outflow is either rotating or is oriented biconically. In either case, it is clearly distinct from the C1 gas component, which is nearly quiescent. This indicates that the gas in the galaxy is nearly stationary, but there is a distinct outflow present. The C2 W$_{80}$ velocity map also shows a significant dispersion, unlike the C1 W$_{80}$ velocity map.

\begin{figure}[ht!]
\begin{center}
    \includegraphics[scale=0.5]{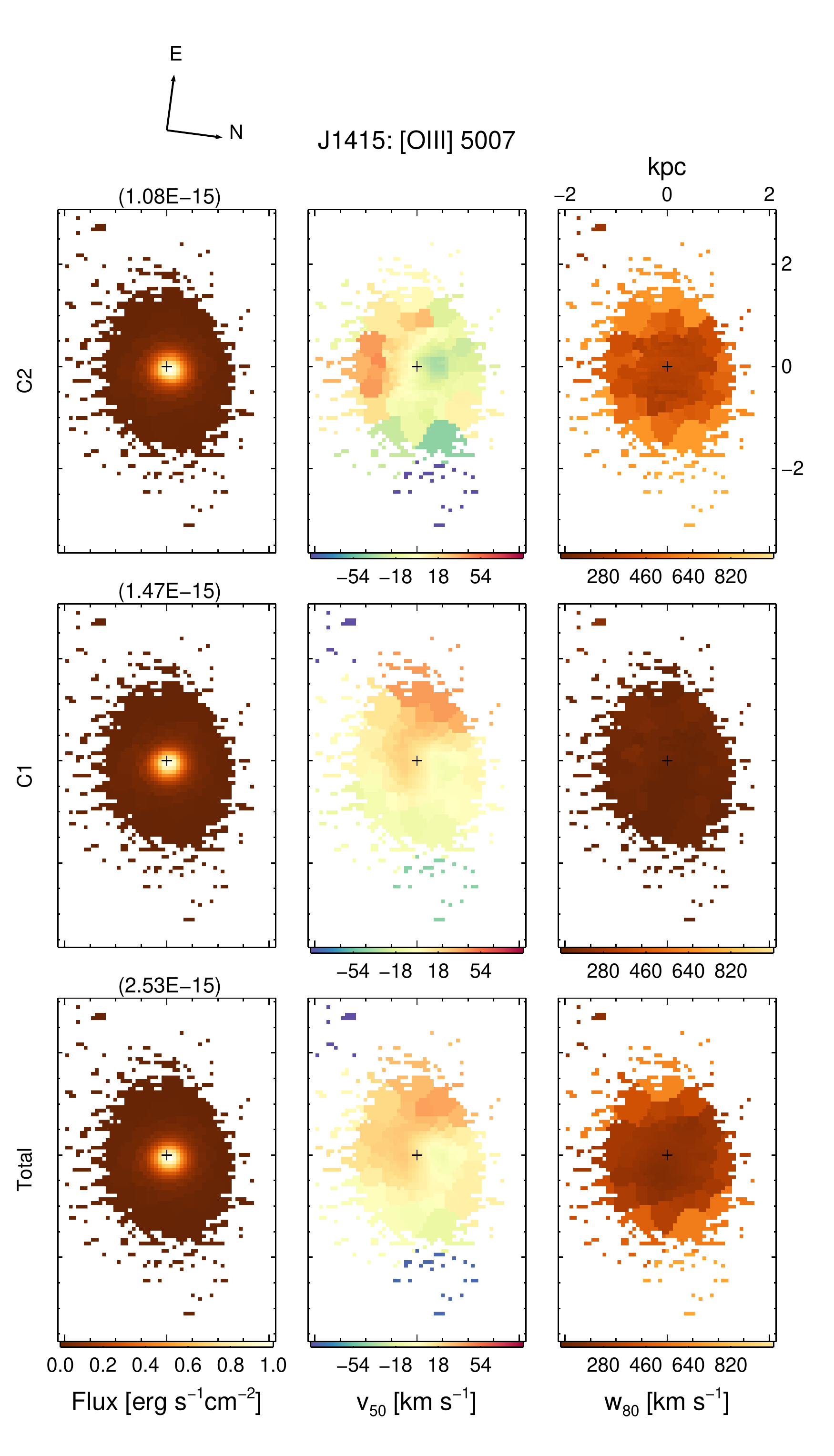}
\end{center}
\caption{ \footnotesize Same as Figure \ref{fig:[OIII5007]_0838} but for J1415. \label{fig:[OIII5007]_1415}}
\end{figure}

\subsection{J1415}
The C2 v$_{50}$ velocity map shows some regions of blue-shift, which could indicate an outflow. It is distinct once again from the C1 v$_{50}$ velocity map, which is mostly quiescent but has some hints of a rotating gas. The C2 W$_{80}$ velocity map also shows significant dispersion, indicating that it is distinct from the C1 quiescent gas.

\subsection{J1622}

\begin{figure}[]
\begin{center}
    \includegraphics[scale=0.5]{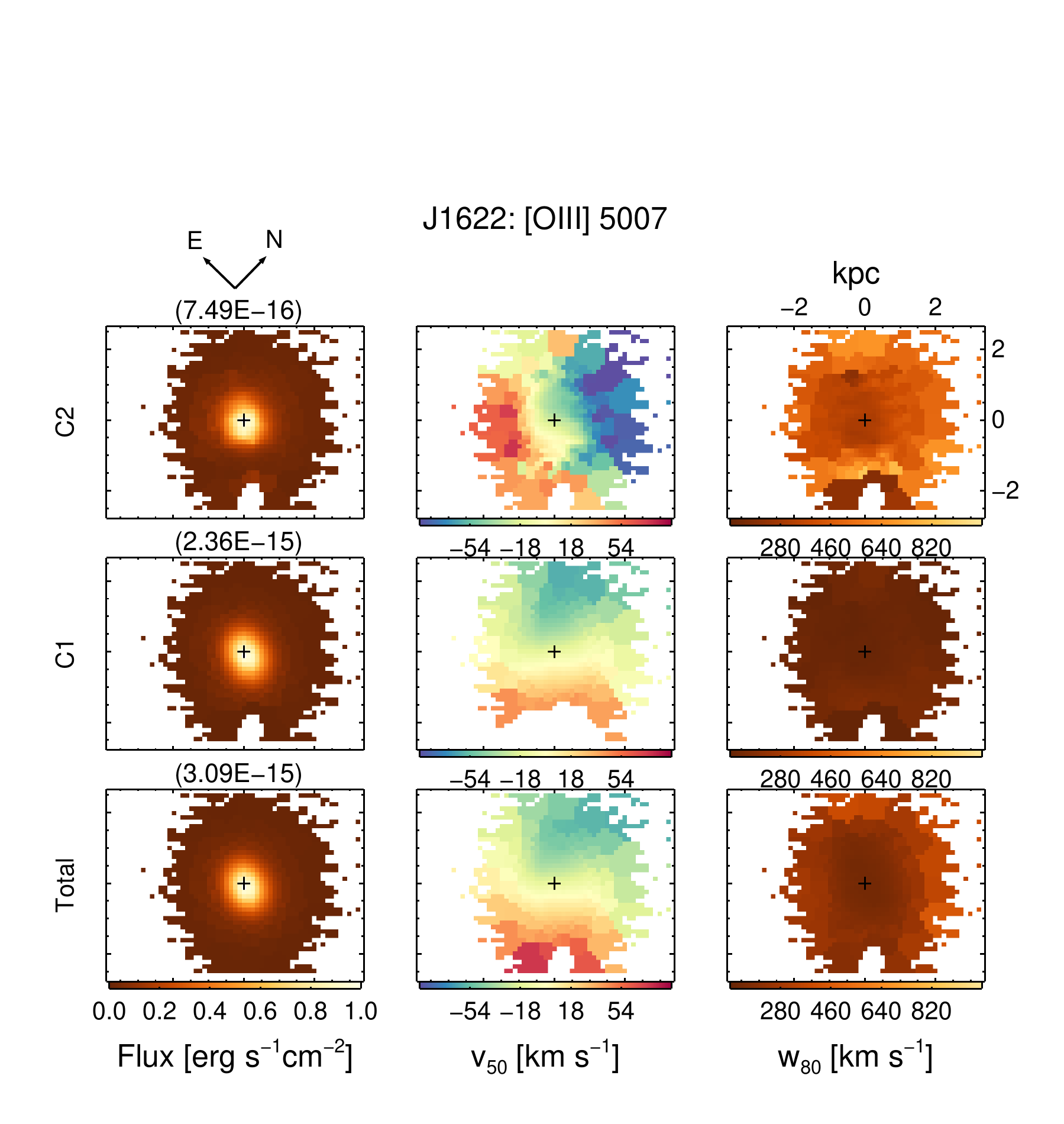}
\end{center}
\caption{ \footnotesize Same as Figure \ref{fig:[OIII5007]_0838} but for J1622. \label{fig:[OIII5007]_1622}}
\end{figure}

The v$_{50}$ velocity flux maps of J1622 show extremely interesting structures going from C1 to C2. The C2 component can constitute an outflow because of the significant blue-shift in the velocities, but they appear to be oriented perpendicular to the rotating gas in the galaxy. The significant velocity dispersion in the C2 W$_{80}$ velocity map is indicative of an outflow. The narrow component does not have a significant dispersion once again.

\begin{figure}
\begin{center}
    \includegraphics[scale=0.45]{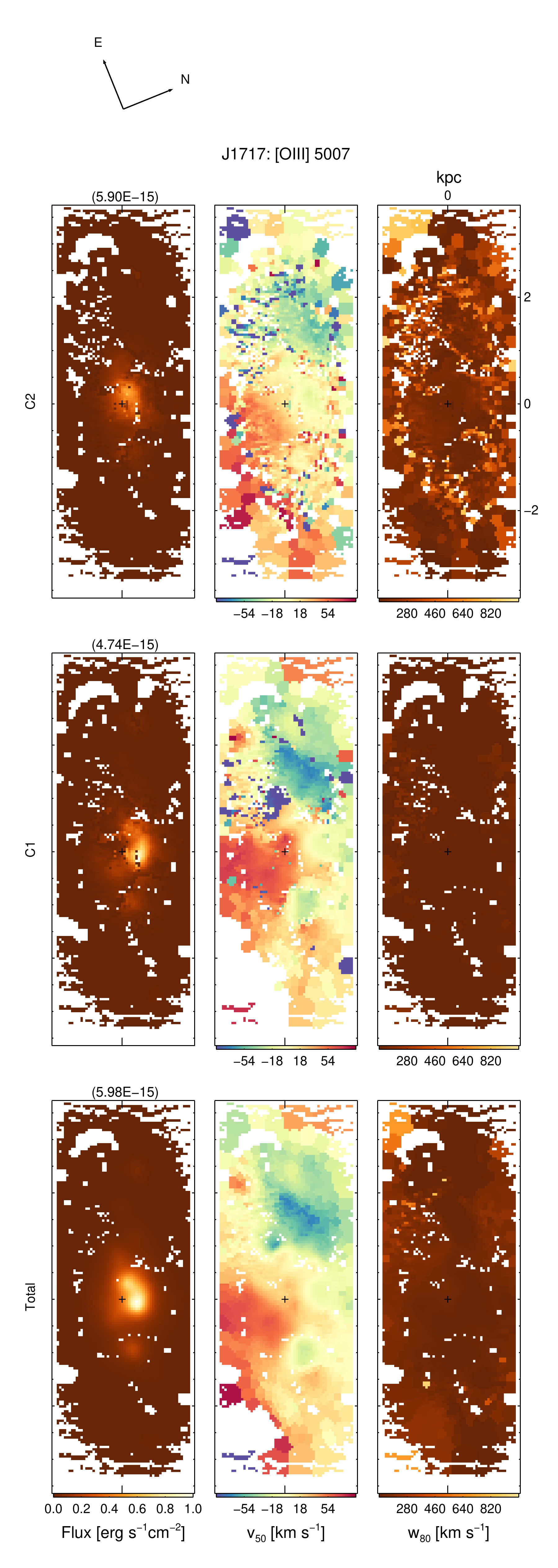}
\end{center}
\caption{ \footnotesize Same as Figure \ref{fig:[OIII5007]_0838} but for J1717. \label{fig:[OIII5007]_1717}}
\end{figure}
\subsection{J1717}

J1717 is at the lowest redshift compared to the rest of the galaxies and thus occupies a larger fraction in the sky. The galaxy appears extended and also has interesting flux maps, as seen in Figure \ref{fig:[OIII5007]_1717}. The C1 flux map indicates the presence of several sources of star-forming regions near the center and one slightly near the bottom, which is only faintly seen in the C2 flux map. The galaxy also appears to have a rotating gas component, as seen in the C1 v$_{50}$ map. The C2 component seems to trace this gas slightly based on the rotating pattern observed, but it has lower values of the v$_{50}$ velocity. There also does not seem to be a prominent dispersion, although the map is significantly noisy. We could not go to higher SNR as we started losing spatial information. There seems to be a slight visual correlation between the flux maps and the velocity maps near the regions of the additional sources, although it is not too apparent from the W$_{80}$ velocity maps. This is different from J1307, where we also had obvious multiple sources but no apparent correlations between the flux and velocity maps (See Figure \ref{fig:[OIII5007]_1307}).

\section{Radial Flux profiles} \label{appendix:radial_profiles}

\begin{figure}[ht!]
\begin{center}
 \includegraphics[scale=0.3]{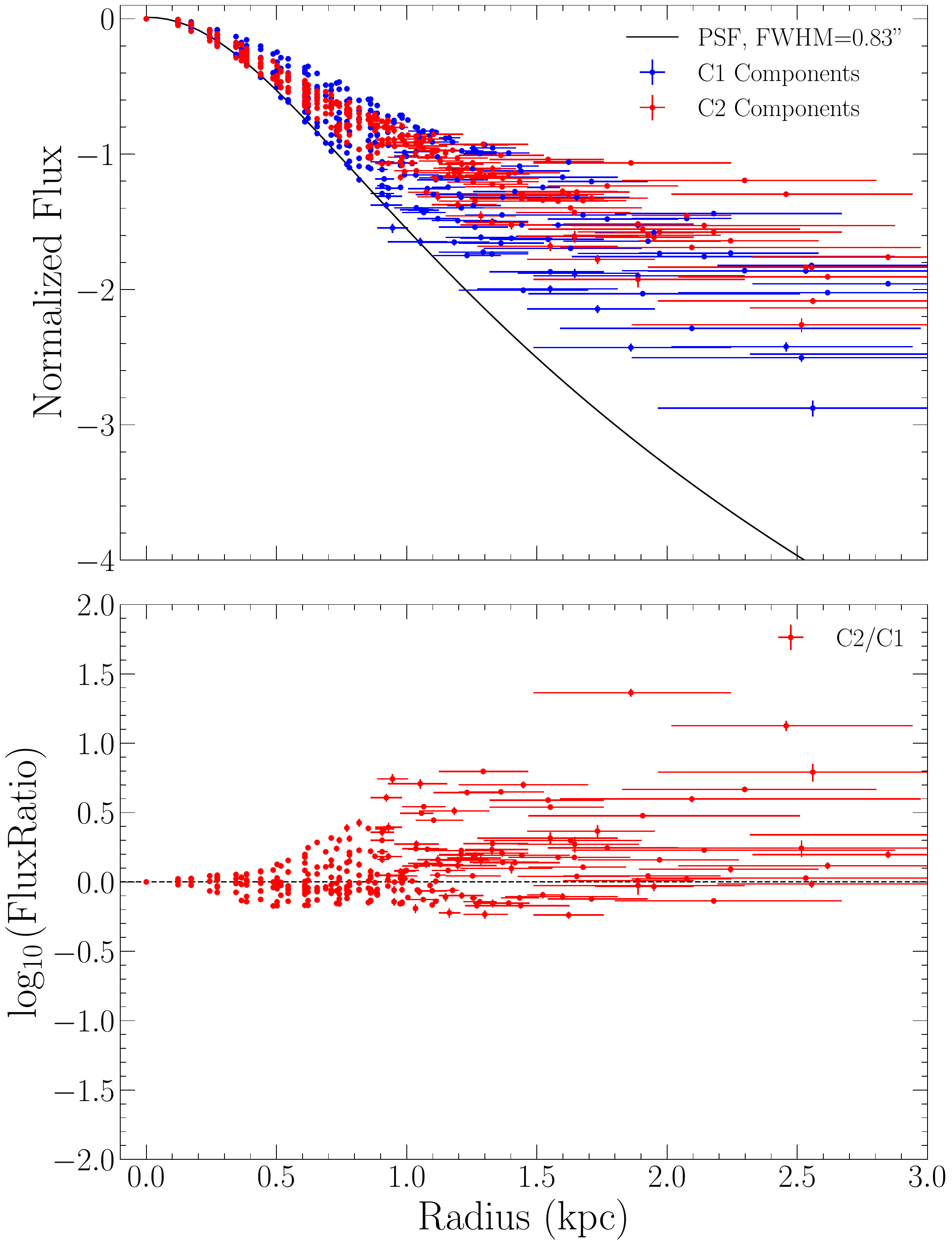}   
\end{center}
\caption{ \footnotesize Top panel: The radial profiles of [O\,III] $\lambda$5007 fluxes for the two velocity components of J0838. Here, the fluxes are normalized to the maximum value of the flux. The PSF profile is obtained by fitting a Moffat profile to the spectrophotometric standard star (see Section \ref{observations} for more details.). Bottom panel: C2/C1 flux ratios plotted on a logarithmic scale as a function of distance from the spaxel marked by the black cross in Figure \ref{fig:[OIII5007]_0838}. The error bars in the x-axis indicate the radial coverage of the spatial bin. \label{0838_radial}}
\end{figure}

\begin{figure*}[]
\gridline{\fig{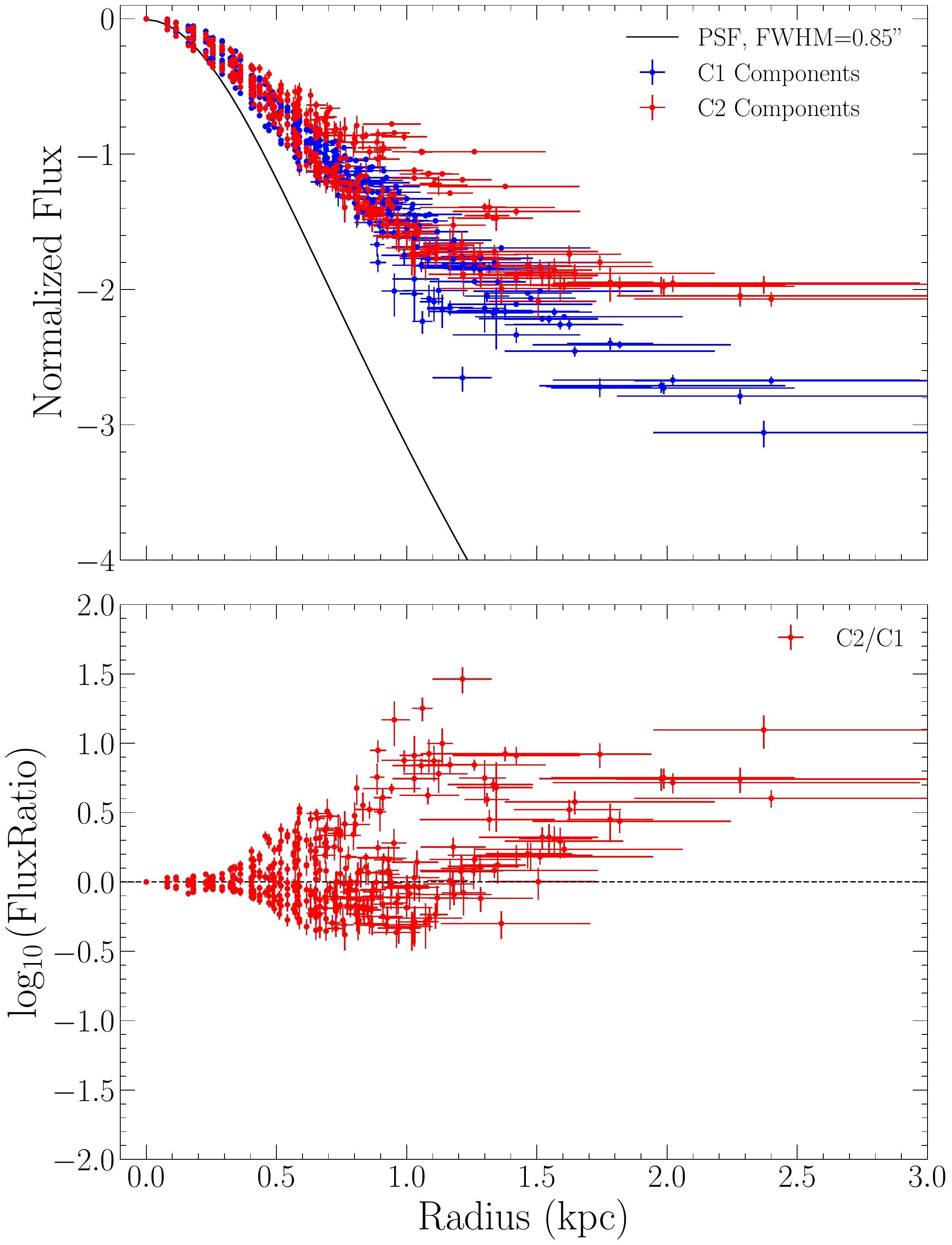}{0.3\textwidth}{(a) J0850}
\fig{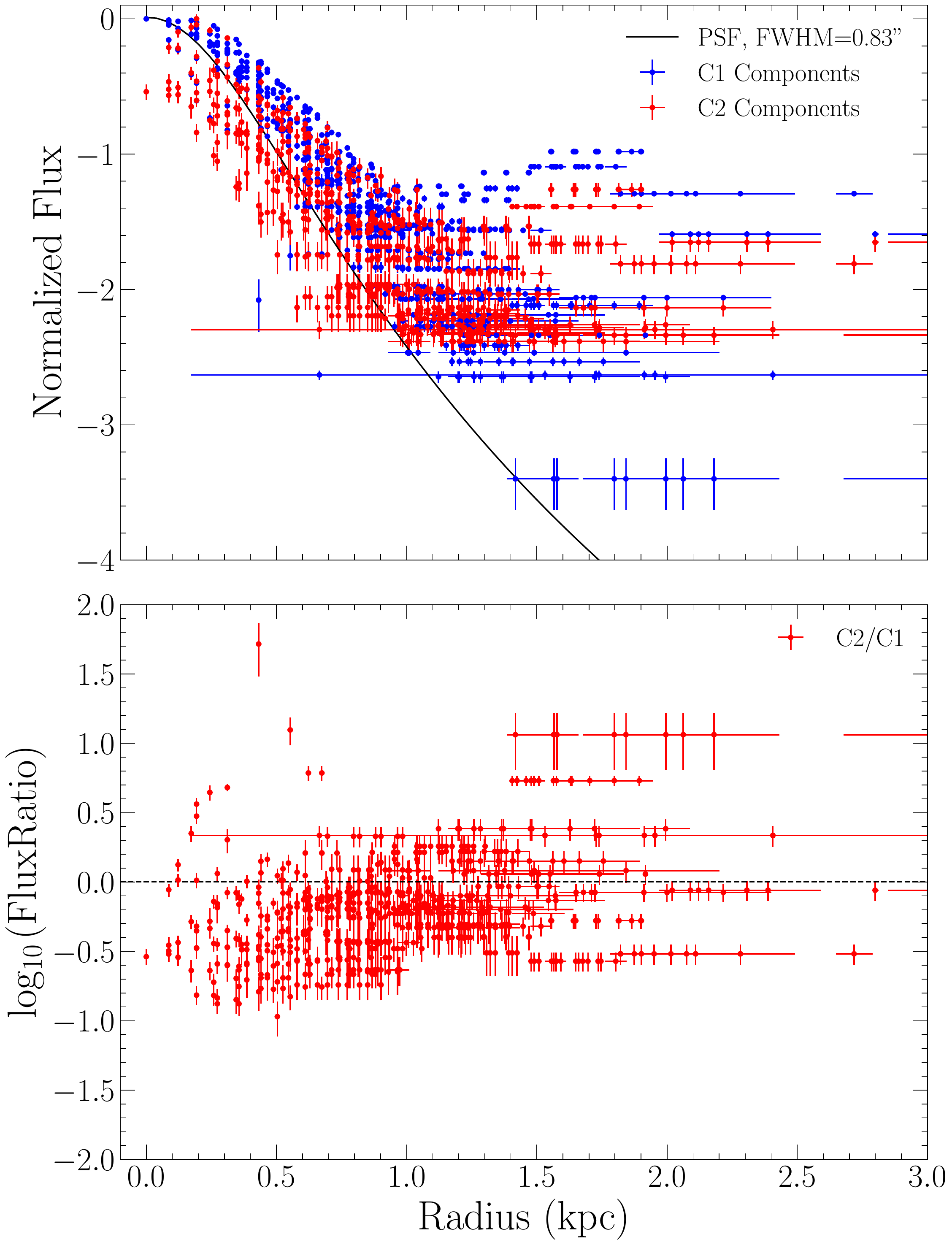}{0.3\textwidth}{(b) J1014}
\fig{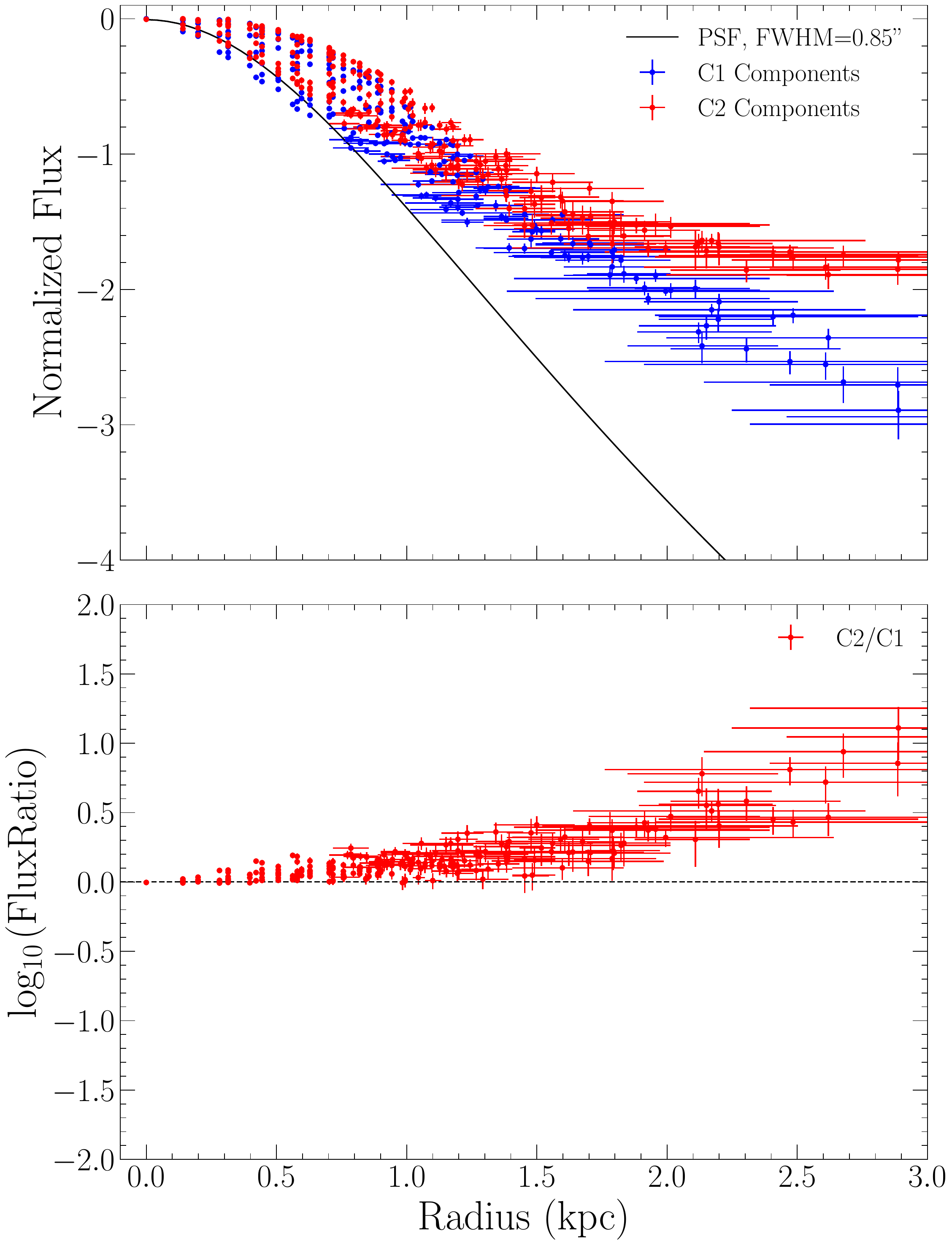}{0.3\textwidth}{(c) J1047}}
\gridline{\fig{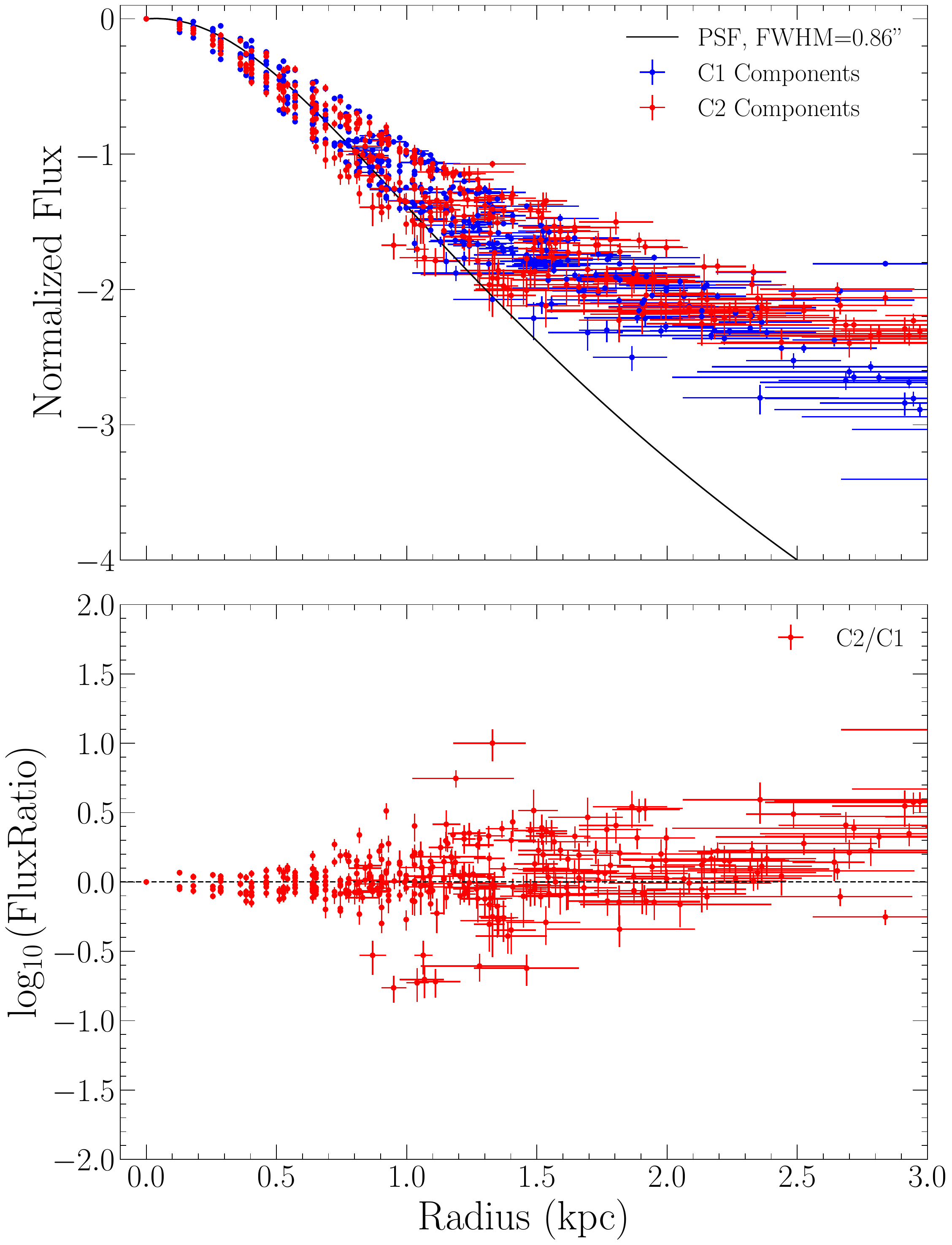}{0.3\textwidth}{(d) J1302}
\fig{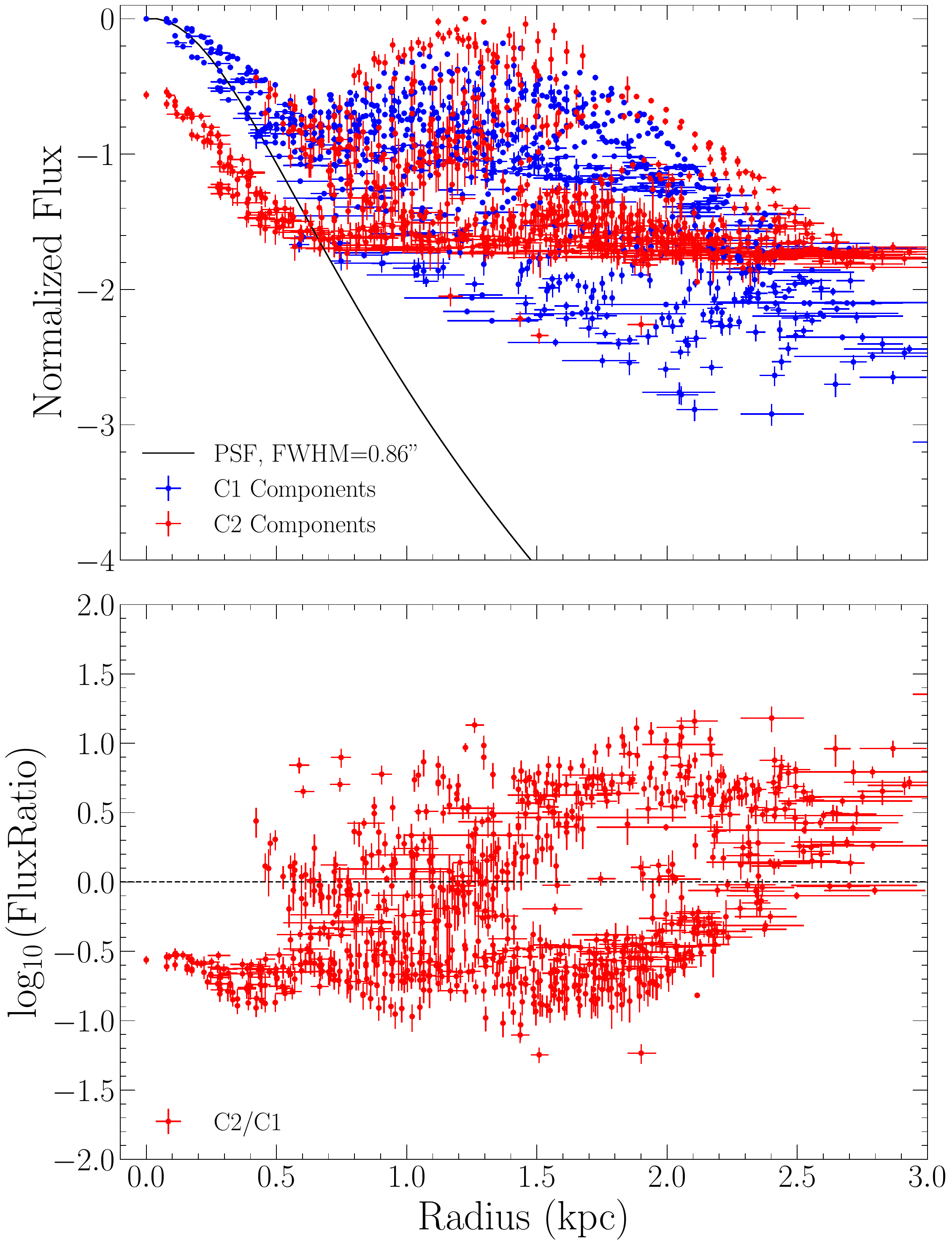}{0.3\textwidth}{(e) J1307}
\fig{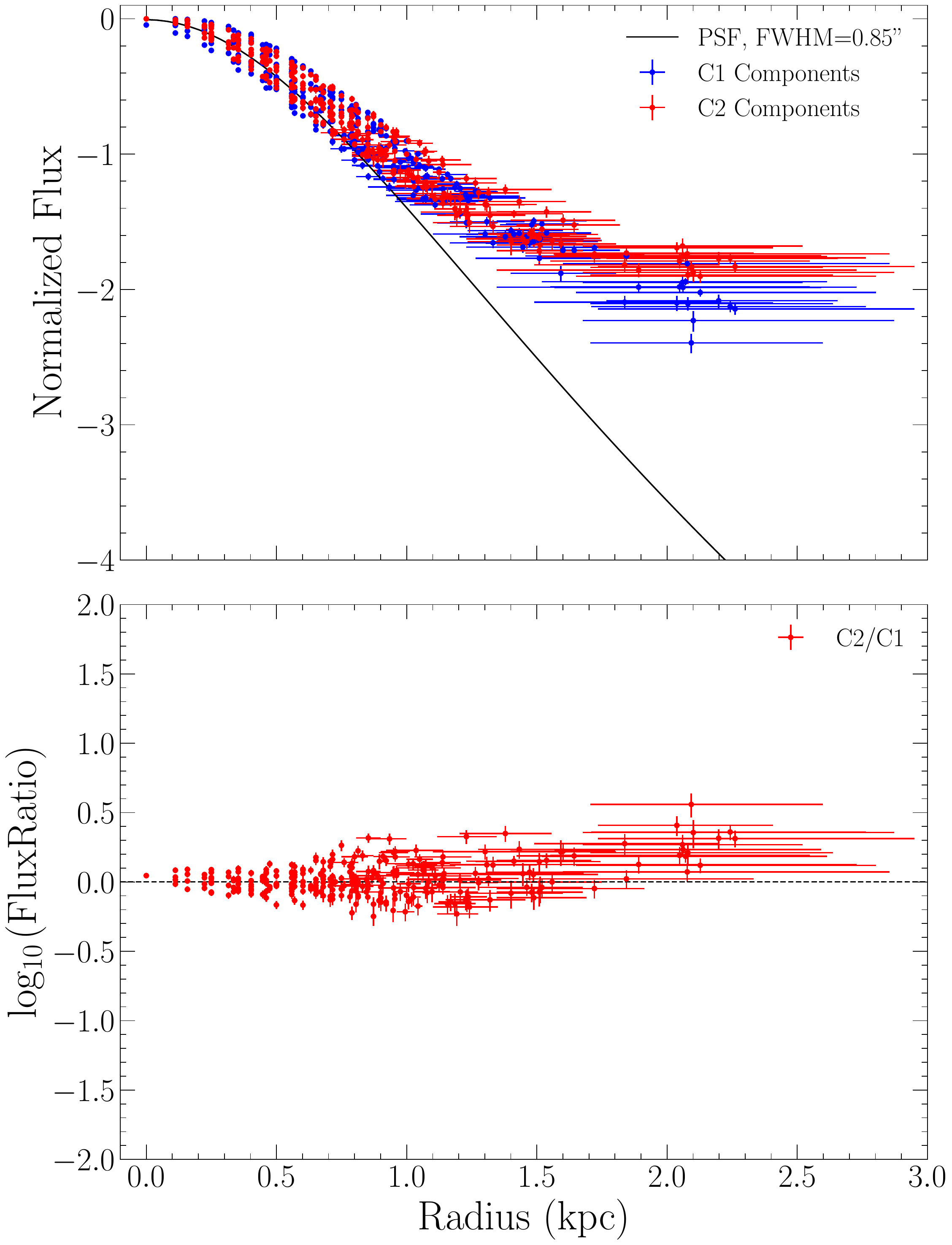}{0.3\textwidth}{(f) J1325}}
\gridline{\fig{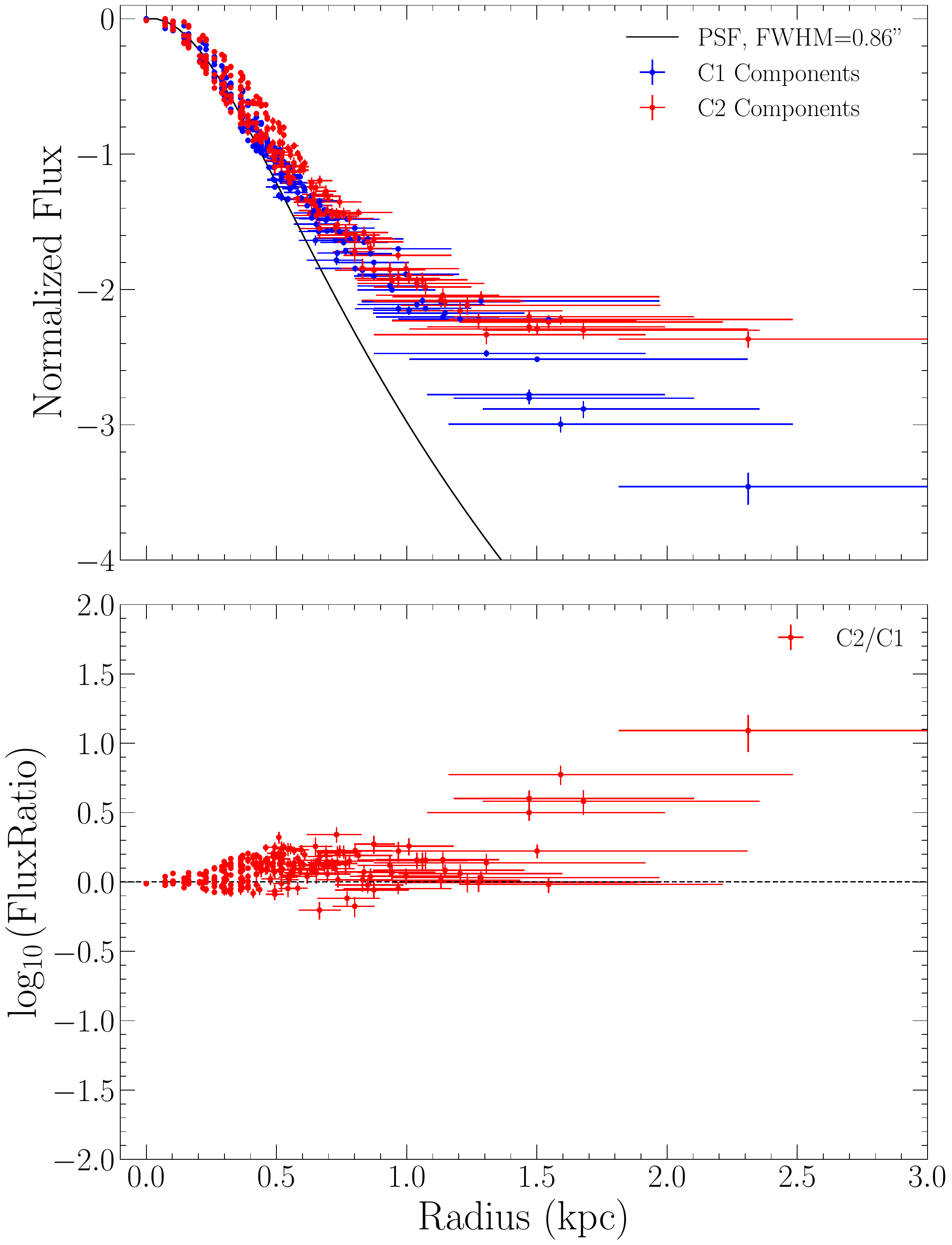}{0.3\textwidth}{(g) J1415}
\fig{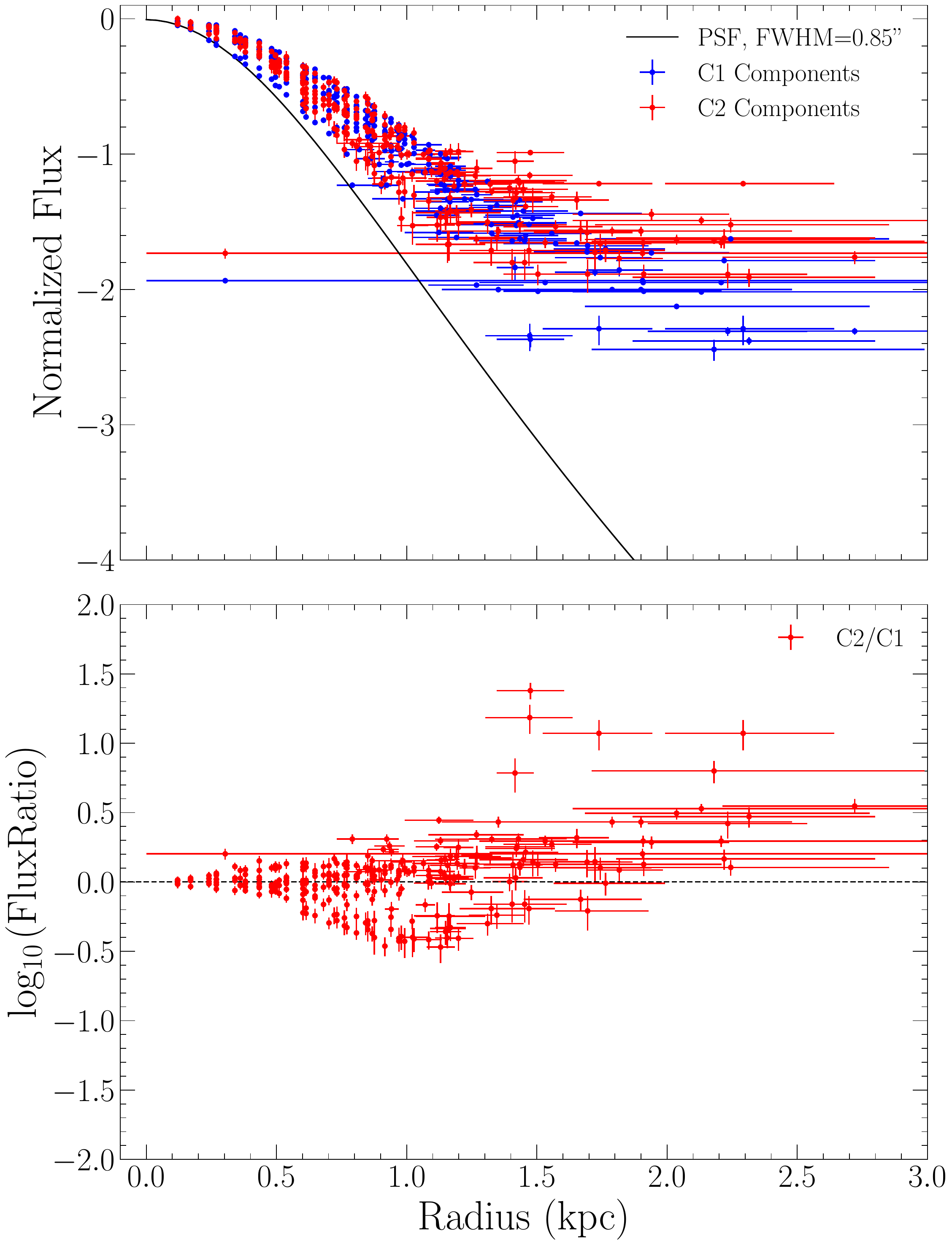}{0.3\textwidth}{(h) J1622}
\fig{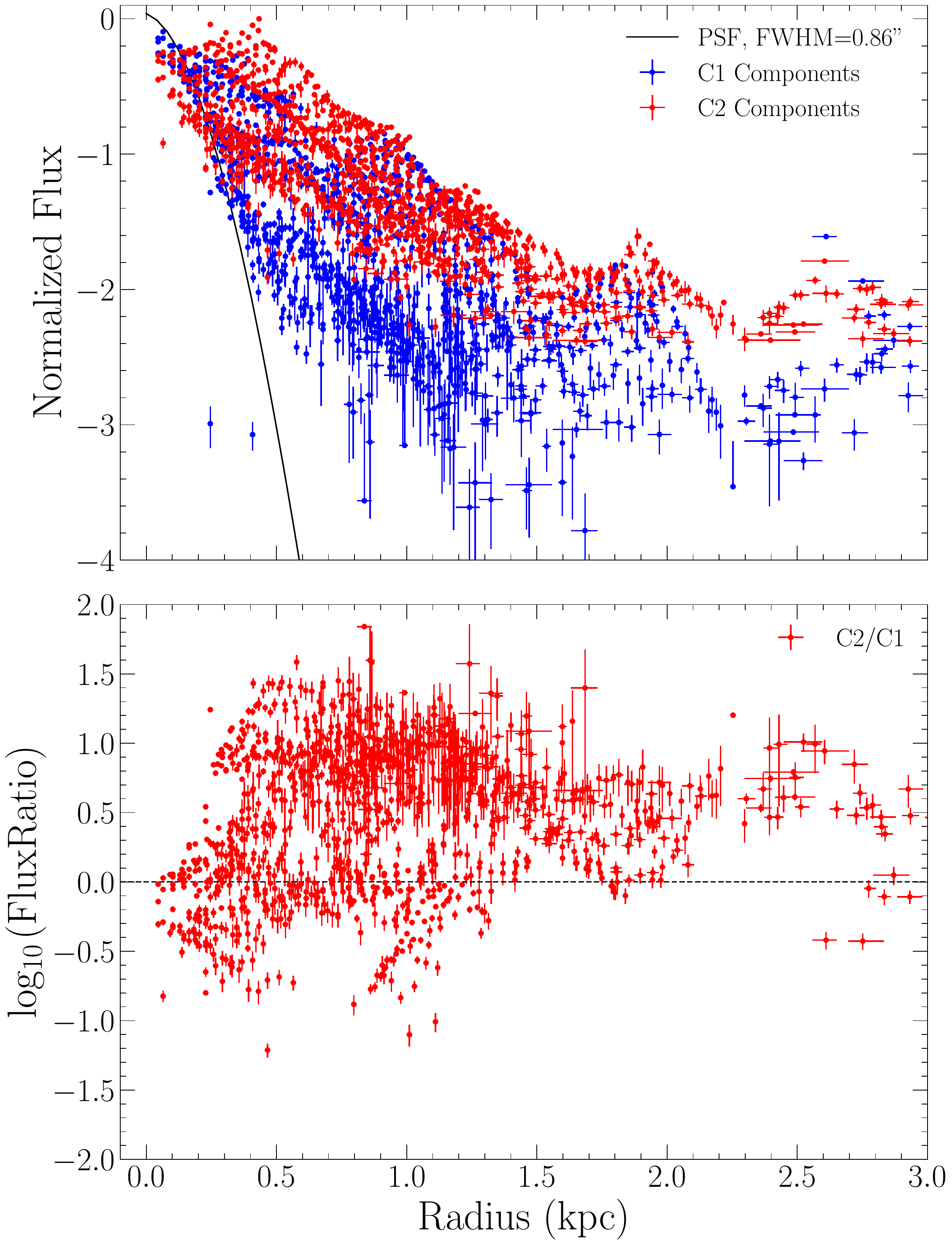}{0.3\textwidth}{(i) J1717}}
\caption{ \footnotesize Same as Figure \ref{0838_radial} but for the rest of the targets. \label{all_radial}}
\end{figure*}

The radial flux profiles of the targets are plotted in Figures \ref{0838_radial} and \ref{all_radial}. The black solid line is the PSF of a standard star observed close to the target to reduce errors due to the atmosphere. We then plot the individual velocity components and see that they are spatially resolved in the KCWI data. The bottom plot shows that the ratio of C2 to C1 flux increases at greater distances, showing that the two components have different flux distributions.

The radial flux profiles of J0850 are plotted in Figure \ref{all_radial} (a). The panels and plots are the same as mentioned in Figure \ref{0838_radial}. There seem to be two distinct streams being traced out by the C2 component, which is seen in both panels. This could be because the thin band spotted in the middle of the C2 flux maps in \ref{fig:[OIII5007]_0838} which indicates that there are two different flux values at the same distance on either side of the central spaxel used for the flux normalization. It is clearly distinct from the C1 component that traces the quiescent gas.

The radial flux profiles of J1307 are plotted in Figure \ref{all_radial} (e). The lower values of the normalized flux near the center indicate that the flux in the broad component peaks at a distance further away from the center of the galaxy. The presence of multiple sources is evident from the peaks present in the profile, which can be seen in both the components, which indicates that the broad component may also trace the multiple sources, which was not very evident in the flux and velocity maps in Figure \ref{fig:[OIII5007]_1307}. The nature of the radial profile is still spatially extended.

The radial flux profile of J1717 in Figure \ref{all_radial} (i)  is crowded due to a large number of pixels. However, the spatially extended profile is visible, and there are also multiple peaks that can be discerned in the profile. The C2 component also seems to trace the peak, which is could indicate that additional sources could also contribute to the broad component.

For the rest of the targets, the [O\,III] $\lambda$5007 flux radial profile  is clearly more extended than the PSF, which is consistent with the spatially resolved velocity gradients seen in both the ionized gas and underlying stellar population.

\newpage
\bibliography{citations}{}
\bibliographystyle{aasjournal}

\end{document}